\documentclass[pdflatex,sn-mathphys]{sn-jnl}

\usepackage{graphicx}%
\usepackage{multirow}%
\usepackage{amsmath,amssymb,amsfonts}%
\usepackage{amsthm}%
\usepackage{mathrsfs}%
\usepackage[title]{appendix}%
\usepackage{xcolor}%
\usepackage{textcomp}%
\usepackage{manyfoot}%
\usepackage{booktabs}%
\usepackage{algorithm}%
\usepackage{algorithmicx}%
\usepackage{algpseudocode}%
\usepackage{listings}%

\usepackage{amsmath, amssymb, amscd, amsthm, amsfonts, graphicx, hyperref, natbib,array, xcolor, geometry, float, scalerel, booktabs, multirow, appendix,setspace, bm, empheq, lipsum}

\jyear{2021}%

\theoremstyle{thmstyleone}%
\theoremstyle{thmstyletwo}%

\theoremstyle{thmstylethree}%

\begin{document}

\title[A dual-structured mathematical model]{
\begin{center}
    Blood lipoproteins shape the phenotype and lipid content of early atherosclerotic lesion macrophages:
\end{center}
\large\textbf{\\ A dual-structured mathematical model}}


\author*[1]{\fnm{Keith L.} \sur{Chambers}}\email{keith.chambers@maths.ox.ac.uk}

\author[2]{\fnm{Mary R.} \sur{Myerscough}}\email{mary.myerscough@sydney.edu.au}

\author[3]{\fnm{Michael G.} \sur{Watson}}\email{michael.watson1@unsw.edu.au}

\author[1,4]{\fnm{Helen M.} \sur{Byrne}}\email{helen.byrne@maths.ox.ac.uk \clearpage}

\affil*[1]{\orgdiv{Wolfson Centre for Mathematical Biology}, \orgname{Mathematical Institute, University of Oxford}, \orgaddress{\street{Andrew Wiles Building, Radcliffe Observatory Quarter, Woodstock Road}, \city{Oxford}, \postcode{OX2 6GG}, \state{Oxfordshire}, \country{United Kingdom}}}

\affil[2]{\orgdiv{School of Mathematics and Statistics}, \orgname{University of Sydney}, \orgaddress{\street{Carslaw Building, Eastern Avenue, Camperdown}, \city{Sydney}, \postcode{2006}, \state{New South Wales}, \country{Australia}}}

\affil[3]{\orgdiv{School of Mathematics and Statistics}, \orgname{University of New South Wales}, \orgaddress{\street{Anita B. Lawrence Centre, University Mall, UNSW, Kensington}, \city{Sydney}, \postcode{2052}, \state{New South Wales}, \country{Australia}}}

\affil[4]{\orgdiv{Ludwig Institute for Cancer Research}, \orgname{University of Oxford}, \orgaddress{\street{Old Road Campus Research Build, Roosevelt Dr, Headington}, \city{Oxford}, \postcode{OX3 7DQ}, \state{Oxfordshire}, \country{United Kingdom}}}

\abstract{Macrophages in atherosclerotic lesions exhibit a spectrum of 
behaviours or \textit{phenotypes}. The phenotypic distribution of monocyte-derived macrophages (MDMs), its correlation with MDM lipid content, 
and relation to blood lipoprotein densities are not well understood. 
Of particular interest is the balance between low density lipoproteins (LDL) and high density lipoproteins (HDL), which carry {\it{bad}} and {\it{good}} cholesterol respectively. 
To address these issues, we have developed a mathematical model for early atherosclerosis in which the MDM population is structured by phenotype and lipid content. The model admits a simpler, closed subsystem 
whose analysis shows how lesion composition becomes more pathological as the blood density of LDL increases relative to the HDL capacity.
%
%
We use asymptotic analysis to 
derive a power-law relationship between MDM phenotype and lipid content at steady-state. This relationship enables us to understand why, for example, lipid-laden MDMs have a more inflammatory phenotype than lipid-poor MDMs when blood LDL lipid density greatly exceeds HDL capacity.  
We show further that the MDM phenotype distribution always attains a local maximum, while the lipid content distribution may be unimodal, adopt a quasi-uniform profile or decrease monotonically. Pathological lesions exhibit a local maximum in both the phenotype and lipid content MDM distributions, with the maximum at an inflammatory phenotype and near the lipid content capacity respectively.
These results illustrate how macrophage heterogeneity arises in early atherosclerosis
and provide a framework for future model validation through comparison with single-cell RNA sequencing data. 
}

\keywords{phenotype, lipid, structured population model, atherosclerosis, discrete, continuum}



\maketitle

\clearpage
\section{Introduction}\label{sec: Intro}

Atherosclerosis is a chronic inflammatory condition of the artery wall \cite{back2019inflammation}. The disease begins with the retention of low-density-lipoprotein (LDL) particles in the artery wall. LDL particles, which carry fatty compounds called lipids, enter the artery wall from the bloodstream and are retained via interactions with extracellular matrix. Retained LDL (rLDL) particles are rapidly modified via oxidation and aggregation. The accumulation of rLDL particles triggers an immune response that attracts monocyte-derived macrophages (MDMs) to the lesion. MDMs are typically the most numerous immune cell type in early atherosclerotic lesions \cite{willemsen2020macrophage}. They play a key role in disease progression by ingesting extracellular lipid 
and offloading lipid to high-density lipoprotein (HDL) particles, which also enter the lesion from the bloodstream \cite{kloc2020role}. Importantly, MDMs may adopt a variety of phenotypes depending on their interaction with the lesion microenvironment \cite{tabas2016macrophage, back2019inflammation}. This includes inflammatory (M1-like) and resolving (M2-like) phenotypes. Over time, sustained inflammation and the death of lipid-laden MDMs may cause the lesion to transition into an atherosclerotic plaque with a large core of extracellular lipid \cite{guyton1996development, gonzalez2017macrophage}. The rupture of this plaque releases the lipid core into the bloodstream, where it promotes blood clot formation and can induce an acute clinical event. Plaque rupture is the most common cause of myocardial infarction \cite{costopoulos2017plaque} and a leading cause of ischaemic strokes \cite{rothwell2007atherothrombosis}. 
Understanding how the \textit{in vivo} distribution of MDM phenotype is influenced by MDM lipid content and  blood LDL/HDL densities are active areas of research.

Macrophages in atherosclerotic lesions exhibit a continuum of inflammatory to resolving phenotypes \cite{leitinger2013phenotypic, back2019inflammation}. This view supersedes the traditional dichotomous M1/M2 classification of macrophage phenotype; M1 and M2 polarisation now typically refer to the extremes of a phenotype continuum \cite{barrett2020macrophages}. Macrophage phenotype modulation appears to be reversible \cite{barrett2020macrophages, lin2021macrophage,wang2014molecular}, and is largely determined by the balance between (pro-)inflammatory and (pro-)resolving mediators \cite{tabas2016macrophage, back2019inflammation}. Following the classification presented in Tabas et al. \cite{tabas2016macrophage}, inflammatory mediators include cytokines such as TNF and IL-1, that are secreted by MDMs upon uptake of modLDL \cite{liu2014oxldl}, and damage associated molecular patterns (DAMPs) that are released upon the secondary necrosis of apoptotic cells \cite{sachet2017immune}. Resolving mediators include the cytokines IL-10 and IL-13, and specialised pro-resolving lipid mediators. Resolving mediators are synthesised by macrophages upon apoptotic cell uptake \cite{decker2021pro} and interaction with HDL \cite{serhan2018resolvins}. LDL and HDL promote the synthesis of opposing mediator types (inflammatory and resolving respectively) and, so, are likely to induce opposing effects on MDM phenotype. 

Mathematical models of atherosclerosis are an emerging field of study \cite{parton2016computational, avgerinos2019mathematical, cai2021mathematical, mc2022modeling}. The existing literature includes (i) models of LDL infiltration \cite{prosi2005mathematical, yang2006modeling, yang2008low}, (ii) mechanical models of plaque growth \cite{fok2012mathematical, watson2018two, fok2018media, watson2020multiphase, fok2021modeling}, and (iii) models that focus on lesion immunology. Lesion immunology has been modelled using ODEs \cite{bulelzai2012long, cohen2014athero, islam2015mathematical, thon2018quantitative, lui2021modelling, xie2022well}, spatial PDEs \cite{calvez2009mathematical,fok2012growth, hao2014ldl, chalmers2015bifurcation, mukherjee2019reaction, mohammad2020integrated, ahmed2023macrophage} and agent-based approaches \cite{corti2020fully, bayani2020spatial}. Importantly, existing models which incorporate macrophage phenotype do so via binary M1/M2 classification rather than a continuum setting \cite{friedman2015mathematical, bezyaev2020model, liu2022macrophage}. 
Macrophage lipid content is also typically treated via a binary distinction between macrophages with little internalised lipid (simply termed `macrophages') and those that are lipid-laden (termed `foam cells') \cite{calvez2009mathematical, hao2014ldl, chalmers2017nonlinear, silva2020modeling}. However, several recent studies capture gradual lipid accumulation in lesion macrophages via structured population modelling \cite{ford2019lipid, meunier2019mathematical, chambers2022lipid, chambers2023new, watson2023lipid}. 

Of particular relevance is the recent lipid-structured model of Chambers et al. \cite{chambers2023new}, which serves as the foundation of the present study. 
By extending this model to account for simultaneous variation in MDM phenotype \textit{and} lipid content, we provide a mechanistic framework to explore the diversity of lipid-associated macrophage states revealed by single-cell RNA sequencing \cite{dib2023lipid}.
Other authors have proposed dual-structured mathematical models (e.g. \cite{bernard2003analysis, doumic2007analysis, laroche2016threshold, hodgkinson2019spatio}, reviewed in \cite{kang2020nonlinear}). 
A key difference between these existing models and ours relates to the time evolution of the structure variables: in most existing models, the structure variables are independent whereas in our model their time evolution is coupled.
%
%
We use our dual-structured model to address the following questions:
\begin{enumerate}
    \item How do blood LDL/HDL levels 
    impact lesion composition? In particular: \\
    a. What are their effects on the time-evolution of lesion composition? \\
    b. How do they impact lesion composition at steady state?
    \item How are phenotype and lipid content distributed amongst MDMs? In particular:  \\
    a. How do MDM phenotype and lipid content evolve over time? \\
    b. Are MDM phenotype and lipid content correlated? \\
    c. What are the qualitative features of the phenotype and lipid content \\
    ${} \, \, \, \,  \, \,$ marginal distributions at steady state?

\end{enumerate}

The remainder of the paper is structured as follows. Sec. \ref{sec: Model development} details the model development, including the derivation of a closed subsystem and discussion of parameter values. Sec. \ref{sec: results} contains the results of our model analysis. Key questions 1 and 2 are addressed in Secs. \ref{sec: lesion composition} and \ref{sec: MDM distribution}, respectively. Finally, we discuss the results and their implications in Sec. \ref{sec: discussion}. 



\section{Model development} \label{sec: Model development}

\begin{figure}
    \centering
    \includegraphics[width = 0.92\textwidth]{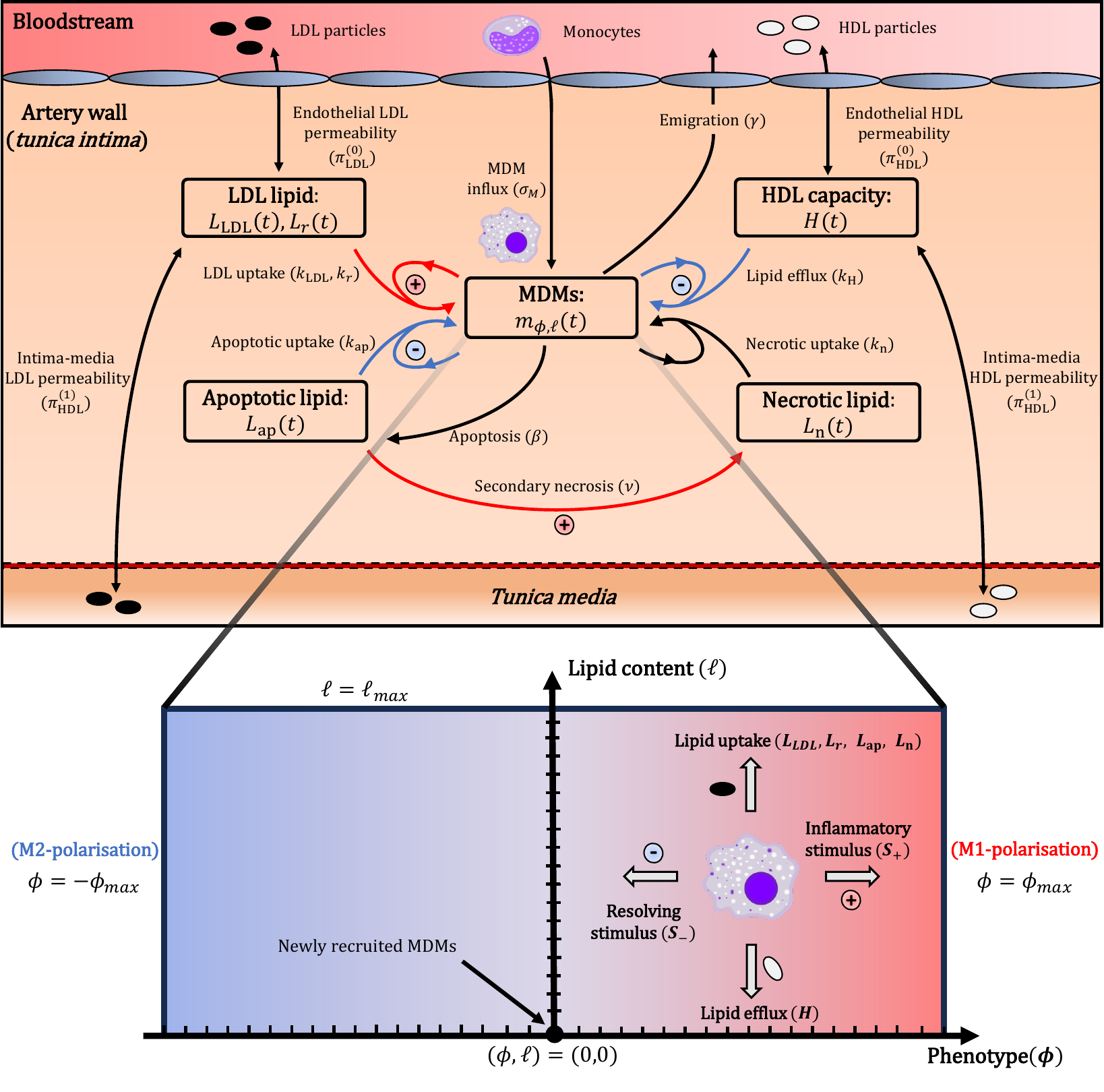}
    \caption{\textbf{A schematic of the MDM-lipid dynamics in the model.} Processes represented with blue arrows stimulate the secretion of resolving mediators by MDMs, while those represented by red arrow stimulate the emission of inflammatory mediators. The lower half illustrates the discrete phenotype-lipid structure space that underpins the MDM dynamics. LDL retention and constitutive mediator production by MDMs are not shown. \\ \\}
    \label{fig: model schematic}
    \includegraphics[width=0.92\textwidth]{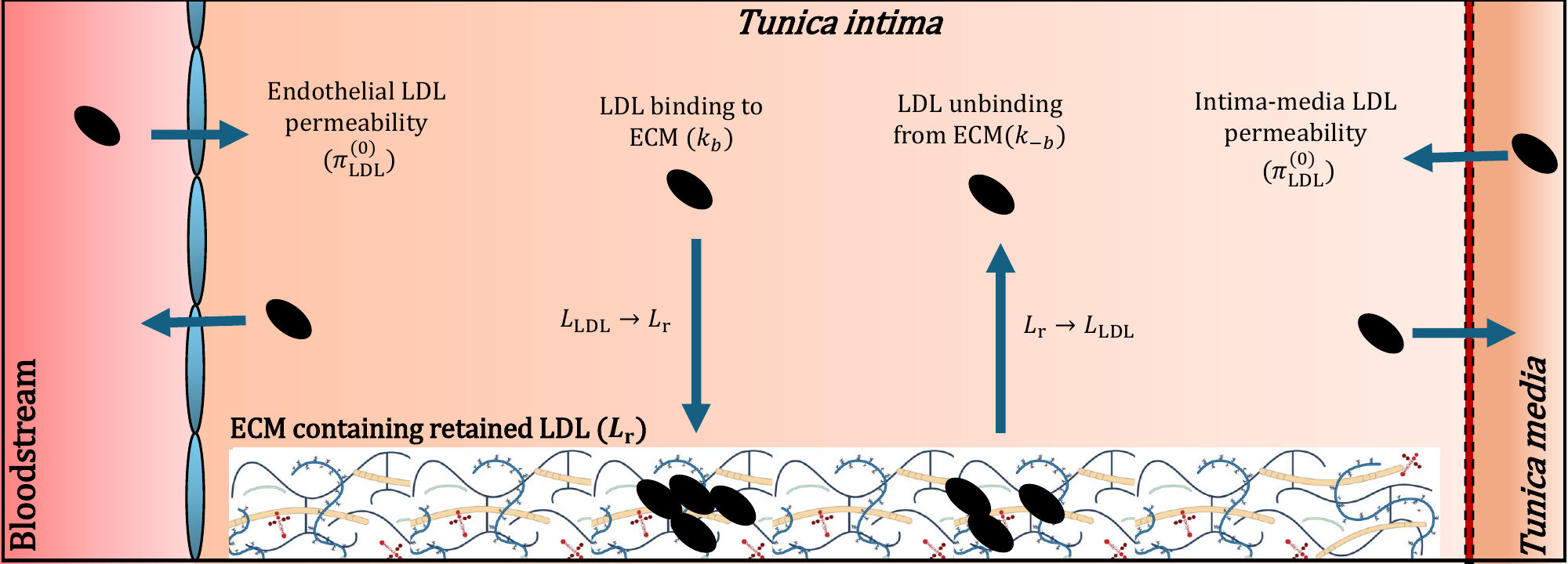}
    \caption{\textbf{Schematic of the LDL kinetics in absence of MDMs. } The model distinguishes between free LDL ($L_{\text{\scaleto{LDL}{3.5pt}}}$), which diffuses freely between the bloodstream, tunica intima and tunica media, and retained LDL ($L_\text{r}$) that is bound to the ECM within the tunica intima. We assume in Eqs. \eqref{eqn: LDL}-\eqref{eqn: rLDL} that the ECM has a finite capacity for rLDL, $K_\text{r}$, and that the rate of binding is proportional to the available capacity: $k_b (K_\text{r} - L_\text{r})$. }
    \label{fig: LDL schematic}
\end{figure}

In this section we present a phenotype-lipid dual-structured model for MDM populations in early atherosclerosis. Model schematics for the MDM dynamics and LDL retention are given in Figs. \ref{fig: model schematic} and \ref{fig: LDL schematic}, respectively.

\subsection{Assumptions and definitions}

We assume for simplicity that macrophage phenotype and lipid content change by finite increments, $\Delta \phi = 1$ and $\Delta a > 0$ respectively. Specifically, we let $m_{\phi, \ell}(t)$ denote the number density of MDMs with phenotype $\phi$ and lipid content $a_0 + \ell \Delta a$ at time $t \geq 0$. The phenotype index runs over both positive and negative integer values: $\phi = 0, \pm 1, \dots, \pm \phi_{\text{max}}$. Macrophages with $\phi > 0$ are pro-inflammatory and have an M1-like phenotype; those with $\phi < 0$ are anti-inflammatory and have an M2-like phenotype. The extreme values $\phi = \pm \phi_{\text{max}}$ can be interpreted as complete M1 and M2 polarisation. The lipid index runs over non-negative values: $\ell = 0, 1, \dots, \ell_{\text{max}},$ so that macrophage lipid content ranges from their endogenous content, $a_0 \geq 0$, to a maximum value, $ a_0 + \kappa$, where $\kappa := \ell_{\text{max}} \Delta a$ is the maximum capacity for ingested lipid.

We also introduce variables to describe the extracellular environment. We denote by $L_{\text{\scaleto{LDL}{3.5pt}}}(t) \geq 0$, $L_{\text{r}}(t) \geq 0$, $L_{\text{ap}}(t) \geq 0$ and $L_{\text{n}}(t) \geq 0$ the mass densities of free LDL lipid, retained LDL lipid, apoptotic lipid and necrotic lipid respectively. We let $H(t)$ be the lipid capacity of HDL particles in the lesion. Finally, we denote by $S_{+}(t) \geq 0$ and $S_{-}(t) \geq 0$ the mass densities of inflammatory and resolving mediators respectively.

\subsection{Model equations}

\noindent \textbf{MDMs.} We propose that the MDM population evolves according to the following ODEs:
\begin{align}
    \begin{split} \label{eqn: m_{phi, l}}
        \frac{d}{dt}m_{\phi, \ell} &= \, \, \, \, \, \underbrace{\bm{k_L} \cdot \bm{L} \, \, \big[ (\ell_\text{max}-\ell + 1) m_{\phi, \ell-1} - (\ell_\text{max} - \ell)m_{\phi, \ell} \big]}_{\text{lipid uptake}} \\
    &\quad + \underbrace{k_H H \, \, \, \, \, \big[ (\ell + 1) m_{\phi, \ell + 1} - \ell m_{\phi, \ell} \big]}_{\text{lipid efflux to HDL}} \\
    &\quad + \underbrace{k_S \chi S_{+} \big[ (\phi_\text{max}-\phi + 1)m_{\phi+1, \ell}- (\phi_\text{max}-\phi)m_{\phi, \ell} \big]}_{\text{inflammatory phenotype modulation}} \\
    &\quad + \underbrace{k_S \chi S_{-} \big[ (\phi_\text{max}+\phi + 1)m_{\phi+1, \ell}- (\phi_\text{max}+\phi)m_{\phi, \ell} \big]}_{\text{resolving phenotype modulation}} \\
    &\quad + \underbrace{R_{\phi, \ell}(t)}_{\text{recruitment}} - \underbrace{(\beta + \gamma)m_{\phi, \ell}}_{\text{apoptosis and egress}},
    \end{split}
\end{align}
and closure conditions:
\begin{align}
    m_{\phi, -1} \equiv m_{\phi, \ell_\text{max}+1} \equiv m_{-\phi_\text{max}-1, \ell} \equiv m_{\phi_\text{max}+1, \ell} \equiv 0,
\end{align}
for every $\phi = 0, \pm 1, \dots, \pm \phi_{\text{max}}$ and $\ell = 0, 1, \dots, \ell_\text{max}$. 

The first term on the right hand side of Eqs.\eqref{eqn: m_{phi, l}} accounts for lipid uptake. Following \cite{chambers2023new}, we model lipid uptake with a mass-action treatment of the following reactions:
\begin{align}
    m_{\phi, \ell}(t) + L_i(t)/\Delta a \xrightarrow[]{k_i ( \ell_\text{max} - \ell )} m_{\phi, \ell + 1}(t), \label{eqn: uptake reaction}
\end{align}
for every $(\phi, \ell)$ and $i \in \{ \text{\scaleto{LDL}{5pt}}, \text{r}, \text{ap}, \text{n} \}$. Reactions \eqref{eqn: uptake reaction} assume that the rate of lipid uptake decreases linearly with lipid content in a manner commensurate with available capacity. For notational brevity in Eqs.\eqref{eqn: m_{phi, l}}, we introduce vectors for the uptake rates, $\bm{k_L}:= ( k_{\text{\scaleto{LDL}{3.5pt}}},  k_{\text{r}}, k_\text{ap}, k_\text{n})$, and the extracellular lipids, $\bm{L} := ( L_{\text{\scaleto{LDL}{3.5pt}}},  L_{\text{r}}, L_\text{ap}, L_\text{n})$, so that $\bm{k_L}\cdot \bm{L} = k_{\text{\scaleto{LDL}{3.5pt}}} L_{\text{\scaleto{LDL}{3.5pt}}} + k_{\text{r}}L_{\text{r}} + k_\text{ap}L_\text{ap} + k_\text{n}L_\text{n}$.

The second term of the RHS of Eqs.\eqref{eqn: m_{phi, l}} accounts for lipid efflux to HDL. Again, following \cite{chambers2023new}, we treat  efflux with mass-action kinetics according to the reactions:
\begin{align}
    m_{\phi, \ell}(t) + H(t)/\Delta a \xrightarrow[]{k_H \ell} m_{\phi, \ell - 1}(t), \label{eqn: efflux reaction}
\end{align}
for every $(\phi, \ell)$. The reactions \eqref{eqn: efflux reaction} assume that the efflux rate increases linearly with lipid content. 

The third and fourth terms of the RHS of Eqs.\eqref{eqn: m_{phi, l}} account for MDM phenotype modulation by inflammatory and resolving mediators \cite{back2019inflammation}, $S_{+}$ and $S_{-}$ respectively. We model phenotype modulation via the reactions:
\begin{align} \label{eqn: S+- reactions}
    \begin{split}
        m_{\phi, \ell}(t) + S_{\pm}(t)/\Delta s \xrightarrow[]{k_S}
    \begin{cases}
        m_{\phi \pm 1, \ell}(t) & \text{w/ probability } p_\phi^{\pm} := \chi (\phi_\text{max} \mp \phi)\\
        m_{\phi, \ell}(t) & \text{w/ probability } 1 - p_\phi^{\pm},
    \end{cases}
    \end{split}
\end{align}
for every $(\phi, \ell)$. Here $k_S$ is the rate of mediator binding to MDM surface receptors and $\Delta s$ is the mediator mass bound per interaction. We assume each mediator binding interaction stimulates a move to a new phenotype class with probability: $p_\phi^{\pm} = \chi (\phi_\text{max} \mp \phi).$ The parameter $0 \leq \chi \leq (2\phi_\text{max})^{-1}$ which modulates this probability can be interpreted as the phenotypic plasticity of the MDMs. We assume that $p_\phi^{\pm}$ decreases linearly to zero as $\phi \rightarrow \pm \phi_\text{max}$, so that it becomes increasingly difficult for MDMs to become more polarised the more polarised they are. Biologically, this property reflects the saturation of intracellular signalling pathways when macrophages are continually exposed to inflammatory or resolving mediators. 

The final terms on the RHS of Eqs.\eqref{eqn: m_{phi, l}} account for MDM recruitment, apoptosis and egress. We assume that newly recruited MDMs enter the lesion from the bloodstream carrying only endogenous lipid and with an uncommitted phenotype:
\begin{align}
        R_{\phi, \ell}(t) := \begin{cases} \label{eqn: recruitment}
            \sigma_M \Big( \frac{S_{+}}{S_{+} + S_{+}^{\text{c50}} + \rho S_{-}} \Big) & \text{if } (\phi, \ell) = (0,0); \\
            0 & \text{if } (\phi, \ell) \neq (0,0).
        \end{cases}
\end{align}
The recruitment rate is a first-order Hill function of the inflammatory mediator density, $S_{+}$, which saturates at the maximum value $\sigma_M$. Resolving mediators, $S_{-}$, inhibit recruitment by linearly increasing the threshold for half-maximal recruitment from the basal value $S_{+}^{\text{c50}}$; the parameter $\rho$ governs the sensitivity of the recruitment rate to $S_{-}$. We assume for simplicity that the rates of MDM apoptosis, $\beta$, and egress, $\gamma$, are constant.\\

\noindent \textbf{Extracellular lipids. } We assume that the densities of free and retained LDL evolve according to the ODEs:
\begin{align}
    \begin{split} \label{eqn: LDL}
        \frac{d L_{\text{\scaleto{LDL}{3.5pt}}}}{dt} &= \underbrace{\pi_{L}^{(0)} (L^{(0)} - L_{\text{\scaleto{LDL}{3.5pt}}})}_{\text{lumenal flux}} - \underbrace{\pi_{L}^{(1)} (L_{\text{\scaleto{LDL}{3.5pt}}} - L^{(1)})}_{\text{tunica media flux}} \\
        &\qquad \qquad \qquad - \underbrace{k_b L_{\text{\scaleto{LDL}{3.5pt}}} (K_{\text{r}} - L_{\text{r}})}_{\text{retention via ECM binding}} + \underbrace{k_{-b} L_{\text{r}}}_{\text{unbinding}} \\
        &\qquad \qquad \qquad \qquad \qquad \qquad  - \underbrace{k_{\text{\scaleto{LDL}{3.5pt}}} \Delta a L_{\text{\scaleto{LDL}{3.5pt}}} \sum_{\phi, \ell} (\ell_\text{max} - \ell) m_{\phi, \ell}}_{\text{uptake by MDMs}},
    \end{split} \\
    \begin{split} \label{eqn: rLDL}
        \frac{d L_{\text{r}}}{dt} &= \qquad \qquad \quad + \quad   k_b L_{\text{\scaleto{LDL}{3.5pt}}} (K_{\text{r}} - L_{\text{r}}) \quad - \quad k_{-b} L_{\text{r}} \\
        &\qquad \qquad \qquad \qquad \qquad \qquad  -k_{\text{r}} \Delta a L_{\text{r}} \sum_{\phi, \ell} (\ell_\text{max} - \ell) m_{\phi, \ell}.
    \end{split}
\end{align}
The first two terms on the RHS of Eq.\eqref{eqn: LDL} account for the flux of free LDL between the lesion and the lumen/tunica media. The parameters $\pi_L^{(0)}$ and $\pi_L^{(1)}$ denote the LDL exchange rate at the endothelium and internal elastic lamina, respectively. The densities of free LDL in the lumen, $L^{(0)}$, and tunica media, $L^{(1)}$, are assumed to be non-negative constants. We assume that free LDL binds to ECM proteoglycans with rate constant $k_b$ in a capacity-limited manner, and unbinds at rate $k_{-b}$; the maximum capacity for LDL retention is $K_{\text{r}}$. The final terms on the RHS of Eqs.\eqref{eqn: LDL} and \eqref{eqn: rLDL} account for MDM uptake of free and bound LDL via reactions \eqref{eqn: uptake reaction}.

We propose that the densities of apoptotic and necrotic lipid satisfy:
\begin{align} 
    \frac{dL_{\text{ap}}}{dt} &= \underbrace{\beta \sum_{\phi, \ell} (a_0 + \ell \Delta a) m_{\phi, \ell}}_{\text{apoptosis}} \,\, - \,\,\nu L_{\text{ap}} - k_\text{ap} \Delta a L_\text{ap} \sum_{\phi, \ell} (\ell_\text{max} - \ell) m_{\phi, \ell}, \label{eqn: Lap} \\
    \frac{dL_\text{n}}{dt} &= {\color{white}\beta \sum (a_0 + \ell \Delta a) m_{\phi, \ell}} \,\, + \underbrace{\nu L_{\text{ap}}}_{\text{necrosis}} - \underbrace{k_\text{n} \Delta a L_\text{n} \sum_{\phi, \ell} (\ell_\text{max} - \ell) m_{\phi, \ell}}_{\text{uptake by MDMs}}. \label{eqn: Ln}
\end{align}
The first term of the RHS of Eq. \eqref{eqn: Lap} accounts for lipid deposition into the extracellular space due to MDM apoptosis. We assume that apoptotic cells undergo secondary necrosis at rate $\nu$, which provide linear sink and source terms in Eqs.\eqref{eqn: Lap} and \eqref{eqn: Ln}, respectively. The final terms on the RHS of Eqs. \eqref{eqn: Lap} and \eqref{eqn: Ln} describe apoptotic and necrotic lipid uptake by MDMs. \\

\noindent \textbf{HDL lipid capacity. } We assume that the lipid capacity of the HDL particles in the lesion evolves according to:
\begin{align} \label{eqn: H}
        \frac{dH}{dt} &= \underbrace{\pi_H^{(0)} (H^{(0)} - H)}_{\text{lumenal flux}} - \underbrace{\pi_H^{(1)} (H - H^{(1)})}_{\text{tunica media flux}} - \underbrace{k_H \Delta a H \sum_{\phi, \ell} \ell m_{\phi,\ell}}_{\text{lipid efflux by MDMs}}.
\end{align}
The first two terms on the RHS of Eq.\eqref{eqn: H} describe the flux of HDL lipid capacity (via HDL particle diffusion) between the lesion and the lumen/tunica media. 
For simplicity, we assume that the diffusivity of HDL particles is independent of their lipid capacity, so that $\pi_H^{(0)}$ and $\pi_H^{(1)}$ represent the common permeability of HDL particles and HDL lipid capacity at the endothelium and internal elastic lamina respectively.
The final term accounts for MDM lipid efflux to HDL particles. \\

\noindent \textbf{Inflammatory and resolving mediators. } We propose that the densities of inflammatory and resolving mediators satisfy the following ODEs:
\begin{align}
    \begin{split} \label{eqn: S+}
        \frac{dS_{+}}{dt} &= \underbrace{\alpha L_{\text{r}}}_{\substack{\text{resident} \\ \text{signals}}} +\mu \Big[ (\underbrace{k_{\text{\scaleto{LDL}{3.5pt}}} L_{\text{\scaleto{LDL}{3.5pt}}} + k_{\text{r}} L_{\text{r}})\Delta a  \sum_{\phi, \ell} (\ell_\text{max} - \ell) m_{\phi, \ell}}_{\text{LDL-stimulated emission}} + \underbrace{\nu L_{\text{ap}}}_{\text{DAMPs}} \Big] \\
        &\quad + \underbrace{ k_c \sum_{\phi, \ell} \Big( 1 + \frac{\phi}{\phi_\text{max}}\Big)m_{\phi, \ell} }_{\text{MDM constitutive production}} - \underbrace{\bigg[k_S \Delta s \sum_{\phi, \ell} m_{\phi, \ell}  + \delta_S \bigg] S_{+} }_{\text{MDM binding and decay}},
    \end{split} \\
    \begin{split} \label{eqn: S-}
        \frac{dS_{-}}{dt} &= \mu \Big[ \underbrace{k_{\text{ap}} \Delta a L_\text{ap} \sum_{\phi,\ell}(\ell_\text{max}-\ell)m_{\phi,\ell}}_{\text{efferocytosis-stimulated emission}} + \underbrace{k_H \Delta a H \sum_{\phi,\ell} \ell m_{\phi,\ell}}_{\text{HDL-stimulated emission}}  \Big]  \\ 
        &\quad + k_c \sum_{\phi, \ell} \Big( 1 - \frac{\phi}{\phi_\text{max}}\Big)m_{\phi, \ell} - \bigg[k_S \Delta s \sum_{\phi, \ell} m_{\phi, \ell}  + \delta_S \bigg] S_{-}.
    \end{split}
\end{align}

The first term on the RHS of Eq.\eqref{eqn: S+} models the release of inflammatory signals by resident cells (e.g. smooth muscle cells, tissue-resident macrophages) \cite{williams2019cytokine, williams2020limited}. The signal cascades which first stimulate MDM recruitment are, as yet, unknown, but are thought to be the result of excessive LDL retention \cite{williams2005lipoprotein}. Hence, we assume for simplicity that the resident cells produce inflammatory mediators at rate proportional to the retained LDL lipid density.

Eqs.\eqref{eqn: S+} and \eqref{eqn: S-} also account for mediator production due to lipid activity. This includes MDM production of inflammatory mediators due to LDL uptake
(both native and modified forms of LDL induce inflammatory responses in macrophages \cite{allen2022ldl, chen2015oxidized}), and MDM production of resolving mediators upon apoptotic lipid uptake \cite{dalli2012specific} and interaction with HDL \cite{serhan2018resolvins}. We also include production of inflammatory DAMPs that are released by apoptotic bodies upon secondary necrosis \cite{sachet2017immune}. We assume for simplicity that the magnitude of mediator production is proportional to the amount of lipid involved in each of the above interactions; the parameter $\mu$ represents the mediator mass produced per unit lipid. 

The remaining source terms in Eqs.\eqref{eqn: S+} and \eqref{eqn: S-} account for constitutive mediator production by MDMs. Indeed, macrophages are potent cytokine emitters (even in the absence of stimulants \cite{chen2015oxidized}) and exhibit a phenotype-dependent secretion profile \cite{kadomoto2021macrophage}. To account for these effects, we assume that MDMs constitutively produce mediators at the constant rate $2k_c$ per cell, but that the ratio of inflammatory to resolving mediator production is skewed linearly according to phenotype. M1-polarised cells with $\phi = \phi_\text{max}$ secrete only inflammatory mediators while M2-polarised cells with $\phi = -\phi_\text{max}$ emit only resolving mediators.

We further assume that mediators bind to MDM surface receptors, according to reaction \eqref{eqn: S+- reactions}, and undergo natural decay at rate $\delta_S$. We use a common decay rate for both inflammatory and resolving mediators since experimentally reported half-lives for inflammatory and resolving cytokines are comparable \cite{liu2021cytokines}. Similarly, we use a common binding rate in the absence of evidence for MDM preferential binding. \\

\noindent \textbf{Initial conditions. } We close Eqs.\eqref{eqn: m_{phi, l}}, \eqref{eqn: LDL}-\eqref{eqn: S-} by supposing that at $t = 0$:
\begin{align}
    \begin{split} \label{eqn: init dim}
        &m_{\phi, \ell} = 0 \quad \text{for every } (\phi, \ell), \\    
        &L_{\text{\scaleto{LDL}{3.5pt}}} = \frac{\pi_L^{(0)} L^{(0)} + \pi_L^{(1)} L^{(1)}}{\pi_L^{(0)} + \pi_L^{(1)}}, \quad \, L_{\text{r}} = \frac{k_b L_{\text{\scaleto{LDL}{3.5pt}}} K_{\text{r}} }{k_b L_{\text{\scaleto{LDL}{3.5pt}}} + k_{-b}}, \quad L_\text{ap} = L_\text{n} = 0, \\
        &H = \frac{\pi_H^{(0)} H^{(0)} + \pi_H^{(1)} H^{(1)}}{\pi_H^{(0)} + \pi_H^{(1)}}, \qquad S_{+} = \frac{\alpha L_{\text{r}}}{\delta_S}, \qquad \qquad \, \, \, \, \, S_{-} = 0.
    \end{split}
\end{align}
The conditions \eqref{eqn: init} describe the atherosclerotic lesion immediately prior to MDM recruitment. These expressions are derived by solving Eqs.\eqref{eqn: m_{phi, l}}, \eqref{eqn: LDL}-\eqref{eqn: S-} at steady state with $m_{\phi, \ell} = 0$ for every $(\phi, \ell)$. We find that free LDL lipid and HDL lipid capacity are balanced by their fluxes at the endothelium and internal elastic lamina. Retained LDL levels reflect a balance of binding/unbinding kinetics and directly scale inflammatory mediator levels. The remaining variables are initially zero. 

\subsection{A closed subsystem}

We can derive a closed subsystem from Eqs.\eqref{eqn: m_{phi, l}}, \eqref{eqn: LDL}-\eqref{eqn: S-} by defining the population variables:
\begin{align}
    &M(t) := \sum_{\phi, \ell} m_{\phi, \ell}(t), \label{eqn: Mdef}\\
    &\hat{\Phi}_M(t) := \frac{1}{M(t)} \sum_{\phi, \ell} \Big( \frac{\phi}{\phi_\text{max} } \Big) m_{\phi, \ell}(t), \label{eqn: Phimdef}\\
    &\hat{L}_M(t) := \frac{1}{ M(t)} \sum_{\phi, \ell} \Big( \frac{\ell}{\ell_\text{max}} \Big) m_{\phi, \ell}(t). \label{eqn: Lmdef}
\end{align}
Here $M(t) \geq 0$ represents the total MDM density. The quantity $-1 \leq \hat{\Phi}_M(t)\leq 1$ is the mean MDM phenotype, normalised such that $\hat{\Phi}_M = 1$ corresponds to M1-polarisation and $\hat{\Phi}_M = -1$ to M2-polarisation. Finally, $0 \leq \hat{L}_M(t) \leq 1$ is the mean MDM lipid content, normalised by the maximal capacity.

By differentiating definitions \eqref{eqn: Mdef}-\eqref{eqn: Lmdef} with respect to time and substituting from Eqs.\eqref{eqn: m_{phi, l}}, we obtain ODEs for $M$, $\hat{\Phi}_M$ and $\hat{L}_M$:
\begin{align}
    \frac{dM}{dt} &= \bigg(\frac{\sigma_M S_+}{S_+ + S_+^{\text{c50}} + \rho S_-}\bigg) - (\beta + \gamma) M, \label{eqn: M}\\
    \frac{d \hat{\Phi}_M}{dt} &= k_s \chi \big[ S_+ (1 - \hat{\Phi}_M)  - S_- (1+\hat{\Phi}_M) \big] - \bigg( \frac{\sigma_M S_+}{S_+ + S_+^{\text{c50}} + \rho S_-} \bigg) \frac{\hat{\Phi}_M}{M}, \label{eqn: Phim} \\
    \frac{d\hat{L}_M}{dt} &= \bm{k_L}\cdot \bm{L} \, \,  \, (1 - \hat{L}_M) \,   - k_H H \hat{L}_M \qquad -  \bigg( \frac{\sigma_M S_+}{S_+ + S_+^{\text{c50}} + \rho S_-} \bigg) \frac{\hat{L}_M}{M}. \label{eqn: Lm}
\end{align}
Sink terms arise in Eqs.\eqref{eqn: Phim} and \eqref{eqn: Lm} because recruited MDMs enter the lesion with $\phi = \ell = 0$, reducing the mean (absolute) phenotype and lipid content. We can also rewrite Eqs.\eqref{eqn: LDL}-\eqref{eqn: S-} as follows:
\begin{align}
\begin{split} \label{eqn: Lldl sub}
        \frac{d L_{\text{\scaleto{LDL}{3.5pt}}}}{dt} &= \quad \pi_L^{(0)}(L^{(0)} - L_{\text{\scaleto{LDL}{3.5pt}}}) - \pi_L^{(1)}(L_{\text{\scaleto{LDL}{3.5pt}}} - L^{(1)})  \\
        &\quad - k_b L_{\text{\scaleto{LDL}{3.5pt}}} (K_{\text{r}} - L_{\text{r}})  \quad \, \, \, \, + k_{-b}L_{\text{r}} \qquad - k_{\text{\scaleto{LDL}{3.5pt}}} \kappa L_{\text{\scaleto{LDL}{3.5pt}}} M(1-\hat{L}_M),
    \end{split} \\
    \begin{split} \label{eqn: Lr sub}
        \frac{d L_{\text{r}}}{dt} &= \quad k_b L_{\text{\scaleto{LDL}{3.5pt}}} (K_{\text{r}} - L_{\text{r}})  \quad \, \, \,  - k_{-b}L_{\text{r}} \qquad \,   - k_{\text{r}} \kappa L_{\text{r}} M(1-\hat{L}_M),
    \end{split} \\
    \frac{dL_\text{ap}}{dt} &= \quad \beta M (1 + \kappa \hat{L}_M) \quad \, \, \, \, \, \, \, \, - \nu L_\text{ap} \qquad \, \, \, \,  - k_\text{ap} \kappa L_\text{ap} M (1 - \hat{L}_M), \label{eqn: Lap sub} \\
    \frac{dL_\text{n}}{dt} &= \qquad \qquad \qquad \qquad \qquad \, \, \, \, \,\nu L_\text{ap} \qquad \, \, \, \, - k_\text{n} \kappa L_\text{n} M(1-\hat{L}_M), \label{eqn: Ln sub} \\
    \frac{dH}{dt} &= \quad \pi_H^{(0)}(H^{(0)} - H) - \pi_L^{(1)}(H - H^{(1)})  \,\, - k_H\kappa H M \hat{L}_M, \label{eqn: H sub} \\
    \begin{split}
        \frac{dS_+}{dt} &= \quad  \alpha L_\text{r} \, \, \, + \mu \Big[ (k_{\text{\scaleto{LDL}{3.5pt}}}  L_{\text{\scaleto{LDL}{3.5pt}}} + k_\text{r} L_\text{r}) \kappa M(1-\hat{L}_M) + \nu L_\text{ap} \Big]  \\
        &\quad + k_c M (1 + \hat{\Phi}_M) - (k_S \Delta s M + \delta_S) S_{+}, \label{eqn: S+ sub}
    \end{split} \\
    \begin{split}
        \frac{dS_-}{dt} &= \quad \qquad \quad \, \, \, \mu \Big[ k_\text{ap} \kappa  L_\text{ap} M (1 - \hat{L}_M) \quad \, +  k_H \kappa H M \hat{L}_M \, \,  \Big] \\
        &\quad + k_c M (1 - \hat{\Phi}_M) - (k_S \Delta s M + \delta_S)S_{-}. \label{eqn: S- sub}
    \end{split}
\end{align}
Eqs.\eqref{eqn: M}-\eqref{eqn: S- sub} with the initial conditions \eqref{eqn: init dim} comprise a closed subsystem that can be solved independently of Eq.\eqref{eqn: m_{phi, l}}.

\subsection{Parameter values}

\begin{table}[]
    \caption{Model parameters} 
    \centering
    \begin{tabular}{cp{3.8cm}lp{1.8cm}}  
    \toprule
    Parameter    & Interpretation & Estimate & Source \\
    \midrule
    $L^{(0)}$ & Lumen LDL lipid density & $0-440$ mg/dL & \cite{lee2012characteristics, orlova1999three}  \\
    $H^{(0)}$ & Lumen HDL lipid capacity & $0-230$ mg/dL & \cite{madsen2017extreme} \\ 
    $\pi_L^{(0)}$ & Endothelial LDL exchange rate & $1.5$ month$^{-1}$ & \cite{nielsen1996transfer, holzapfel2005determination}\\
    $\pi_H^{(0)}$ & Endothelial HDL exchange rate & $3.0$ month$^{-1}$ &  \cite{stender1981transfer}, $\approx 2\pi_L^{(0)}$ \\
    $L^{(1)}$ & Tunica media LDL lipid density & $0$ mg/dL & \cite{smith1982plasma}  \\
    $H^{(1)}$ & Tunica media HDL lipid capacity & $0$ mg/dL &   \cite{smith1982plasma}, est. \\ 
    $\pi_L^{(1)}$ & Internal elastic lamina LDL exchange rate & $4.5$ month$^{-1}$ & \cite{penn1994relative}, $\approx 3 \pi_L^{(0)}$\\
    $\pi_H^{(1)}$ & Internal elastic lamina HDL exchange rate & $9.0$ month$^{-1}$ &  \cite{penn1994relative}, $\approx 3 \pi_H^{(0)}$ \\
    $K_r$ & LDL retention capacity & $15-7500$ mg/dL & \cite{wight2018role, guyton1989lipid, liu2023co}  \\
    $k_b$ & LDL retention rate & $0.008$ dL/mg month$^{-1}$ & \cite{bancells2009high, smith1982plasma}  \\
    $k_{-b}$ & LDL unbinding rate & $1.8$ month$^{-1}$ & \cite{bancells2009high, smith1982plasma} \\
    $\sigma_M$ & Maximum MDM entry rate & $19000$ mm$^{-3}$ month$^{-1}$ & \cite{williams2009transmigration, nelson2014immunobiology, lee2019sirt1} \\    
    $\beta$ & MDM apoptosis rate & $1.0$ month$^{-1}$ & \cite{yona2013fate, williams2020limited} \\
    $\gamma$ & MDM egress rate & $0.2$ month$^{-1}$ & \cite{williams2018limited, lee2019sirt1}, est. \\
    $\nu$ & Secondary necrosis rate & $37$ month$^{-1}$ & \cite{saraste2000morphologic} \\
    $a_0$ & MDM endogenous lipid & $55$pg &  \cite{sokol1991changes, cooper2022cell}\\
    $\kappa$ & MDM lipid capacity & $1600$pg &  \cite{ford2019efferocytosis}, $\approx 29 a_0$ \\
    $k_{\text{\scaleto{LDL}{3.5pt}}}$ & LDL lipid uptake rate & $0.00034$ dL/mg month$^{-1}$ &  \cite{sanda2021aggregated} \\
    $k_{\text{r}}$ & rLDL lipid uptake rate & $0.023$ dL/mg month$^{-1}$ &  \cite{sanda2021aggregated} \\
    $k_{\text{ap}}$ & Apoptotic lipid uptake rate & $0.060$ dL/mg month$^{-1}$ &  \cite{taruc2018quantification, schrijvers2005phagocytosis} \\
    $k_{\text{n}}$ & Necrotic lipid uptake rate & $0.015$ dL/mg month$^{-1}$ &  \cite{brouckaert2004phagocytosis}, $\approx  k_{\text{ap}}/4$ \\
    $k_{H}$ & Lipid efflux rate & $0.34$ dL/mg month$^{-1}$ &  \cite{kritharides1998cholesterol, woudberg2018pharmacological} \\
    $S_{+}^{\text{c50}}$ & Mediator density for half-maximum MDM recruitment & $5$ ng/mL &  \cite{o2015pro, pugin1993lipopolysaccharide} \\
    $\rho$ & Sensitivity of MDM recruitment to resolving mediators  & $0.40$ &  \cite{sha2015interleukin} \\
    $\delta_S$ & Mediator natural decay rate  & $1600$ month$^{-1}$ &  \cite{liu2021cytokines} \\
    $\Delta s$ & Mediator mass  & $17$ kDa &  \cite{atzeni2013tumor} \\
    $k_s$ & Mediator binding rate to MDM surface receptors  & $8800$ mL/ng month$^{-1}$ &  \cite{watanabe1988continuous, niitsu1988analysis} \\
    $k_c$ & Constitutive mediator production rate by MDMs  & $1.5$ pg month$^{-1}$ &  \cite{schutte2009cytokine} \\
    $\mu$ & Mediator production per unit lipid stimulus  & $0.042$ &  \cite{schutte2009cytokine} \\
    $\alpha$ & Resident inflammatory mediator production per rLDL lipid  & $3.9 \times 10^{-5}$ month$^{-1}$ & \cite{williams2020limited}   \\
    $\chi$ & MDM phenotypic plasticity  & $6.0\times 10^{-6}$ &  \cite{tarique2015phenotypic} \\
    $\Delta a$ & MDM uptake/efflux increment of lipid & $16$ pg &  \cite{kontush2007preferential, taefehshokr2021rab}, \, \, \, \, $\in(75\text{kDa}, a_0)$ \\
    $\ell_\text{max}$ & Maximum MDM lipid capacity per uptake/efflux increment & $100$ &  $= \kappa/\Delta a$ \\
    $\phi_\text{max}$ & Half the maximum number of MDM phenotype classes & $50$ & $< (2\chi)^{-1}$ \\
    \bottomrule
    \end{tabular}
        \label{tab:parameters}
\end{table}

The parameters that appear in Eqs. \eqref{eqn: m_{phi, l}},\eqref{eqn: M}-\eqref{eqn: S- sub} are summarised in Table \ref{tab:parameters}. 


The blood densities of LDL lipid, $L^{(0)}$, and HDL lipid capacity, $H^{(0)}$, are key parameters in our model. Since these quantities are sensitive to modifiable lifestyle factors such as diet and exercise \cite{schoeneck2021effects}, we explore a range of plausible values in our analysis. The range for $L^{(0)}$ is based on the human serum LDL cholesterol distribution reported in \cite{lee2012characteristics}, in which $99.8\%$ of the subjects had LDL cholesterol below $280$mg/dL. We multiply this figure by $78/50$ to account for LDL phospholipid to obtain an upper estimate of $480$mg/dL \cite{orlova1999three}. 
We take $H^{(0)} = 230$mg as a conservative upper bound, which exceeds the highest recorded HDL cholesterol concentrations (~193mg/dL) in a sample of 116508 individuals from the general population \cite{madsen2017extreme}.


We consider a range of values for the LDL retention capacity, $K_r$, since it varies between artery wall sections  \cite{lewis2023capacity}. As LDL retention is driven by LDL-proteoglycan binding, we estimate $K_r$ by considering the artery wall proteoglycan density. Artery wall extracellular matrix prior to atherosclerosis-induced collagen degradation consists of $~4\%$ proteoglycan and $~40\%$ collagen \cite{wight2018role}. Using $0.01-3$ mg/mL as an estimate for the collagen density ($0.75$ mg/mL is used to replicate the tunica intima in culture models \cite{liu2023co}), we obtain a proteoglycan density of $0.001-0.3$ mg/mL. Assuming each proteoglycan molecule (est. mass $~800$kDa \cite{yoneda2002biosynthesis}) can support a single aggregate of LDL (typical diameter $75$nm, corresponding to $0.0002$ pg lipid for spherical droplets \cite{guyton1989lipid}), we estimate $K_\text{r} \approx 15-4500$ mg/dL.

The values of the remaining parameters are fixed. See Appendix \ref{sec: appendix parameters} for further details on these choices of parameter values.


\subsection{Non-dimensionalisation}

We recast the model in terms of the following dimensionless variables:
\begin{align}
    \begin{split} \label{eqn: nondim}
        &\Tilde{t} := \beta t, \qquad \qquad \qquad \quad \Tilde{m}_{\phi, \ell}(\Tilde{t}) := \frac{\beta}{\sigma_M} m_{\phi, \ell}(t), \quad \, \, \, \, \, \Tilde{M}(\Tilde{t}) := \frac{\beta}{\sigma_M} M(t), \\
        &\Tilde{\bm{L}}(\Tilde{t}) := \frac{\beta}{a_0 \sigma_M} \bm{L}(t), \quad \, \, \, \, \, \Tilde{H}(\Tilde{t}) := \frac{\beta}{a_0 \sigma_M} H(t), \qquad \quad   \Tilde{S}_{\pm}(\Tilde{t}) := \frac{1}{S_{+}^\text{c50}} S_{\pm}(t).
    \end{split}
\end{align}
This scaling measures time in units of mean MDM lifespan, $\beta^{-1} \approx 1 \text{month}$, and MDM densities relative to the maximum influx per MDM lifespan, $\sigma_M \beta^{-1} \approx 17000$ mm$^{-3}$. We express the extracellular lipid densities and HDL lipid capacity relative to the maximum influx of MDM endogenous lipid per MDM lifespan, $a_0 \sigma_M \beta^{-1} \approx 45$ mg/dL. Mediator densities are measured relative to the density for half-maximal MDM recruitment, $S_{+}^{\text{c50}} \approx 5$ ng/mL. We also introduce a number of dimensionless parameters in Table \ref{tab: nondim parameters}.

\begin{table}[]
    \caption{Dimensionless parameters in the rescaled Eqs. \eqref{eqn:mphil nondim}-\eqref{eqn: S- nondim}}. 
    \centering
    \begin{tabular}{clp{6cm}c}  
    \toprule
    Parameter & Definition  & Interpretation & Estimate \\
    \midrule
    $\Tilde{L}^\star$ & $\frac{\beta }{a_0 \sigma_M} L^{(0)}  $ & Lumen LDL lipid density & $0-10$ \\
    $\Tilde{H}^\star$ & $\frac{ \beta }{a_0 \sigma_M} H^{(0)}$ & Lumen HDL lipid capacity & $0-5$ \\
    $\Tilde{{K}}_\text{r}$ & $\frac{\beta}{a_0 \sigma_M} K_r $ & LDL retention capacity & $0.3-100$ \\ 
    $\Tilde{\pi}_L^{(0)}$ & $\frac{1}{\beta} \pi_{L}^{(0)}$ & Endothelial LDL exchange rate & $1.5$ \\
    $\Tilde{\pi}_H^{(0)}$ & $\frac{1}{\beta} \pi_{H}^{(0)}$ & Endothelial HDL exchange rate & $3.0$ \\
    $\Tilde{\pi}_L^{(1)}$ & $\frac{1}{\beta} \pi_{L}^{(1)}$ & Internal elastic lamina LDL exchange rate & $4.5$ \\
    $\Tilde{\pi}_H^{(1)}$ & $\frac{1}{\beta} \pi_{H}^{(1)}$ & Internal elastic lamina HDL exchange rate & $9.0$ \\
    $\Tilde{k}_b$ & $\frac{a_0 \sigma_M}{\beta^2} k_b$ & LDL retention rate & $2.7$ \\
    $\Tilde{k}_{-b}$ & $\frac{1}{\beta} k_{-b}$ & LDL unbinding rate & $1.8$ \\
    $\Tilde{\gamma}$ & $\frac{1}{\beta} \gamma$ & MDM egress rate & $0.2$ \\
    $\Tilde{\nu}$ & $\frac{1}{\beta} \nu$ & Secondary necrosis rate & $37$ \\
    $\Tilde{\kappa}$ & $\frac{1}{a_0} \kappa$ & MDM lipid capacity per unit endogenous lipid & $29$ \\
    $\Tilde{k}_{\text{\scaleto{LDL}{3.5pt}}}$ & $\frac{a_0 \sigma_M}{\beta^2} k_{\text{\scaleto{LDL}{3.5pt}}}$ & LDL uptake rate & $0.016$ \\
    $\Tilde{k}_{\text{r}}$ & $\frac{a_0 \sigma_M}{\beta^2} k_{\text{r}}$ & rLDL uptake rate & $1.1$ \\
    $\Tilde{k}_{\text{ap}}$ & $\frac{a_0 \sigma_M}{\beta^2} k_{\text{ap}}$ & Apoptotic lipid uptake rate & $5.5$ \\
    $\Tilde{k}_{\text{n}}$ & $\frac{a_0 \sigma_M}{\beta^2} k_{\text{n}}$ & Necrotic lipid uptake rate & $1.4$ \\
    $\Tilde{k}_H$ & $\frac{a_0 \sigma_M}{\beta^2} k_{H}$ & Lipid efflux rate & $16$ \\
    $\Tilde{{\rho}}$ & $\rho$ & MDM influx sensitivity to resolving mediators & $0.4$ \\
    $\Tilde{k}_S$ & $\frac{\sigma_M \Delta_S}{\beta^2}  k_S $  & Mediator binding rate to MDM receptors & $47$ \\
    $\Tilde{\delta}_S$ & $\frac{1}{\beta} \delta_S$  & Mediator natural decay rate & $1600$ \\
    $\Tilde{k}_c$ & $\frac{\sigma_M}{S_{+}^{\text{c50}}\beta^2} k_c$  & Constitutive MDM mediator production rate & $5100$ \\
    $\Tilde{\mu}$ & $\frac{a_0 \sigma_M}{S_{+}^{\text{c50}}\beta} \mu$  & Lipid-stimulated mediator production & $9200$ \\
    $\Tilde{\alpha}$ & $\frac{a_0 \sigma_M}{S_{+}^{\text{c50}}\beta^2} \alpha$  & Resident mediator production per rLDL lipid & $8.5$ \\
    $\Tilde{\chi}$ & $\frac{k_S S_{+}^{\text{c50}}}{\sigma_M} \chi$  & MDM phenotypic plasticity & $0.28$ \\
    $\phi_\text{max}$ & $\phi_\text{max}$ & MDM phenotype resolution & $50$ \\
    $\ell_\text{max}$ & $\ell_\text{max}$ & MDM maximum lipid capacity & $100$ \\
    \bottomrule
    \end{tabular} \label{tab: nondim parameters}
\end{table}

Applying the non-dimensionalisation \eqref{eqn: nondim} and definitions of Table \ref{tab: nondim parameters}, and dropping the tildes for notational convenience, we obtain the following dimensionless ODEs for the MDM population:
\begin{align}
    \begin{split} \label{eqn:mphil nondim}
        \frac{d}{dt}m_{\phi, \ell} &= \, \, \, \, \, \bm{k_L}\cdot \bm{L} \big[ (\ell_\text{max} - \ell + 1) m_{\phi, \ell - 1} - (\ell_\text{max} - \ell) m_{\phi, \ell} \big] \\
        &\quad + k_H  H \, \, \, \big[ (\ell + 1) m_{\phi, \ell + 1} - \ell m_{\phi, \ell} \big] \\
        &\quad + \, \chi  S_{+} \, \, \, \, \big[ (\phi_\text{max} - \phi + 1) m_{\phi - 1, \ell} - (\phi_\text{max} - \phi)m_{\phi, \ell} \big] \\
        &\quad + \, \chi  S_{-} \, \, \, \, \big[ (\phi_\text{max} + \phi + 1) m_{\phi + 1, \ell}- (\phi_\text{max} + \phi) m_{\phi, \ell} \big] \\
        &\quad + R_{\phi, \ell} - (1 + \gamma) m_{\phi, \ell}, 
    \end{split} 
\end{align}
where:
\begin{align}
    &R_{\phi, \ell} := \begin{cases} \label{eqn: R nondim}
            \frac{S_{+}}{S_{+} + 1 + \rho S_{-}} & \text{if } (\phi, \ell) = (0,0); \\
            0 & \text{if } (\phi, \ell) \neq (0,0),
        \end{cases} \\
    & m_{\phi, -1} \equiv m_{\phi, \ell_\text{max}+1} \equiv m_{-\phi_\text{max}-1, \ell} \equiv m_{\phi_\text{max}+1, \ell} \equiv 0,
\end{align}
for every $\phi = 0, \pm 1, \dots, \pm \phi_\text{max}$ and $\ell = 0, 1, \dots, \ell_\text{max}$. The remaining variables solve the closed subsystem:
\begin{align}
    \frac{dM}{dt} &= \frac{S_{+}}{S_{+} + 1 + \rho S_{-}} - (1 + \gamma) M, \label{eqn:M nondim} \\
    \frac{d \hat{\Phi}_M}{dt} &= \chi \big[ S_{+}(1 - \hat{\Phi}_M) - S_{-}(1 + \hat{\Phi}_M) \big] \, \,  - \Big( \frac{S_{+}}{S_{+} + 1 + \rho S_{-}} \Big) \frac{\hat{\Phi}_M}{M}, \label{eqn:Phim nondim} \\
    \frac{d \hat{L}_M}{dt} &= \bm{k_L}\cdot \bm{L} (1 - \hat{L}_M) - k_H H \hat{L}_M \qquad -  \Big( \frac{S_{+}}{S_{+} + 1 + \rho S_{-}} \Big) \frac{\hat{L}_M}{M}, \label{eqn:Lm nondim} \\
    \begin{split} \label{eqn: Lldl nondim}
        \frac{d L_{\text{\scaleto{LDL}{3.5pt}}}}{dt} &= \quad \pi_L^{(0)}(L^\star - L_{\text{\scaleto{LDL}{3.5pt}}}) \quad \, \, \, \, \, - \pi_L^{(1)}L_{\text{\scaleto{LDL}{3.5pt}}}  \\
        &\quad \, - k_b L_{\text{\scaleto{LDL}{3.5pt}}} (K_{\text{r}} - L_{\text{r}})  \quad \, \, \, + k_{-b}L_{\text{r}} \qquad - k_{\text{\scaleto{LDL}{3.5pt}}} \kappa L_{\text{\scaleto{LDL}{3.5pt}}} M(1-\hat{L}_M),
    \end{split} \\
    \begin{split} \label{eqn: Lr nondim}
        \frac{d L_{\text{r}}}{dt} &= \quad k_b L_{\text{\scaleto{LDL}{3.5pt}}} (K_{\text{r}} - L_{\text{r}})  \quad \, \, \,  - k_{-b}L_{\text{r}} \qquad \,   - k_{\text{r}} \kappa L_{\text{r}} M(1-\hat{L}_M),
    \end{split} \\
    \frac{dL_\text{ap}}{dt} &= \quad M (1 + \kappa \hat{L}_M) \qquad \quad - \nu L_\text{ap} \qquad \, \, \, \,  - k_\text{ap} \kappa L_\text{ap} M (1 - \hat{L}_M), \label{eqn: Lap nondim} \\
    \frac{dL_\text{n}}{dt} &= \qquad \qquad \qquad \qquad \qquad \, \, \, \, \,\nu L_\text{ap} \qquad \, \, \, \, - k_\text{n} \kappa L_\text{n} M(1-\hat{L}_M), \label{eqn: Ln nondim} \\
    \frac{dH}{dt} &= \quad \pi_H^{(0)}(H^{\star} - H) \qquad \, \, \, \, \, - \pi_L^{(1)}H  \qquad - k_H\kappa H M \hat{L}_M, \label{eqn: H nondim} \\
    \begin{split}
        \frac{dS_+}{dt} &= \quad  \alpha L_\text{r} \, \, \, + \mu \Big[ (k_{\text{\scaleto{LDL}{3.5pt}}}  L_{\text{\scaleto{LDL}{3.5pt}}} + k_\text{r} L_\text{r}) \kappa M(1-\hat{L}_M) + \nu L_\text{ap} \Big]  \\
        &\quad \, + k_c M (1 + \hat{\Phi}_M) - (k_S  M + \delta_S) S_{+}, \label{eqn: S+ nondim}
    \end{split} \\
    \begin{split}
        \frac{dS_-}{dt} &= \quad \qquad \quad \, \, \, \mu \Big[ k_\text{ap} \kappa  L_\text{ap} M (1 - \hat{L}_M) \quad \, +  k_H \kappa H M \hat{L}_M \, \,  \Big] \\
        &\quad \,  + k_c M (1 - \hat{\Phi}_M) - (k_S  M + \delta_S)S_{-}. \label{eqn: S- nondim}
    \end{split}
\end{align}
Finally, we assume that at $t = 0$:
\begin{align}
    \begin{split} \label{eqn: init}
        &m_{\phi, \ell} = 0 \quad \text{for every } (\phi, \ell), \\    
        &L_{\text{\scaleto{LDL}{3.5pt}}} = \frac{\pi_L^{\star} L^{\star}}{\pi_L^{(0)} + \pi_L^{(1)}}, \qquad \quad \, L_{\text{r}} = \frac{k_b L_{\text{\scaleto{LDL}{3.5pt}}} K_{\text{r}} }{k_b L_{\text{\scaleto{LDL}{3.5pt}}} + k_{-b}}, \qquad L_\text{ap} = L_\text{n} = 0, \\
        &H = \frac{\pi_H^{(0)} H^{\star}}{\pi_H^{(0)} + \pi_H^{(1)}}, \qquad \qquad \, \, S_{+} = \frac{\alpha L_{\text{r}}}{\delta_S}, \qquad \qquad \qquad  S_{-} = 0.
    \end{split}
\end{align}

\subsection{Numerical solutions and timescales} \label{sec: reduction}

The dimensionless parameters in Table \ref{tab: nondim parameters} span several orders of magnitude. In particular, the mediator parameters $\delta_S$, $k_S$ and $\mu$ are considerably larger than the other constants. Numerical solutions of Eqs.\eqref{eqn:mphil nondim}-\eqref{eqn: S- nondim} consequently require small timesteps to maintain stability. We address this issue in our numerical solutions, computed with Wolfram Mathematica, by using the routine \textit{NDSolve} with the ``StiffnessSwitching" option. 

Another way to reduce numerical stiffness is to approximate the mediator dynamics via separation of timescales. With $\delta_S^{-1} \ll 1$ and assuming that $k_S$, $\mu = \mathcal{O}(\delta_S)$, it is straightforward to show that $S_{\pm}$ satisfy the following uniformly-valid quasi-steady state approximations:
\begin{align}
    \begin{split}
        S_{+} &\sim \hat{\alpha} L_\text{r} + \hat{\mu} \big[ ((k_{\text{\scaleto{LDL}{3.5pt}}}  L_{\text{\scaleto{LDL}{3.5pt}}} + k_\text{r} L_\text{r}) \kappa M(1-\hat{L}_M) + \nu L_\text{ap} \big] \\
        &\qquad \quad + \hat{k}_c M(1 + \hat{\Phi}_M),
    \end{split} \label{eqn: S+ approx} \\
    S_{-} &\sim  \, \hat{\mu} \kappa M \big[ k_\text{ap} L_\text{ap}(1-\hat{L}_M) + k_H H \hat{L}_M \big]  + \hat{k}_c M(1 - \hat{\Phi}_M), \, \text{as } \delta_S^{-1} \rightarrow 0, \label{eqn: S- approx}
\end{align}
where $\hat{\alpha} := \alpha/\delta_S$, $\hat{\mu} := \mu/\delta_S$ and $\hat{k}_c := k_c/\delta_S$. Although approximations \eqref{eqn: S+ approx}-\eqref{eqn: S- approx} are not used for the simulations presented in Sec. \ref{sec: results}, they reveal that, at leading order, the mediator densities are proportional to their net rates of production.

\section{Results} \label{sec: results}

We present the results in two sections. In Sec. \ref{sec: lesion composition} we analyse the subsystem \eqref{eqn:M nondim}-\eqref{eqn: S- nondim} to generate insight into lesion composition. In Sec. \ref{sec: MDM distribution} we focus on the MDM phenotype-lipid distribution, $m_{\phi, \ell}$. 

\subsection{Lesion composition} \label{sec: lesion composition}

We begin our analysis of lesion composition by computing time-dependent solutions of the subsystem \eqref{eqn:M nondim}-\eqref{eqn: S- nondim}; the results are presented in Sec. \ref{sec: composition dynamics}. In Sec. \ref{sec: composition steady} we then analyse the impact of the key parameters: $L^\star$, $H^\star$ and $K_\text{r}$ on the steady state values.

\subsubsection{Time evolution} \label{sec: composition dynamics}

\begin{figure}
    \centering
    \includegraphics[width = 0.9\textwidth]{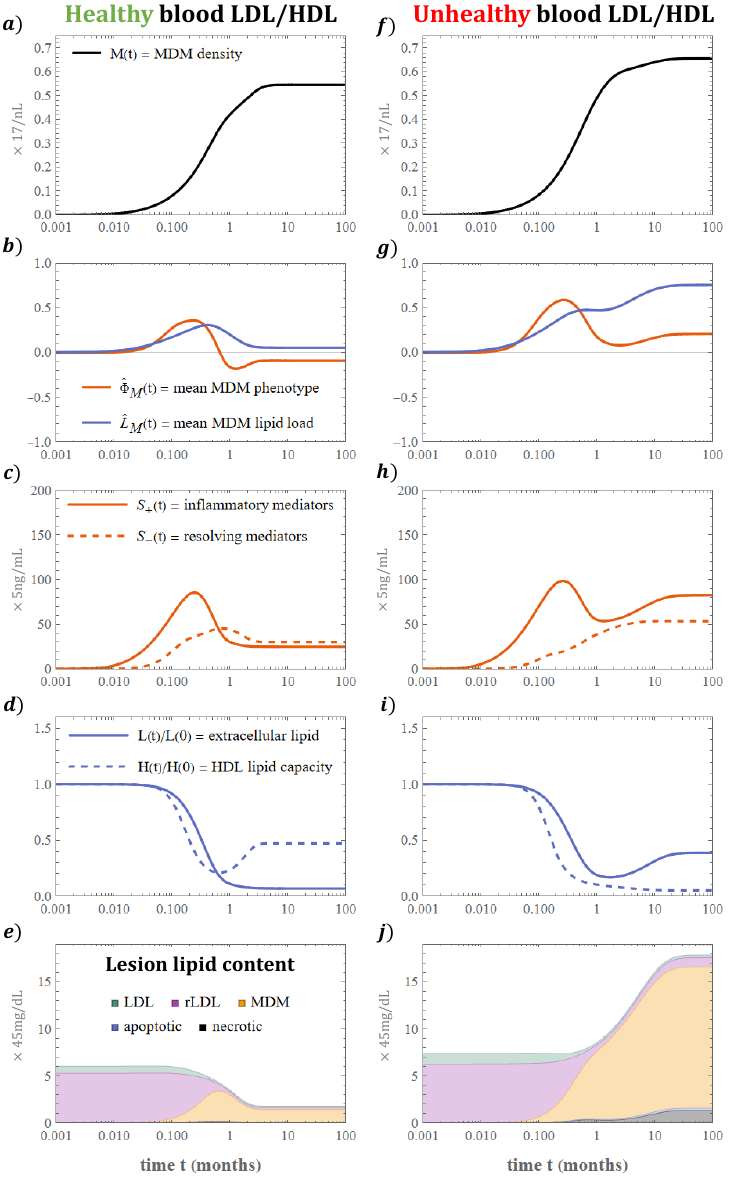}
    \caption{\textbf{Time evolution of lesion composition.} The plots (a)-(j) show numerical solutions of the subsystem \eqref{eqn:M nondim}-\eqref{eqn: S- nondim} for a case with healthy LDL-HDL balance: $ L^\star= 3$, $H^\star = 2.5$ (left), and unhealthy LDL-HDL balance: $L^\star= 4.5$, $H^\star = 1$ (right). The system tends to a non-zero steady state as $t \rightarrow \infty$ with values that depend sensitively on $L^\star$ and $H^\star$. We set $K_\text{r} = 10$ for both cases.}
    \label{fig:composition_dynamics}
\end{figure}

Two typical numerical solutions of the subsystem \eqref{eqn:M nondim}-\eqref{eqn: S- nondim} are shown in Figure \ref{fig:composition_dynamics}. The left and right solutions respectively correspond to healthy ($L^\star = 3$, $H^\star = 2.5$) and unhealthy ($L^\star = 4.5$, $H^\star = 1$) blood levels of LDL lipid and HDL capacity. We set $K_\text{r} = 10$ for both cases. The quantity $L(t)$ in plots d) and i) is the total extracellular lipid density:
\begin{align}
    L(t) := L_{\text{\scaleto{LDL}{3.5pt}}}(t) + L_\text{r}(t) + L_\text{ap}(t) + L_\text{n}(t), \label{eqn: L(t) def}
\end{align}
whereas plots e) and j) show the components of the total lesion lipid content:
\begin{align}
    L_\text{tot}(t) := L_{\text{\scaleto{LDL}{3.5pt}}}(t) + L_\text{r}(t) + \big[1 + \kappa \hat{L}_M(t) \big] M(t) + L_\text{ap}(t) + L_\text{n}(t), \label{eqn: L_tot(t) def}
\end{align}
which includes MDM lipid. The model dynamics can be broadly divided into three phases that we detail below.

The first weeks after initial MDM influx ($0 < t < 0.2$) are characterised by a decline in lesion LDL and rLDL content (Fig. \ref{fig:composition_dynamics}e, j). Both reductions are driven by MDM uptake of rLDL, which also promotes binding of free LDL by restoring the LDL retention capacity. The rise in $\hat{L}_M$ is accompanied by increases in $\hat{\Phi}_M$ and $S_{+}$ since rLDL uptake stimulates production of inflammatory mediators that drive inflammatory phenotype modulation (see Fig. \ref{fig:composition_dynamics}b, c, g, h). 

For $0.2 < t < 2$, the growth of $\hat{L}_M$ and $S_{+}$ slows and $\hat{\Phi}_M$ decreases (see Fig. \ref{fig:composition_dynamics}b, c, g, h). These behaviours arise because the MDM population has ingested enough lipid for the lipid efflux rate to become comparable to that of uptake. For both simulations, the efflux rate exceeds the uptake rate at $t \approx 0.5$, where $\hat{L}_M$ attains a local maximum. The corresponding increase in resolving mediators, $S_{-}$, relative to inflammatory mediators, $S_{+}$, promotes resolving phenotype modulation. 

For $2 < t < \infty$, the model tends to a non-zero steady state in a manner sensitive to parameter values. For healthier balances of blood LDL lipid and HDL capacity $(L^\star, H^\star)$: mean MDM lipid content declines to a small non-zero value and mean MDM phenotype remains negative (i.e. resolving) (Fig. \ref{fig:composition_dynamics}b), resolving mediators outbalance inflammatory mediators (Fig. \ref{fig:composition_dynamics}c), extracellular lipid levels are low while HDL capacity remains substantial (Fig. \ref{fig:composition_dynamics}d), and lesion lipid content decreases to densities below those prior to MDM influx (Fig. \ref{fig:composition_dynamics}e). By contrast, when $L^\star$ is sufficiently high relative to $H^\star$ (quantified in Subsection \ref{sec: composition steady}): MDM densities are greater (Fig. \ref{fig:composition_dynamics}f), mean MDM lipid content increases to a large value and mean MDM phenotype is positive (i.e. inflammatory) (Fig. \ref{fig:composition_dynamics}g), inflammatory mediators outbalance resolving mediators (Fig. \ref{fig:composition_dynamics}h), HDL capacity is largely exhausted while extracellular lipid levels remain substantial (Fig. \ref{fig:composition_dynamics}i), and lesion lipid densities increase to values higher than those prior to initial MDM influx (Fig. \ref{fig:composition_dynamics}j).

\begin{figure}
    \centering
    \includegraphics[width = 0.99\textwidth]{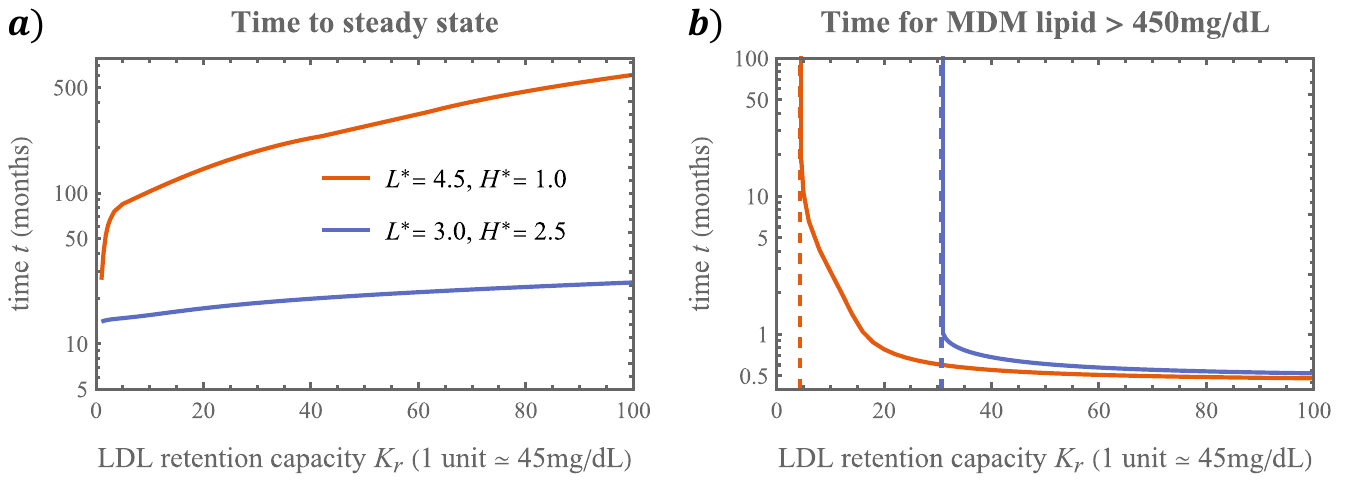}
    \caption{\textbf{LDL retention capacity modulates the timescale of lesion development.} Plot a) shows how time to steady state increases with $K_\text{r}$. Plot b) shows how the time for MDM lipid to exceed $450$mg/dL, a proxy for fatty streak onset, decreases with $K_\text{r}$.}
    \label{fig:steadytime}
\end{figure}
In Fig. \ref{fig:steadytime} we show how the LDL retention capacity, $K_\text{r}$, impacts the timescale of lesion development. The left plot depicts the time to steady state, defined numerically as the smallest time $t$ for which $    \sqrt{\sum_{i}\big(\frac{1}{y_i}\frac{dy_i}{dt}\big)^2}  \leq 10^{-8}$, where the sum is taken over all subsystem variables: $\bm{y} = (M, \hat{\Phi}_M, \hat{L}_M, H, \bm{L}, S_\pm)$. The time to steady state increases with $K_\text{r}$. The trend is more pronounced in the unhealthy case ($L^\star = 4.5$, $H^\star=1$), which exhibits a 20-fold increase over $0.3 \leq K_\text{r} \leq 100$. The right plot shows the time, $t_\text{M}$, for MDM lipid to exceed $450$mg/dL (the smallest $t$ satisfying $(1+\kappa \hat{L}_M(t))M(t) > 10$), which we use as a proxy for fatty streak onset. We find that increases to $K_\text{r}$ yield smaller values of $t_\text{M}$; regions with higher LDL retention capacity develop fatty streaks earlier. Moreover, if $K_\text{r}$ is sufficiently small, the MDM lipid density never exceeds the $450$mg/dL threshold; fatty streaks will not develop in regions of sufficiently low LDL retention capacity.

\subsubsection{Steady state solutions} \label{sec: composition steady}

The results of Sec. \ref{sec: composition dynamics} show that the model tends to a non-zero equilibrium as $t \rightarrow \infty$. Steady state solutions to the subsystem \eqref{eqn:M nondim}-\eqref{eqn: S- nondim} are computed numerically via the Mathematica \textit{FindRoot} routine.
\begin{figure}
    \centering
    \includegraphics[width = 0.99\textwidth]{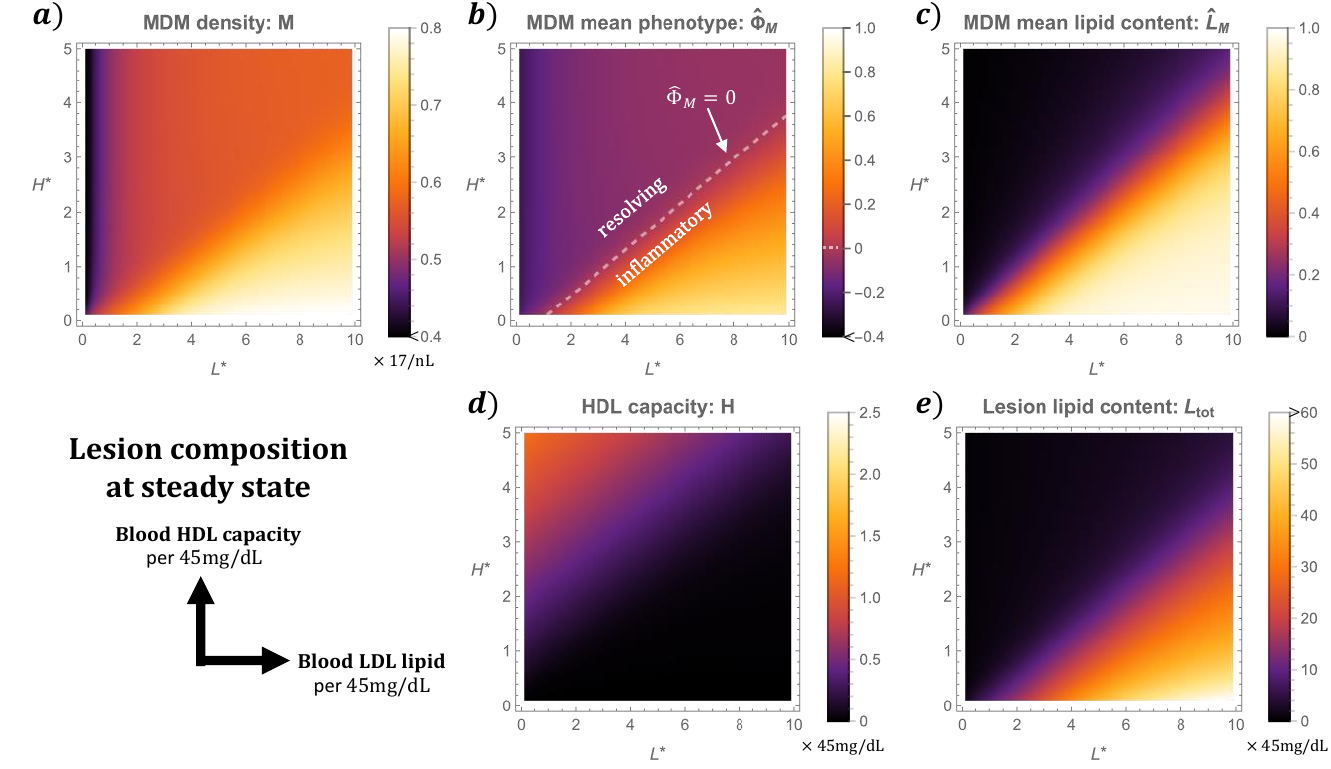}
    \caption{\textbf{Lesion composition at steady state depends on the bloodstream balance of LDL lipid ($L^\star)$ to HDL capacity ($H^\star$).} The plots depict solutions to Eqs. \eqref{eqn:M nondim}-\eqref{eqn: S- nondim} at equilibrium across a uniform grid of values for $(L^\star, H^\star)$. We use a grid resolution of $0.1$ and assume $K_\text{r} = 10$. Note that the markers of pathology: $M$, $\hat{\Phi}_M$, $\hat{L}_M$ and $L_\text{tot}$, each increase with $L^\star$ and decrease with $H^\star$. 
    }
    \label{fig:eq_LH}
\end{figure}

Fig. \ref{fig:eq_LH} illustrates how blood levels of LDL lipid, $L^\star$, and HDL capacity, $H^\star$, impact the model lesion at steady state. Lesion composition becomes more pathological as $L^\star$ increases and $H^\star$ decreases; MDM density, mean phenotype, lipid content and total lipid content each monotonically increase with $L^\star$ and decrease with $H^\star$, while HDL capacity monotonically decreases with $L^\star$ and increases with $H^\star$. We note that all plots exhibit regions in which the contours are approximately linear. For the case $K_\text{r} = 10$ shown, these contours take the form $0.4L^\star - H^\star = $ constant. The greater weighting on $H^\star$ in the linear combination reflects the higher value of the dimensionless lipid efflux rate, $k_H$, relative to the uptake rates in Table \ref{tab: nondim parameters}. Values of $(L^\star, H^\star)$ above the $\hat{\Phi}_M = 0$ contour ($0.4 L^\star - H^\star \approx 0.4$) yield healthy lesions with smaller MDM densities, resolving mean phenotypes and small mean lipid loads. Lesion lipid content is low and HDL capacity remains substantial. Values of $(L^\star, H^\star)$ below the $\hat{\Phi}_M = 0$ contour yield pathological lesions with higher MDM densities, inflammatory mean phenotypes and higher mean lipid loads. Lesion lipid content is also large. These markers of pathology each increase with $0.4 L^\star - H^\star$. By contrast the HDL capacity is exhausted in this region. 


\begin{figure}
    \centering
    \includegraphics[width = 0.99\textwidth]{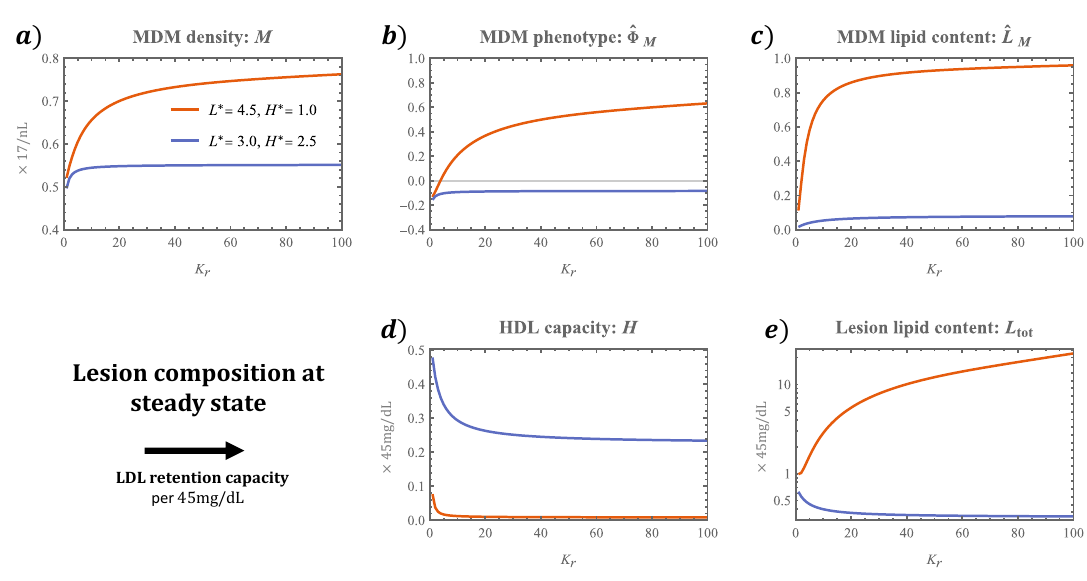}
    \caption{\textbf{Lesion composition at steady state varies with LDL retention capacity ($K_\text{r}$).} The plots show solutions to Eqs. \eqref{eqn:M nondim}-\eqref{eqn: S- nondim} at steady state across $0.3 \leq K_\text{r} \leq 100$. We show a case with healthy LDL-HDL balance: $L^\star = 3.0$, $H^\star = 2.5$, and one with unhealthy LDL-HDL balance: $L^\star = 4.5$, $H^\star = 1.0$.}
    \label{fig:eq_Kr}
\end{figure}
Fig. \ref{fig:eq_Kr} shows how the LDL retention capacity, $K_\text{r}$, impacts lesion composition at steady state. We find that MDM density, MDM mean phenotype and MDM lipid content increase non-linearly with $K_\text{r}$. These trends are amplified when the values $(L^\star, H^\star)$ are more pathological (i.e. $0.4L^\star - H^\star$ is larger). HDL capacity instead decreases with $K_\text{r}$. Moreover, total lesion lipid content increases or decreases with $K_\text{r}$ if the values $(L^\star, H^\star)$ are pathological or healthy respectively. This difference arises because higher LDL retention capacities give rise to higher MDM densities, due partially to elevated signalling by resident cells. In healthy cases, HDL capacity remains substantial so that MDMs provide a net reduction in lesion lipid content by consuming extracellular lipid and efficiently offloading to HDL. In pathological cases, HDL capacity is exhausted and lipid taken up by MDMs can only leave the lesion via MDM egress. This makes MDMs net contributors to lesion lipid content; on average, they remove less lipid from the lesion than they supply via their endogenous lipid content.

\subsection{MDM phenotype-lipid distribution} \label{sec: MDM distribution}

We now turn to Eqs. \eqref{eqn:mphil nondim} to study how phenotype and lipid content are distributed amongst lesion MDMs. After presenting numerical solutions for $m_{\phi, \ell}(t)$ in Sec. \ref{sec: MDM time evolution}, we derive a continuum approximation of Eqs. \eqref{eqn:mphil nondim} in Sec. \ref{sec: continuum approximation} to make analytical progress. In Sec. \ref{sec: MDM trajectories} we show how analysis of the leading-order advective dynamics enables us to characterise the expected trajectories of MDMs through phenotype-lipid space. Finally, in Subsection \ref{sec: MDM steady} we analyse the MDM distribution at steady state.


\subsubsection{Time evolution} \label{sec: MDM time evolution}

\begin{figure}
    \centering
    \includegraphics[width = 0.96\textwidth]{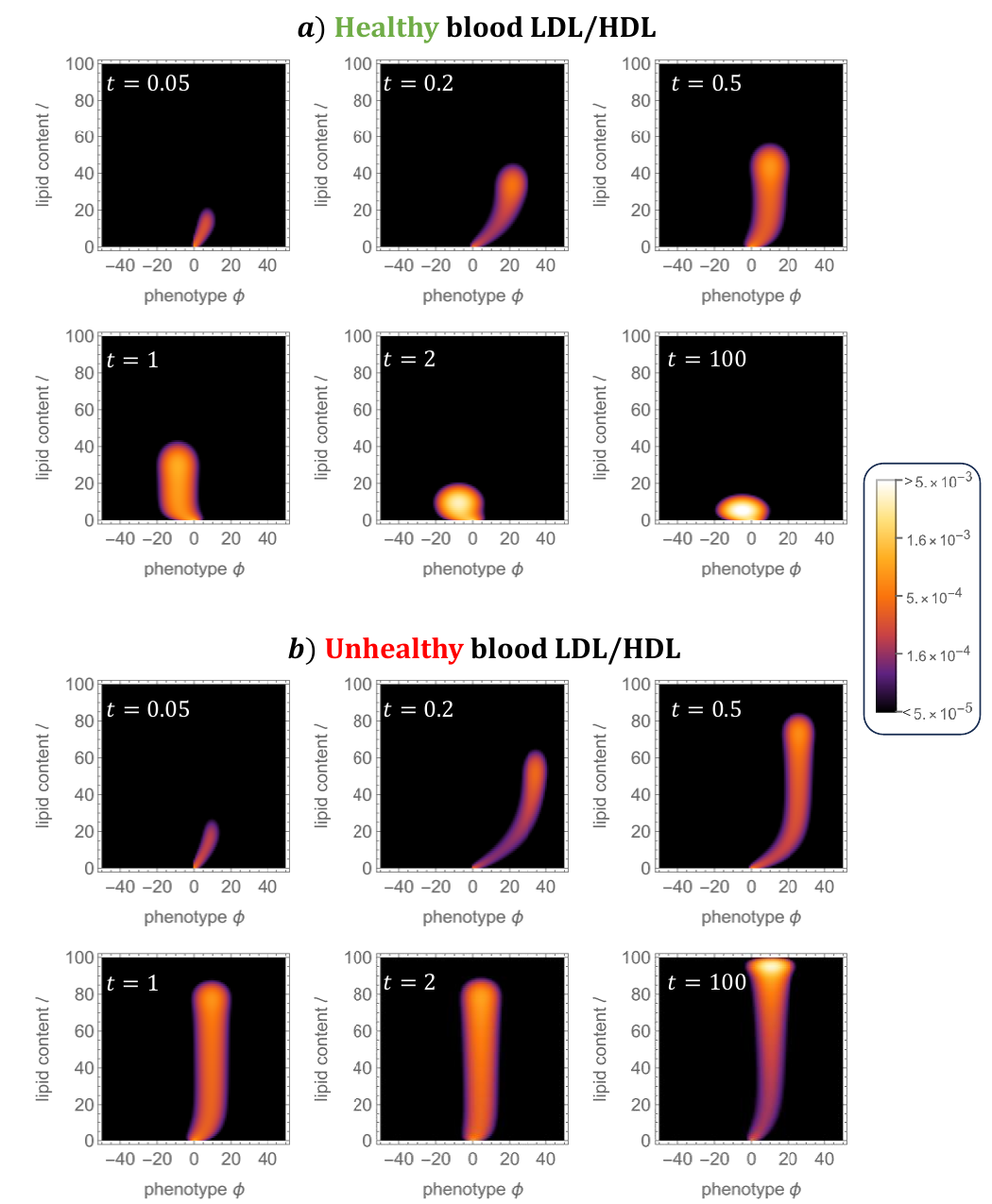}
    \caption{\textbf{MDM phenotype-lipid distribution dynamics for two typical simulations.} The plots show numerical solutions of Eqs. \eqref{eqn:mphil nondim} for $m_{\phi, \ell}(t)$ for a case with healthy LDL/HDL balance (a): $ L^\star= 3$, $H^\star = 2.5$, and unhealthy LDL-HDL balance (b): $L^\star= 4.5$, $H^\star = 1$. We use $K_\text{r} = 10$ for both simulations.
    At steady state (see $t = 100$), the distribution may skew towards resolving (case a) or inflammatory (case b) phenotypes.}
    \label{fig:MDM_dynamics}
\end{figure}
In Figure \ref{fig:MDM_dynamics} we present the dynamics of the MDM phenotype-lipid distribution, $m_{\phi, \ell}(t)$, for the cases shown in Fig. \ref{fig:composition_dynamics}). At early times ($ 0 \leq t \leq 0.2$, the distribution evolves in a wave-like manner, from $(\phi, \ell) = (0,0)$, in the direction of increasing $\phi$ and $\ell$ due to the initial phase of LDL uptake noted in Subsection \ref{sec: composition dynamics}. At later times ($0.2 \leq t \leq 2$), the distribution moves leftwards, becoming concentrated at lower values of $\phi$. This resolving phenotype modulation reflects greater production of resolving mediators, $S_{-}$, relative to inflammatory mediators, $S_{+}$, in this phase (c.f. Fig. \ref{fig:composition_dynamics}e,h). At longer times, the distribution settles to a steady state. In the healthy case, lipid loads gradually reduce to equilibrium as HDL capacity increases and the extracellular lipid densities decrease to their equilibrium values (c.f. Fig. \ref{fig:composition_dynamics}d,e). In the unhealthy case, lipid loads increase towards equilibrium as the extracellular lipid densities increase and HDL capacity declines to their steady state values (c.f. Fig. \ref{fig:composition_dynamics}i,j).

We note that the MDM distribution, in both cases and at all times, is concentrated about a central curve which begins at the origin $(\phi, \ell) = (0,0)$ and terminates at an interior point in $(\phi, \ell)$ space. This feature indicates that phenotype and lipid content are non-linearly correlated. At steady state, this correlation is monotonic but may be negative (as in case a) or positive (as in case b) depending on model parameters. 


It is not straightforward to understand how the MDM distributions in Fig. \ref{fig:MDM_dynamics} arise by directly considering Eqs. \eqref{eqn:mphil nondim}. In order to make progress, we consider a continuum approximation of Eqs. \eqref{eqn:mphil nondim} in the analysis below.

\subsubsection{Continuum approximation} \label{sec: continuum approximation}

We associate the discretely structured MDM distribution $m_{\phi, \ell}(t)$, $\phi = 0, \pm 1, \dots, \pm \phi_\text{max}$, $\ell = 0, 1, \dots, \ell_\text{max}$, with a function $m(\Tilde{\phi}, \Tilde{\ell}, t) \geq 0$ of two continuous structure variables, $\Tilde{\phi}$ and $\Tilde{\ell}$, and time $t$. Specifically, we make the identifications:
\begin{align}
    &\Tilde{\phi} \sim \frac{\phi}{\phi_\text{max}}, & &\Tilde{\ell} \sim \frac{\ell}{\ell_\text{max}}, & &m(\Tilde{\phi}, \Tilde{\ell}, t) \sim  \phi_\text{max} \ell_\text{max} \, m_{\phi, \ell}(t). \label{eqn: identifications}
\end{align}
Here $\Tilde{\phi} \in [-1,1]$ and $\Tilde{\ell} \in [0,1]$, so that $\Tilde{\phi} = -1$ corresponds to fully M2-polarised MDMs, $\Tilde{\phi} = 1$ to fully M1-polarised MDMs, $\Tilde{\ell} = 0$ to MDMs with only endogenous lipid, and $\Tilde{\ell} = 1$ to MDMs at maximal lipid content. The function $m(\Tilde{\phi}, \Tilde{\ell}, t)$ can be interpreted as the number density of MDMs across a continuous phenotype-lipid structure space, with scaling such that we may identify:
\begin{align}
    M(t) = \sum_{\phi, \ell} m_{\phi, \ell}(t) \sim \int_0^1 \int_{-1}^1 m(\Tilde{\phi}, \Tilde{\ell}, t) \, d\Tilde{\phi} \, d\Tilde{\ell} . \label{eqn: integral relation} 
\end{align}


We seek an expansion of Eqs.\eqref{eqn:mphil nondim} in the limit $\ell_\text{max} \sim \phi_\text{max} \rightarrow \infty$. More precisely, we set $\epsilon := \ell_\text{max}^{-1} \ll 1$ and assume that $\theta := \phi_\text{max}/\ell_\text{max} = \mathcal{O}(1)$. Substituting the identifications \eqref{eqn: identifications} into Eqs.\eqref{eqn:mphil nondim}, Taylor-expanding about $\epsilon = 0$, and neglecting $\mathcal{O}(\epsilon^2)$ terms, yields the advection-diffusion PDE:
\begin{align}
    \frac{\partial m}{\partial t} &= \nabla \cdot (\bm{D} \nabla m - \bm{v} m) - (1 + \gamma) m, \label{eqn: advdif}
\end{align}
where $\nabla = \Big( \frac{\partial}{\partial \Tilde{\phi}}, \frac{\partial}{\partial \Tilde{\ell}} \Big)^T$, and the velocity vector, $\bm{v}$, and diffusivity matrix, $\bm{D}$, are respectively:
\begin{align}
    \bm{v} &= 
    \begin{bmatrix}
        \chi (S_{+} - S_{-}) - \chi (S_{+} + S_{-}) \tilde{\phi} \\
        \bm{k_L}\cdot \bm{L} - (\bm{k_L}\cdot \bm{L} + k_H H) \Tilde{\ell}
    \end{bmatrix} + \frac{\epsilon}{2}
    \begin{bmatrix}
        \theta \chi (S_{+} - S_{-}) \\
         \bm{k_L}\cdot \bm{L} - k_H H
    \end{bmatrix}, \label{eqn: v}\\
    \bm{D} &= \frac{\epsilon}{2}
    \begin{bmatrix}
        \theta \chi \big[ (S_{+} + S_{-}) -  (S_{+} - S_{-}) \tilde{\phi} \big] & 0 \\
        0 &  \bm{k_L}\cdot \bm{L} -  (\bm{k_L}\cdot \bm{L} - k_H H) \Tilde{\ell}
    \end{bmatrix}. \label{eqn: D}
\end{align}
We note that Eq.\eqref{eqn: advdif} is advection-dominant since $v = \mathcal{O}(1)$ and $D = \mathcal{O}(\epsilon)$. 

We derive boundary conditions for Eq.\eqref{eqn: advdif} by requiring that the dynamics of the MDM population, $M(t)$, as defined by Eq. \eqref{eqn:M nondim}, are consistent with those of the continuous MDM distribution \eqref{eqn: integral relation}. Integrating Eq. \eqref{eqn: advdif} over $(\Tilde{\phi}, \Tilde{\ell}) \in \mathcal{R} := [-1,1] \times [0,1]$ and applying the divergence theorem yields:
\begin{align}
    \frac{dM}{dt} = \oint_{\partial \mathcal{R}} (D \nabla m - v m ) \cdot \mathbf{n} \, ds - (1+\gamma)M, \label{eqn: div thm}
\end{align}
where $\mathbf{n}$ is the outwards pointing normal vector. We set:
\begin{align}
    &(D \nabla m - v m ) \cdot (\pm1,0)^T = 0 \quad \text{on } \Tilde{\phi} = \pm 1, \label{eqn: bconds1} \\
    &(D \nabla m - v m ) \cdot (0,+1)^T = 0 \quad \text{on } \Tilde{\ell} = 1, \label{eqn: bconds2} \\
    &(D \nabla m - v m ) \cdot (0,-1)^T = R_{0,0} \cdot \delta_0(\Tilde{\phi}) \quad \text{on } \Tilde{\ell} = 0. \label{eqn: bcond recruit}
\end{align}
Eqs. \eqref{eqn: bconds1} and \eqref{eqn: bconds2} are no-flux conditions so that MDMs cannot enter or leave the system by exceeding the phenotype bounds $\Tilde{\phi} = \pm 1$ or maximal lipid content $\Tilde{\ell} = 1$ respectively. We use the Dirac-delta distribution, $\delta_0$, in Eq.\eqref{eqn: bcond recruit} to ensure that MDMs enter the lesion at the origin $(\Tilde{\phi}, \Tilde{\ell}) = (0,0)$.

\subsubsection{Advection at leading order} \label{sec: MDM trajectories}

Eq.\eqref{eqn: advdif} reduces to a hyperbolic PDE at leading order as $\epsilon \rightarrow 0$:
\begin{align}
\begin{split}
    \frac{\partial m}{\partial t} + \chi \frac{\partial}{\partial \Tilde{\phi}} \Big[ \Big( (S_{+} &-S_{-}) -(S_{+}+S_{-})\Tilde{\phi} \Big) m \Big] \\
    &+ \frac{\partial}{\partial \Tilde{\ell}} \Big[ \Big( \bm{k_L} \cdot \bm{L} - (\bm{k_L} \cdot \bm{L} + k_H H) \Tilde{\ell} \Big) m \Big] 
    = -(1+\gamma)m. \label{eqn: advection} 
\end{split}
\end{align}
Eq.\eqref{eqn: advection} describes the advection, in $(\Tilde{\phi},\Tilde{\ell})$ space, of MDMs according to a time-dependent velocity field:
\begin{align}
    \frac{d \Tilde{\phi}}{d t} &= \chi \big[ (S_{+} - S_{-}) - (S_{+} + S_{-}) \Tilde{\phi} \big], \label{eqn: char phi}\\
    \frac{d \Tilde {\ell}}{d t} &=  \bm{k_L} \cdot \bm{L} - (\bm{k_L} \cdot \bm{L} + k_H H) \Tilde{\ell}, \label{eqn: char l}
\end{align}
where the coefficients: $S_\pm(t)$, $\bm{L}(t)$ and $H(t)$, are time-dependent solutions of the subsystem \eqref{eqn:M nondim}-\eqref{eqn: S- nondim}. 

\begin{figure}
    \centering
    \includegraphics[width = 0.8\textwidth]{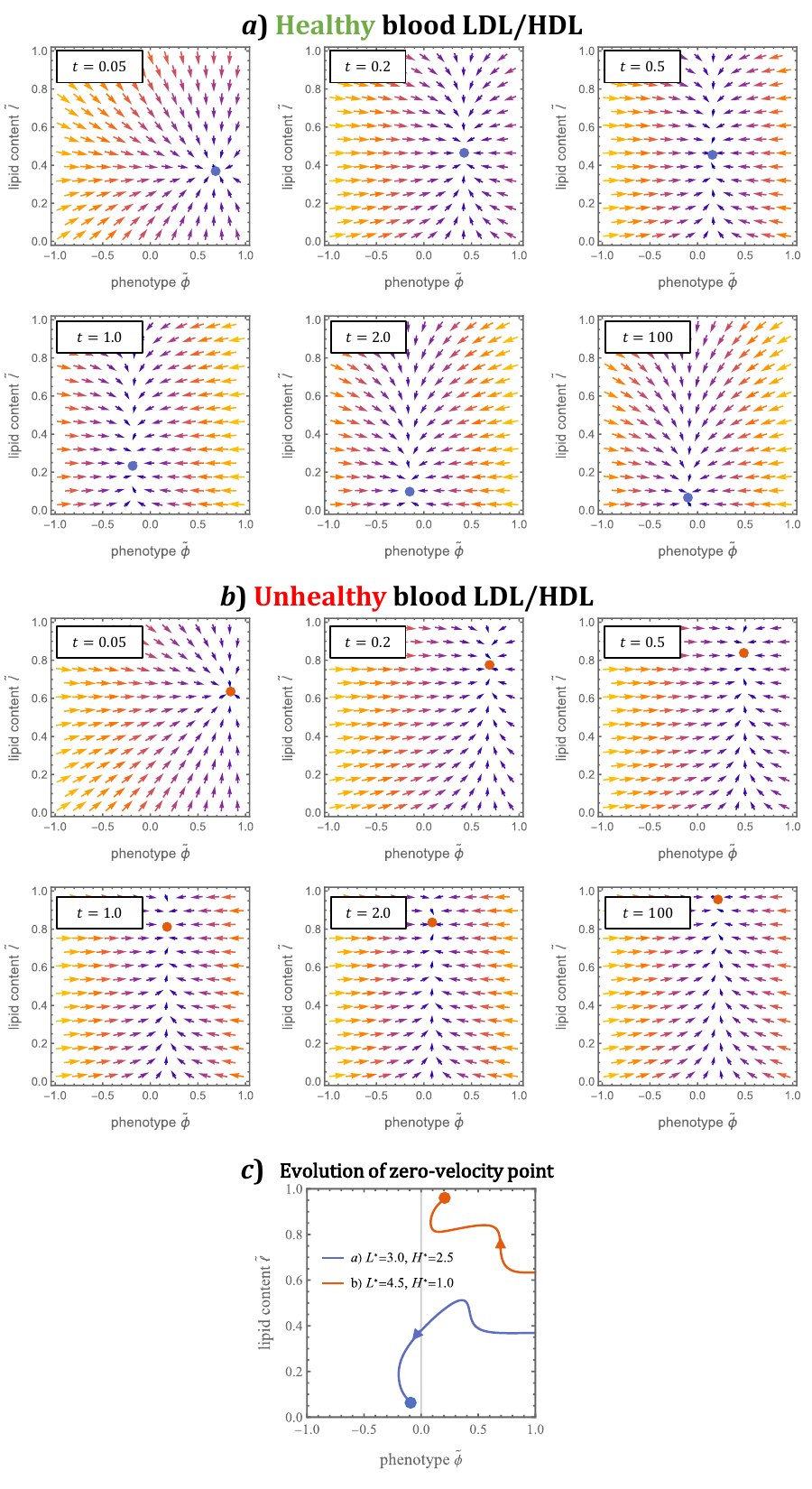}
    \caption{\textbf{Dynamics of the MDM phenotype-lipid velocity field. } The vector field plots illustrate the right hand side of Eqs.\eqref{eqn: char phi}-\eqref{eqn: char l} using the numerical solutions of the subsystem \eqref{eqn:M nondim}-\eqref{eqn: S- nondim} shown in Fig. \ref{fig:composition_dynamics}. The velocity magnitude is indicated by larger arrows and brighter colours. The point of instantaneous zero velocity is indicated by a circle at each time point in the vector field plots, and plotted against time in the bottom plot. We fix $K_\text{r} = 10$ and use $L^\star = 4.5$, $H^\star = 1.0$ for case a) and $L^\star = 3.0$, $H^\star = 2.5$ for case b).
    }
    \label{fig:velocity_field}
\end{figure}
The dynamics of the velocity field defined by Eqs.\eqref{eqn: char phi}-\eqref{eqn: char l} are illustrated in Fig. \ref{fig:velocity_field}. A striking feature of the plots is that,  at all times, the vector field has a single point $(\Tilde{\phi}_\times, \Tilde{\ell}_\times)$ of zero instantaneous velocity. This time-dependent point represents the target phenotype and lipid content that MDMs are driven towards by advection. Setting time derivatives to zero in Eqs.\eqref{eqn: char phi}-\eqref{eqn: char l}, and solving for $\Tilde{{\phi}}$ and $\Tilde{\ell}$ shows that:
\begin{align}
    \Tilde{\phi}_\times(t) :&= \frac{S_{+}(t)-S_{-}(t)}{S_{+}(t)+S_{-}(t)}, & \Tilde{\ell}_\times(t) :&= \frac{\bm{k_L} \cdot \bm{L}(t)}{\bm{k_L} \cdot \bm{L}(t) + k_H H(t)}. \label{eqn: zero pt}
\end{align}
The time-evolution of the zero velocity point, given by Eqs. \eqref{eqn: zero pt}, is shown in Fig. \ref{fig:velocity_field} c). We find that $(\Tilde{\phi}_\times, \Tilde{\ell}_\times)$ first moves from an inflammatory position towards resolving phenotypes. This initial transition occurs because resolving mediator production by MDMs opposes the initially pure inflammatory environment due to rLDL-stimulated resident cells. When the LDL/HDL balance is healthy, the target phenotype becomes resolving and the target lipid content decreases towards steady state. When the LDL/HDL balance is unhealthy, the target phenotype remains inflammatory and the target lipid content increases as the system evolves to its steady state. We note further that $(\Tilde{\phi}_\times, \Tilde{\ell}_\times)$ aligns well with the MDM distribution end points shown in Fig. \ref{fig:MDM_dynamics}.

\begin{figure}
    \centering
    \includegraphics[width = 0.95\textwidth]{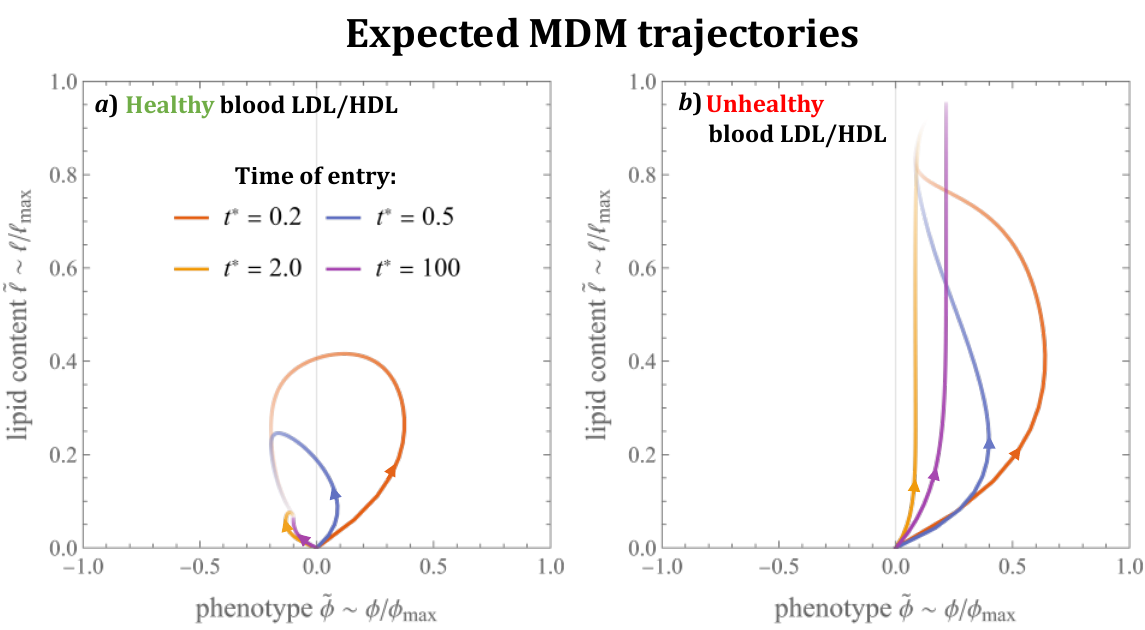}
    \caption{\textbf{Expected trajectory through phenotype-lipid content space of MDMs that enter the lesion at time $t = t^\star$.} The curves are numerical solutions to equations \eqref{eqn: char phi}-\eqref{eqn: char init}. The opacity of each point on the curves represents the probability that MDMs travel through that point before either dying or emigrating from the lesion. Note how MDMs which enter at later times have simple monotonic trajectories, in contrast to those that enter earlier.}
    \label{fig:MDM_trajectories}
\end{figure}

Since MDMs enter the model lesion with minimal lipid content and a neutral phenotype, we are particularly interested in solutions to Eqs. \eqref{eqn: char phi}-\eqref{eqn: char l} which satisfy the initial conditions:
\begin{align}
    \Tilde{\phi}(t^\star) &= 0, & \Tilde{\ell}(t^\star) &= 0. \label{eqn: char init}
\end{align}
Such solutions represent the trajectory of MDMs that enter the lesion at time $t = t^\star$ in the limit $\epsilon \rightarrow 0$. They can be expressed in terms of the subsystem variables as follows:
\begin{align}
    &\Tilde{\phi}(t) = I_1(t)^{-1} \int_{t^\star}^{t} \chi \big( S_{+}(\tau) - S_{-}(\tau) \big) I_1(\tau) d\tau, \label{eqn: traj sol1}\\
    &\Tilde{\ell}(t) = I_2(t)^{-1} \int_{t^\star}^{t} \bm{k_L} \cdot \bm{L}(\tau) I_2(\tau) d\tau, & t\geq t^\star, \label{eqn: traj sol2}
\end{align}
where $I_1(t)$ and $I_2(t)$ are integrating factors:
\begin{align}
    &I_1(t) := e^{\int_{t^\star}^{t} \chi \big( S_{+}(\tau) + S_{-}(\tau) \big) d\tau}, & &I_2(t) := e^{\int_{t^\star}^{t} \big( \bm{k_L} \cdot \bm{L}(\tau) + k_H H(\tau) \big) d\tau}. \nonumber
\end{align}
We use numerical solutions of the subsystem \eqref{eqn:M nondim}-\eqref{eqn: S- approx} to evaluate Eqs.\eqref{eqn: traj sol1}-\eqref{eqn: traj sol2} in Fig. \ref{fig:MDM_trajectories}. As MDMs travel along the trajectories shown, they emigrate from the lesion and die via apoptosis at the constant rate $(1 + \gamma)$. The probability that MDMs which enter the lesion at time $t = t^\star$ are still alive and in the lesion at time $t \geq t^\star$ is given by $e^{-(1+\gamma)(t-t^\star)}$. This probability is represented in Fig. \ref{fig:MDM_trajectories} by the opacity of the curves. The plots illustrate that MDMs which enter the lesion early (e.g. at $t^\star = 0.2$) can be expected to first transition from phenotypic neutrality to an inflammatory state with moderate lipid loads, before evolving to either a resolving phenotype with low lipid load (case a), or a milder inflammatory phenotype with high lipid load (case b). The trajectories of MDMs that enter the lesion at later times are typically monotonic, in contrast to the looping trajectories of MDMs that enter at earlier times. We note in particular that trajectories with $t^\star = 100$ (near steady state) align well with the centre line of the MDM distributions (c.f. Fig. \ref{fig:MDM_dynamics} at $t = 100$). 

\subsubsection{Steady state solutions} \label{sec: MDM steady}

\begin{figure}
    \centering
    \includegraphics[width = 0.88\textwidth]{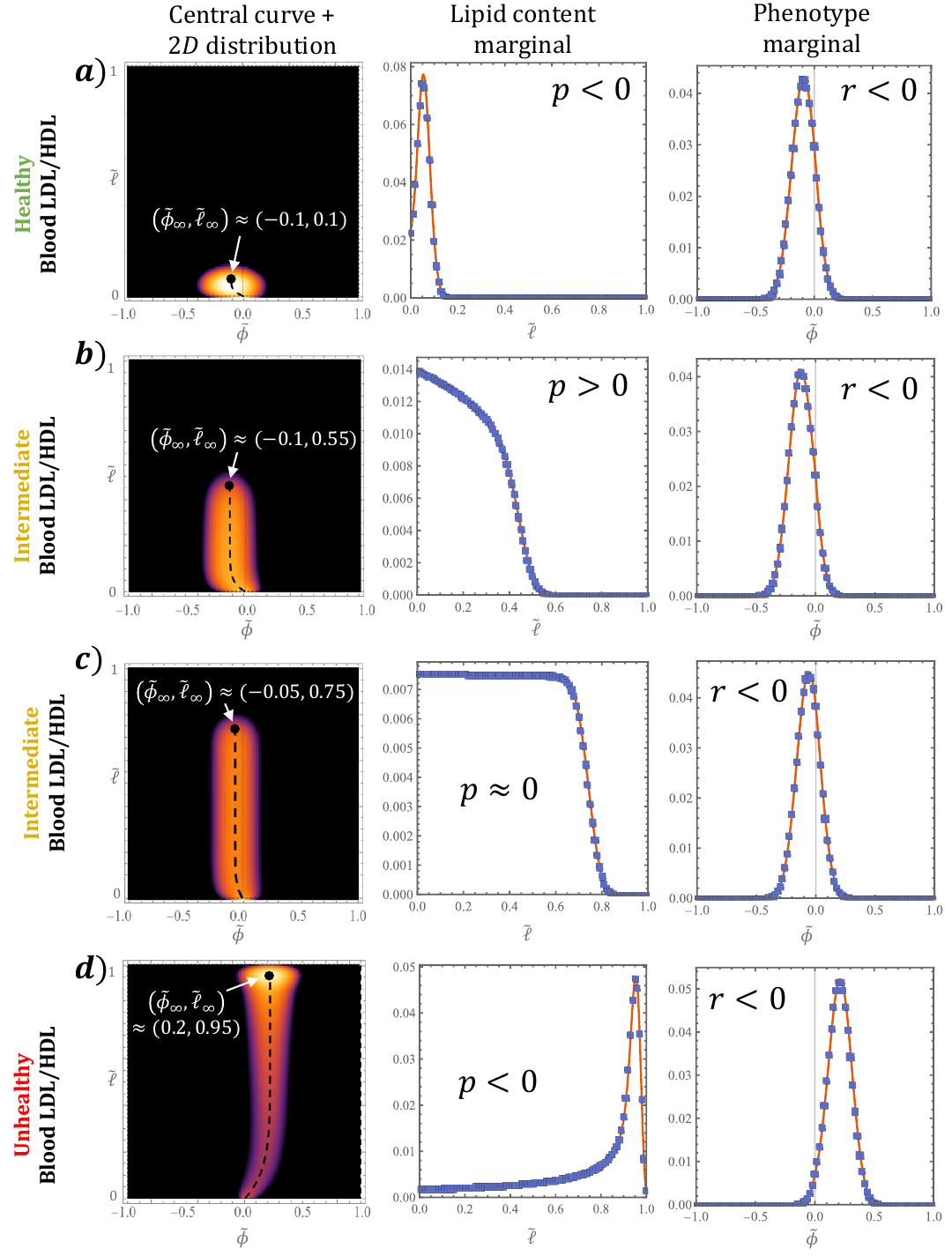}
    \caption{\textbf{Comparison of continuum analytical results to numerical solutions for the discrete MDM distribution, $m_{\phi, \ell}$, at steady state.} Cases a)-d) use the values $(L^\star, H^\star) = $ $(3, 2.5)$, $(1.7, 0.8)$, $(1.9, 0.6)$ and $(4.5, 1)$ and are ordered roughly by pathology. The first column overlays Eq. \eqref{eqn: crit curve} for the central curve with $m_{\phi, \ell}$ (same colour legend as in Fig. \ref{fig:MDM_dynamics}). The second and third columns compare the exact solutions for the lipid content and phenotype marginals, given by Eqs. \eqref{eqn: W exact} and \eqref{eqn: Vpm sol} respectively, with the discrete marginals. We set $K_\text{r} = 10$ for all cases. }
    \label{fig:MDM_eq_distributions}
\end{figure}

\begin{figure}
    \centering
    \includegraphics[width = 0.95\textwidth]{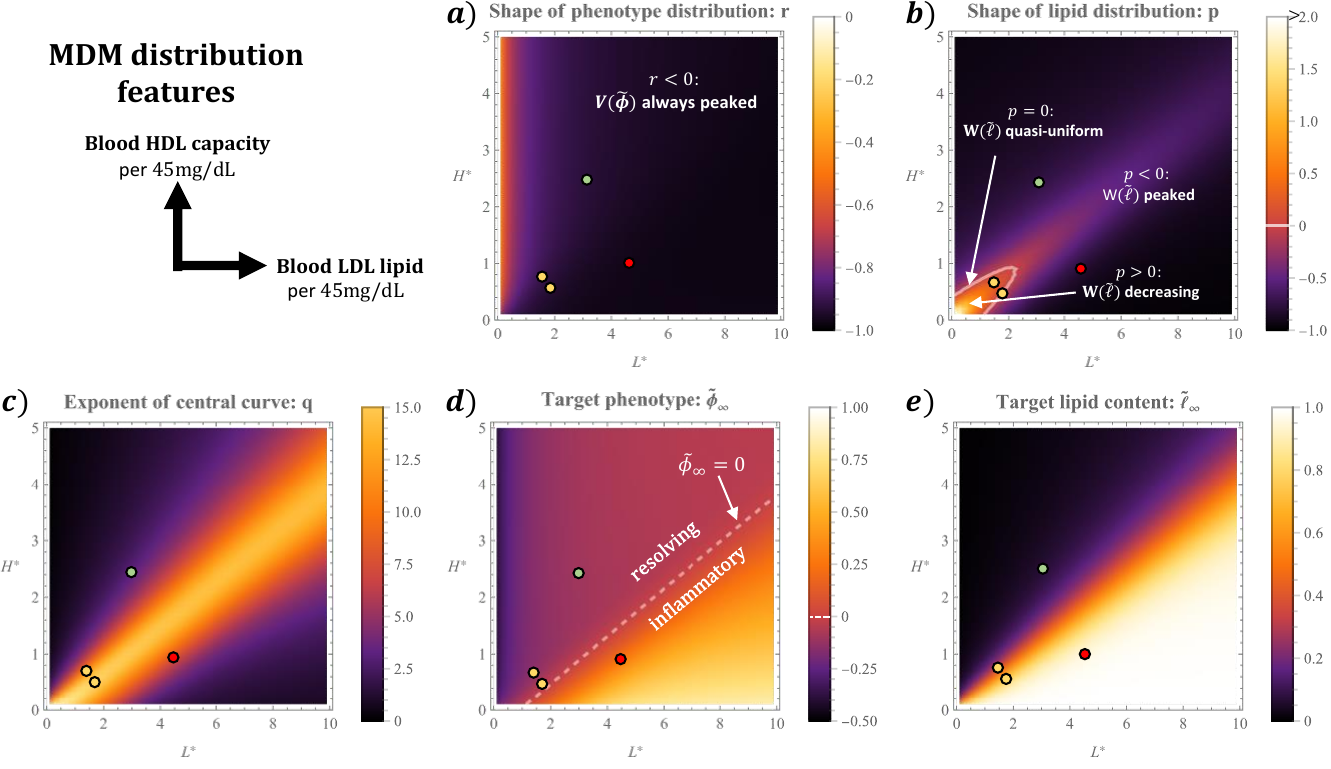}
    \caption{\textbf{Impact of blood LDL lipid ($L^\star$) and HDL capacity ($H^\star$) on the qualitative features of the MDM distribution.} The values were computed by combining the results shown in Fig. \ref{fig:eq_LH} with the formulae \eqref{eqn: par groupings} and $r = -1 + \frac{p+1}{q}$. The four cases presented in Fig. \ref{fig:MDM_eq_distributions} are also indicated: $(L^\star, H^\star) = $ $(3, 2.5)$, $(1.7, 0.8)$, $(1.9, 0.6)$ and $(4.5, 1)$.}
    \label{fig:mdist_features_LH}
\end{figure}

It is straightforward to show that $m^\star(\Tilde{\phi}, \Tilde{\ell})$, the steady state solution of Eq. \eqref{eqn: advdif}, satisfies:
\begin{align}
\begin{split}
    \frac{\epsilon \theta q}{2} \frac{\partial^2}{\partial \Tilde{\phi}^2} \Big[ (1 - &\Tilde{\phi}_\infty \Tilde{\phi}) m^\star \Big] + \frac{\epsilon}{2} \frac{\partial^2}{\partial \Tilde{\ell}^2} \Big[ \Big( \Tilde{\ell}_\infty -(2\Tilde{\ell}_\infty - 1) \Tilde{\ell} \Big) m^\star \Big] \\ &- q (\Tilde{\phi}_\infty - \Tilde{\phi}) \frac{\partial m^\star}{\partial \Tilde{\phi}} - (\Tilde{\ell}_\infty - \Tilde{\ell}) \frac{\partial m^\star}{\partial \Tilde{\ell}} - (p-q) m^\star = 0,
\end{split} \label{eqn: mdist equilibrium}
\end{align}
where the constants $\Tilde{\phi}_\infty$, $\Tilde{\ell}_\infty$, $q$ and $p$ are defined in terms of model parameters and the equilibrium values of the subsystem variables:
\begin{align}
    \begin{split}
        &\Tilde{\phi}_\infty := \frac{S_{+} - S_{-}}{S_{+} + S_{-}}, \qquad \quad \,   \Tilde{\ell}_\infty := \frac{\bm{k_L} \cdot \bm{L}}{\bm{k_L} \cdot \bm{L} + k_H H }, \\
        &q := \frac{\chi (S_{+} + S_{-})}{\bm{k_L} \cdot \bm{L} + k_H H}, \qquad p := - 1 + \frac{1+\gamma}{\bm{k_L} \cdot \bm{L} + k_H H}.
    \end{split} \label{eqn: par groupings}
\end{align}
We note that $\Tilde{\phi}_\infty \in (-1,1)$ and $\Tilde{\ell}_\infty \in (0,1)$ coincide exactly with the target phenotype and lipid content, $\Tilde{\phi}_\times$ and $\Tilde{\ell}_\times$ respectively (c.f. Eqs.\eqref{eqn: zero pt} at steady state). We note also that $p \in (-1, \infty)$ and $q \in (0, \infty)$. The boundary conditions \eqref{eqn: bconds1}-\eqref{eqn: bcond recruit} become:
\begin{align}
    &J_{\phi}\vert_{\Tilde{\phi} = -1} = 0, & &J_{\phi}\vert_{\Tilde{\phi} = 1} = 0, & &J_{\ell}\vert_{\Tilde{\ell} = 1} = 0, & &J_{\ell}\vert_{\Tilde{\ell} = 0} = \Big(\frac{\Tilde{\ell}_\infty R_{0,0}}{\bm{k_L} \cdot \bm{L}} \Big)  \cdot  \delta_0(\Tilde{\phi}), \label{eqn: equilibrium bconds}
\end{align}
where the dimensionless fluxes are given by:
\begin{align}
    J_{\phi} &= q(\Tilde{\phi}_\infty - \Tilde{\phi}) m^\star - \frac{\epsilon \theta q}{2} \frac{\partial}{\partial \Tilde{\phi}} \Big[ (1 - \Tilde{\phi}_\infty \Tilde{\phi}) m^\star \Big], \\
    J_{\ell} &= (\Tilde{\ell}_\infty - \Tilde{\ell})m^\star - \frac{\epsilon}{2} \frac{\partial}{\partial \Tilde{\ell}} \Big[\big( \Tilde{\ell}_\infty - (2 \Tilde{\ell}_\infty - 1) \Tilde{\ell} \big) m^\star \Big].
\end{align}
Rather than searching for a closed-form solutions to Eq. \eqref{eqn: mdist equilibrium} in full generality, we characterise the solutions via asymptotic analysis in the limit $\epsilon \rightarrow 0$. More specifically, we derive equations for the central curve of the MDM distribution and marginal distributions with respect to lipid content, $W(\Tilde{\ell})$, and phenotype, $V(\Tilde{\phi})$, where:
\begin{align}
    &W(\Tilde{\ell}) := \int_{-1}^{1} m^\star (\Tilde{\phi}, \Tilde{\ell}) \, d \Tilde{\phi}, & &V(\Tilde{\phi}) := \int_{0}^{1} m^\star (\Tilde{\phi}, \Tilde{\ell}) \, d \Tilde{\ell}. \label{eqn: def marginals}
\end{align}
Finally, we consider the impact of the  LDL lipid and HDL capacity blood densities, $L^\star$ and $H^\star$ respectively. The results are summarised in Figs. \ref{fig:MDM_eq_distributions} and \ref{fig:mdist_features_LH}. 
\\

\noindent \textbf{Central curve.} We determine the curve about which $m^\star( \Tilde{\phi}, \Tilde{\ell})$ is centred by considering Eq.\eqref{eqn: mdist equilibrium} at leading order. Indeed, when $\epsilon = 0$, Eq.\eqref{eqn: mdist equilibrium} reduces to a first order hyperbolic PDE, with characteristics that satisfy the equation:
\begin{align}
    \frac{d \Tilde{\phi}}{d\Tilde{\ell}} = \frac{q(\Tilde{\phi}_\infty - \Tilde{\phi})}{\Tilde{\ell}_\infty - \Tilde{\ell}}. \label{eqn: steady_char}
\end{align}
Solving Eq.\eqref{eqn: steady_char} subject to $\Tilde{\phi}(\Tilde{\ell} = 0) = 0$ yields:
\begin{align}
    \Tilde{\phi} = \Tilde{\phi}_c( \Tilde{\ell} ) := \Tilde{\phi}_\infty \bigg[ 1 - \bigg( 1 - \frac{\Tilde{\ell}}{\Tilde{\ell}_\infty} \bigg)^q \bigg]. \label{eqn: crit curve}
\end{align}
In the first column of Fig. \ref{fig:MDM_eq_distributions}, the solution \eqref{eqn: crit curve} is superimposed on numerical solutions for $m_{\phi, \ell}$ at steady state, showing good agreement with the centre of the MDM distributions. 

Eq.\eqref{eqn: crit curve} confirms that, at steady state, phenotype and lipid content are monotonically correlated via a power law. The correlation is positive when $\Tilde{\phi}_\infty > 0$ and negative when $\Tilde{\phi}_\infty < 0$. Equivalently, using the first of Eqs.\eqref{eqn: par groupings}, if inflammatory mediators dominate resolving ones ($S_{+} > S_{-}$) then MDMs with higher lipid loads have a more inflammatory phenotype; if, however, resolving mediators dominate ($S_{+} > S_{-}$) then MDMs with higher lipid loads have a more resolving phenotype. The nonlinearity of the correlation is determined by the constant $q$, which measures the relative amount of mediator-MDM activity to lipid-MDM activity.
\\

\noindent \textbf{Distribution of lipid content: } Integrating Eqs. \eqref{eqn: mdist equilibrium} and \eqref{eqn: equilibrium bconds} with respect to $\Tilde{\phi} \in [-1,1]$ yields the following ODE for $W(\Tilde{\ell})$:
\begin{align}
    \frac{\epsilon}{2} \frac{d^2}{d \Tilde{\ell}^2} \Big[ \big( \Tilde{\ell}_\infty - (2\Tilde{\ell}_\infty - 1) \, \Tilde{\ell} \, \big) W \Big] - (\Tilde{\ell}_\infty - \Tilde{\ell}) \frac{dW}{d \Tilde{\ell}} - pW = 0, \label{eqn: W ODE}
\end{align}
and the boundary conditions:
\begin{align}
    - \frac{\epsilon}{2} \frac{d}{d\Tilde{\ell}} \Big[ \big( \Tilde{\ell}_\infty - (2 \Tilde{\ell}_\infty - 1) \, \Tilde{\ell} \big) \, W \Big] + (\Tilde{\ell}_\infty - \Tilde{\ell}) W = 
    \begin{cases}
        \frac{\Tilde{\ell}_\infty R_{0,0}}{\bm{k_L} \cdot \bm{L}}  & \text{at } \Tilde{\ell} = 0, \\
        0 & \text{at } \Tilde{\ell} = 1. 
    \end{cases} \label{eqn: W bconds}
\end{align}
Eq. \eqref{eqn: W ODE} admits an exact solution (found via the Mathematica \textit{DSolve} routine):
\begin{align}
    W(\Tilde{\ell}) = e^{-f(\Tilde{\ell})} \Big[ K_1 U_p^{2-f(\Tilde{\ell}_\infty)}(f(\Tilde{\ell})) + K_2 \mathcal{L}_{-p}^{1-f(\Tilde{\ell}_\infty)}(f(\Tilde{\ell})) \Big], \label{eqn: W exact}
\end{align}
where $f$ is the function:
\begin{align}
    f(\Tilde{\ell}) := \frac{2}{(1-2\Tilde{\ell}_\infty)\epsilon} \bigg[ \Tilde{\ell} + \bigg( \frac{\Tilde{\ell}_\infty}{1-2\Tilde{\ell}_\infty} \bigg) \bigg],
\end{align}
$U_a^b(z)$ is the confluent hypergeometric function and $\mathcal{L}_a^b(z)$ the generalised Laguerre polynomial. The constants $K_1$ and $K_2$ are determined by substituting Eq. \eqref{eqn: W exact} into the boundary conditions \eqref{eqn: W bconds}; exact expressions are readily obtained via the Mathematica \textit{Solve} routine, but are too involved to be insightful and so are omitted here for brevity. We compare the solution \eqref{eqn: W exact} to numerical solutions of the discrete model in the second column of Fig. \ref{fig:MDM_eq_distributions}, showing excellent agreement. 

The form of Eq. \eqref{eqn: W exact} makes it difficult to understand how $W(\Tilde{\ell})$ depends on the constants defined in Eqs. \eqref{eqn: par groupings}. Nonetheless, analysis of the leading order ``outer" solution shows how the qualitative behaviour of $W(\Tilde{\ell})$ depends on $p$ and $\Tilde{\ell}_\infty$. Here we follow the asymptotic analysis in \cite{chambers2023new}. Setting $\epsilon = 0$ in Eq. \eqref{eqn: W ODE} admits the following solution:
\begin{align}
    W_\text{outer}(\Tilde{\ell}) = K_3 \bigg( 1 - \frac{\Tilde{\ell}}{\Tilde{\ell}_\infty} \bigg)^p, \label{eqn: W outer}
\end{align}
where $K_3 \geq 0$ is a constant of integration. Eq. \eqref{eqn: W outer} admits three possible behaviours as $\Tilde{\ell} \rightarrow \Tilde{\ell}_\infty$ from below. For $p>0$, $W_\text{outer}$ decreases monotonically to zero (with non-zero derivative if $p < 2$); for $p = 0$, $W_\text{outer}$ is constant; and for $p < 0$, $W_\text{outer}$ diverges to $+\infty$. The full solution does not exhibit such discontinuities at $\Tilde{\ell} = \Tilde{\ell}_\infty$ due to the regularising effects of the second derivative term. The corresponding behaviour for $W(\Tilde{\ell})$ is as follows. For $p > 0$, $W$ decreases monotonically and smoothly to zero near $\Tilde{\ell} = \Tilde{\ell}_\infty$ (Fig. \ref{fig:MDM_eq_distributions}b); for $p = 0$, $W$ takes a quasi-uniform sigmoidal profile with a rapid decrease to zero near $\Tilde{\ell} = \Tilde{\ell}_\infty$ (Fig. \ref{fig:MDM_eq_distributions}c); and for $p < 0$, $W$ increases monotonically before attaining a local maximum near $\Tilde{\ell} = \Tilde{\ell}_\infty$ (Fig. \ref{fig:MDM_eq_distributions}a,d). In the above, ``near $\Tilde{\ell} = \Tilde{\ell}_\infty$" means an $\mathcal{O}(\epsilon^{\frac{1}{2}})$ neighbourhood of $\Tilde{\ell} = \Tilde{\ell}_\infty$. This scaling can be derived by searching for an inner variable $\hat{\ell} =(\Tilde{\ell} - \Tilde{\ell}_\infty)/ \epsilon^n$ for some $\hat{\ell} = \mathcal{O}(1)$ in Eq. \eqref{eqn: W ODE}; an exponent $n = \frac{1}{2}$ is required to bring the second derivative into the dominant balance. 
\\

\noindent \textbf{Distribution of phenotype: } Integrating Eqs. \eqref{eqn: mdist equilibrium} and \eqref{eqn: equilibrium bconds} with respect to $\Tilde{\ell} \in [0,1]$ gives the following boundary value problem for $V(\Tilde{\phi})$:
\begin{align}
    \frac{\epsilon \theta}{2} \frac{d^2}{d \Tilde{\phi}^2} \big[ (1 - \Tilde{\phi}_\infty \Tilde{\phi}) V \big] - (\Tilde{\phi}_\infty - \Tilde{\phi}) \frac{dV}{d \Tilde{\phi}} - r V = -\frac{\Tilde{\ell}_\infty R_{0,0}}{q \bm{k_L} \cdot \bm{L}} \cdot \delta_0 (\Tilde{\phi}), \label{eqn: V ODE} \\
     - \frac{\epsilon \theta }{2} \frac{\partial}{\partial \Tilde{\phi}} \Big[ (1 - \Tilde{\phi}_\infty \Tilde{\phi}) V \Big] + (\Tilde{\phi}_\infty - \Tilde{\phi}) V = 0 \quad \text{at } \Tilde{\phi} = \pm 1, \label{eqn: V bconds}
\end{align}
where $r := -1 + \frac{p+1}{q}$. As in the derivation of a Green's function, we first recast the Dirac-delta source, which describes MDM recruitment, as a jump condition. That is, we seek a solution of the form:
\begin{align}
    V(\Tilde{\phi}) = 
    \begin{cases}
        V_{-}(\Tilde{\phi}), & -1 \leq \Tilde{\phi} \leq 0; \\
        V_{+}(\Tilde{\phi}), & \,\,\,\, \, 0 \leq \Tilde{\phi} \leq 1,
    \end{cases}
\end{align}
where $V_\pm$ solve the ODE:
\begin{align}
    \frac{\epsilon \theta}{2} \frac{d^2}{d \Tilde{\phi}^2} \big[ (1 - \Tilde{\phi}_\infty \Tilde{\phi}) V_{\pm} ] \big] - (\Tilde{\phi}_\infty - \Tilde{\phi}) \frac{dV_\pm}{d \Tilde{\phi}} - r V_\pm = 0, \label{eqn: Vpm ODE}
\end{align}
and satisfy:
\begin{align}
    &V_{-}(0) = V_{+}(0) \label{eqn: Vpm continuity}\\
    &V_{-}'(0) = V_{+}'(0) + \frac{2 \Tilde{\ell}_\infty R_{0,0}}{\epsilon \theta q \bm{k_L} \cdot \bm{L}} \label{eqn: Vpm jump} \\
    -&\frac{\epsilon \theta }{2} \Big[ (1 - \Tilde{\phi}_\infty \Tilde{\phi}) V_{\pm} \Big]' + (\Tilde{\phi}_\infty - \Tilde{\phi}) V_{\pm} = 0 \quad \text{at } \Tilde{\phi} = \pm 1. \label{eqn: Vpm bconds}
\end{align}
The exact solution for Eq. \eqref{eqn: Vpm ODE} can be written as:
\begin{align}
    V_{\pm}(\Tilde{\phi}) &= J_{\pm, 1} U_{-r}^{1+g(\Tilde{\phi}_\infty)}(g(\Tilde{\phi})) + J_{\pm, 2} \mathcal{L}_r^{g(\Tilde{\phi}_\infty)}(g(\Tilde{\phi})), \label{eqn: Vpm sol}
\end{align}
where:
\begin{align}
    g(\Tilde{\phi}) := \frac{2(\Tilde{\phi}_\infty \Tilde{\phi} - 1)}{\epsilon \theta \Tilde{\phi}_\infty^2}.
\end{align}
Expressions for the constants $J_{\pm, 1}$, $J_{\pm, 2}$ can be found by substituting Eq. \eqref{eqn: Vpm sol} into conditions \eqref{eqn: Vpm continuity}-\eqref{eqn: Vpm bconds}. As for $W(\Tilde{\ell})$, they are too complicated to be insightful and are omitted for brevity. The comparison of solutions \eqref{eqn: Vpm sol} with the discrete model output in column 3 of Fig. \ref{fig:MDM_eq_distributions} shows good agreement. We note that the jump condition amounts to a small reduction in slope at $\Tilde{\phi} = 0$ that is almost imperceptible for the parameter values in Table \ref{tab:parameters}. 

Following the analysis of $W(\Tilde{\ell})$, we compute the leading order ``outer" solution to Eq. \eqref{eqn: V ODE}:
\begin{align}
    V_{\text{outer}} &= J_3 \Big( 1 - \frac{\Tilde{\phi}}{\Tilde{\phi}_\infty} \Big)^r. \label{eqn: V outer}
\end{align}
Eq. \eqref{eqn: V outer} is of the same form as Eq. \eqref{eqn: W outer}; $\Tilde{\phi}_\infty$ and $r$ play the same roles for $V$ as $\Tilde{\ell}_\infty$ and $p$ play for $W$. The main difference, as shown in Fig. \ref{fig:MDM_eq_distributions}, is that only the case $r < 0$ manifests in numerical solutions. Consequently, $V(\Tilde{\phi})$ always attains a local maximum in a small (again $\mathcal{O}(\epsilon^{\frac{1}{2}})$) neighbourhood of $\Tilde{\phi} = \Tilde{\phi}_\infty$. This observation is supported by the numerical results below.
\\

\noindent \textbf{Impact of blood LDL lipid and HDL capacity levels. } We conclude our steady state analysis by examining how the qualitative form of the MDM distribution changes as $L^\star$ and $H^\star$ vary. The results are summarised in Figure \ref{fig:mdist_features_LH} where we plot the parameter groupings in Eqs. \eqref{eqn: par groupings} against $L^\star$ and $H^\star$; the four cases shown in Fig. \ref{fig:MDM_eq_distributions} are also indicated. Recall that $p$ and $r$ determine the overall shape (e.g. whether they attain a local maximum) of the lipid and phenotype marginal distributions, $W(\Tilde{\ell})$ and $V(\Tilde{\phi})$ respectively, while $\Tilde{\phi}_\infty$, $\Tilde{\ell}_\infty$ and $q$ determine the central curve.

We note first how $r$ and $p$ vary with $L^\star$ and $H^\star$. The value of $r$ decreases monotonically with $L^\star$ and is less sensitive to $H^\star$. Importantly, we find that $r < 0$ for all $L^\star \in (0,10)$ and $H^\star \in (0,5)$, indicating that the phenotype marginal attains a local maximum near $\Tilde{\phi} = \Tilde{\phi}_\infty$ across the range of physiologically plausible values for blood LDL lipid and HDL capacity. By contrast, $p \geq 0$ for values of $(L^\star, H^\star)$ sufficiently close to the origin. This suggests that monotone decreasing and sigmoidal lipid marginals only occur when blood levels of LDL lipid and HDL capacity are both sufficiently low; otherwise the lipid marginal attains a local maximum near $\Tilde{\ell} = \Tilde{\ell}_\infty$. 

The target phenotype, $\Tilde{\phi}_\infty$, and lipid content, $\Tilde{\ell}_\infty$, both increase with $L^\star$ and decrease with $H^\star$. In particular, their plots exhibit the linear contours prominent in Fig. \ref{fig:eq_LH}; increases to the relative level of LDL lipid to HDL capacity raise both these markers of pathology. 

Finally, we note that the exponent $q$ of the central curve is largest when $L^\star$ and $H^\star$ are comparable. This causes the central curve to appear straighter (i.e. $\Tilde{\phi}_c(\Tilde{\ell}) \approx \Tilde{\phi}_\infty$ except for the smallest values of $\Tilde{\ell}$) in cases of intermediate pathology. Indeed, in Fig. \ref{fig:MDM_eq_distributions} the intermediate cases b) and c) are notably straighter than the unhealthy case d). 

\section{Discussion} \label{sec: discussion}

In this paper we have developed a mathematical model for early atherosclerosis in which the MDM population is structured by phenotype, $\phi$, and lipid content, $\ell$. This framework allows for incremental changes in phenotype and lipid content, which contrasts their treatment as binary variables in much of the existing modelling literature. The model couples the MDM dynamics to the densities of: free LDL lipid $L_{\text{\scaleto{LDL}{3.5pt}}}(t)$, retained LDL lipid $L_{\text{r}}(t)$, apoptotic lipid, $L_\text{ap}(t)$, necrotic lipid, $L_\text{n}(t)$, HDL capacity, $H(t)$, inflammatory mediators, $S_{+}(t)$, and resolving mediators, $S_{-}(t)$. These variables form a closed subsystem of ODEs when coupled with the MDM density, $M(t)$, mean phenotype, $\hat{\Phi}_M(t)$, and mean lipid content, $\hat{L}_M(t)$. This subsystem can be solved independently of the structured MDM population, $m_{\phi,\ell}(t)$.

We parameterised the model using data from the biological literature. Where possible, we used human \textit{in vivo} data (e.g. blood measurements for $L^{(0)}$ and $H^{(0)}$) or \textit{ex vivo} data (e.g. surgical data for $\pi_{L}^{(0)}$ and postmortem data for $L^{(1)}$ and $H^{(1)}$). However, the majority of the model parameters are calibrated to \textit{in vitro} experiments. We prioritised studies with human cell lines (e.g. for $a_0$, $k_{\text{\scaleto{LDL}{3.5pt}}}$, $k_\text{r}$, $k_H$, $S_{+}^{\text{c50}}$, $\rho$), and used data from nonhuman cell lines when necessary (e.g. murine data for $\kappa$, $k_{\text{ap}}$ and $k_\text{n}$). Since the point estimates in Table \ref{tab:parameters} are likely to carry high degrees of uncertainty, we have cautiously interpreted the results of the current study by focusing on trends rather than precise quantitative outputs. 

Our model analysis focused on the impact of three dimensionless parameters: $L^\star$, $H^\star$ and $K_\text{r}$. The quantities $L^\star$ and $H^\star$ are proportional to the blood densities of LDL lipid and HDL capacity, respectively. These vary considerably according to genetic and lifestyle factors. We explored values in the range $L^\star \in (0,10)$ and $H^\star \in (0,5)$; the upper bounds correspond to densities of $450$ and $225$mg/dL, respectively. The quantity $K_\text{r}$ is proportional to the capacity for LDL lipid retention. This varies according to the specific region of artery wall under consideration. We explored values $K_\text{r} \in (0.3, 100)$, which span athero-resistant regions with capacity $15$mg/dL to athero-prone regions with capacity $7500$mg/dL. 

We discuss our findings below in relation to the key questions posed in Sec. \ref{sec: Intro} . \\

\noindent \textbf{Q1.a. Time-evolution of lesion composition. } Time-dependent numerical solutions revealed that the model lesion evolves in three phases. The first phase consists of the initial influx of MDMs and corresponding decline in lesion LDL and rLDL content. MDM lipid loads are small but increasing and MDM mean phenotype is inflammatory. The second phase is characterised by modulation towards resolving MDM phenotypes and a slower rise of MDM lipid content. 
These phases are consistent with observations of macrophage behaviour during acute inflammation. Macrophages first adopt inflammatory phenotypes and transition to resolving phenotypes during a phase of inflammatory resolution and tissue repair \cite{perez2022macrophage}. To the best of our knowledge, the present model is the first to capture this transition as an emergent property of the dynamics. Rather than completely resolving, however, the model lesion enters a final phase in which the dynamics tend to a nonzero steady state. If blood LDL lipid is low relative to HDL capacity, the lesion tends to a healthy state with low lipid burden and resolving phenotypes. This behaviour suggests that the well-documented spontaneous regression of many early atherosclerotic lesions \cite{insull2009pathology} may simply be the natural and expected progression under a healthy blood lipoprotein balance. By contrast, when blood LDL lipid is high relative to HDL capacity, MDM lipid loads increase to equilibrium and the model lesion accumulates necrotic lipid. 
Overall, the three-phase dynamics support the idea that chronic inflammation in atherosclerosis can be understood as an acute inflammatory response to LDL retention with incomplete resolution \cite{sansbury2016resolution}. 

We also studied the impact of LDL retention capacity on the timescale of atherosclerosis development. In particular, we computed the time for MDM lipid to exceed a density of $450$mg/dL; we use this lipid density as a proxy for fatty streak formation, which is characterised by the appearance of foam cells \cite{daskalopoulos2015role}. Our results indicate that this time decreases with LDL retention capacity; fatty streak onset in the model lesion occurs earlier for regions of high LDL retention. This finding is consistent with observations of the murine aortic arch in which regions of lower retention capacity (e.g. central zone of the arch) developed atherosclerosis later than regions of high retention (e.g. dorsal and ventral zones) \cite{lewis2023capacity}. \\

\noindent \textbf{Q1.b. Lesion composition at steady state. } Analysis of steady state solutions revealed how equilibrium lesion composition depends on the parameters $L^\star$, $H^\star$ and $K_\text{r}$. The results indicate that the degree of pathology is largely determined by a linear combination of the form: $0.4 L^\star - H^\star$; it is the (weighted) \textit{relative} value of LDL lipid to HDL capacity in the blood that matters. The greater weighting on $H^\star$ in the linear combination reflects the higher value of the dimensionless lipid efflux rate, $k_H$, relative to the uptake rates in Table \ref{tab: nondim parameters}. Hence, the increased amount of lipid efflux promoted by a rise in blood HDL lipid capacity is greater than the increase in lesion lipid content by an equal rise in blood LDL lipid density. Overall, the model predicts that early atherosclerotic lesions regress upon blood LDL lipid density increases or blood HDL capacity increases. The degree of pathology also generally increases with $K_\text{r}$. Regions of higher LDL retention capacity exhibit greater MDM densities, more inflammatory MDM phenotypes and higher MDM lipid loads. \\

\noindent \textbf{Q2.a. Time evolution of MDM phenotype and lipid content. } We analysed the MDM distribution by deriving a continuum analogue of the discrete equations \eqref{eqn:mphil nondim} for $m_{\phi, \ell}(t)$. 
Analysis of the continuum model showed that the time-evolution of phenotype and lipid content for individual MDMs depends on the time of entry into the lesion. MDMs which enter the lesion at early times are expected to first transition from phenotypic neutrality to an inflammatory state, and then to a resolving state (if the blood LDL-HDL balance is healthy) or to a milder inflammatory state (if the LDL-HDL balance is unhealthy). The trajectories of MDMs which enter the lesion at later times are monotonic (in phenotype-lipid structure space) and follow the central curve of the equilibrium MDM distribution. These results suggest that MDMs which enter the lesion during early stages of lesion development experience a greater amount of phenotype modulation throughout their lifespan than those entering at later times. \\

\noindent \textbf{Q2.b. Are phenotype and lipid content correlated? } The asymptotic analysis presented in Subsection \ref{sec: MDM steady} showed that MDM phenotype and lipid content are correlated via a power law at steady state. If LDL lipid density dominates HDL capacity in the blood, lipid-laden MDMs have a more inflammatory phenotype than lipid-poor MDMs, while if blood LDL lipid density is sufficiently low, lipid-laden MDMs have a more resolving phenotype than MDMs with a lower lipid burden. The non-linearity of the correlation is determined by a constant, $q$, that measures the relative amount of lipid activity to mediator activity in the lesion (made precise by Eq. \eqref{eqn: par groupings}). Although we did not pursue a time-dependent mathematical analysis, numerical solutions (e.g. Fig. \ref{fig:MDM_dynamics}) demonstrated that the MDM phenotype-lipid distribution is always concentrated about a central curve, indicating that a monotone correlation between phenotype and lipid content holds for all times. These findings are consistent with the recent discovery of PLIN2$^{\text{hi}}$/TREM1$^{\text{hi}}$ macrophages in human lesions, for which the transcriptional signatures of lipid loading and inflammation are coupled \cite{dib2023lipid}.\\


\noindent \textbf{Q2.c. Features of the marginal distributions at steady state. } Further analysis in Subsection \ref{sec: MDM steady} showed that the phenotype marginal distribution always attains a single local maximum. The location of the maximum, which represents the most common phenotype in the lesion, is a close approximation to the ``target" phenotype that MDMs are driven towards by the extracellular environment over their lifetime. By contrast, the lipid marginal distribution varies more in shape as $L^\star$ and $H^\star$ are varied; it may exhibit a local maximum, adopt a quasi-uniform profile, or decrease monotonically according to the value of $p$, a constant which quantifies the amount of MDM-lipid activity in the lesion. 
\\

\noindent \textbf{Future directions and conclusions. } There is considerable scope to extend our model. In the present model, phenotype influences the MDM dynamics via its impact on mediator production; macrophage phenotype is largely characterised by the profile of cytokines/effector molecules produced by the cell \cite{brown2009macrophage}. In practice, phenotype is also correlated with other behaviours, including: phagocytic ability \cite{schulz2019depth}, lipid efflux to HDL \cite{lin2021macrophage} and migratory propensity \cite{cui2018distinct}. These effects could be incorporated by respectively allowing the rates of lipid uptake, $ k_{\text{\scaleto{LDL}{3.5pt}}}, k_\text{r}, k_\text{ap}, k_\text{n}$, lipid efflux, $k_H$, and MDM egress, $\gamma$, to depend on $\phi$.

Lipid-dependent MDM behaviours could also be included by allowing the corresponding rates to depend on $\ell$. Lipid-dependent apoptosis, emigration and proliferation were analysed in the work of Watson et al. \cite{watson2023lipid}, who found that such dependencies substantially altered total lesion lipid content, and the distribution of lipid amongst MDMs, apoptotic cells and the necrotic core. Importantly, some experimental studies indicate that lipid loading can inhibit macrophage pro-inflammatory responses \cite{leitinger2013phenotypic, kim2018transcriptome}. This effect could be accounted for in our model by allowing $p_{\phi}^{+}$ to decrease with $\ell$ in reaction \eqref{eqn: S+- reactions}. Extending our model to account for phenotype and lipid-dependent rates produces coupling in the ODEs that would not admit a closed subsystem; analysis of these effects would rely heavily on numerical solutions. 

Finally, we could extend the model to allow for spatial heterogeneity. 
Based on previous studies using spatially-resolved structured population models \cite{celora2023spatio, fiandaca2022phenotype, pan2022propagation, boulouz2022spatially,hu2019spatial, liu2015hopf}, we anticipate that analysis of a spatial extension would rely heavily on numerical simulations and bifurcation analysis. 

In conclusion, in this paper we have presented a new mathematical model for early atherosclerosis in which the MDM population is structured according to phenotype and lipid content. 
The model indicates that lesion composition depends sensitively on the relative density of LDL lipid to HDL capacity in the blood, and the LDL retention capacity of the artery wall. 
Numerical and analytical results at steady state show that MDM phenotype and lipid content are monotonically correlated via a power law, the phenotype marginal distribution is unimodal, and the lipid content distribution may attain a unimodal, quasi-uniform or decreasing profile. These findings develop the current understanding of macrophage heterogeneity in early atherosclerosis.

\backmatter





\bmhead{Acknowledgments}

KLC acknowledges support from the Oxford Australia Scholarships Fund and Clarendon Scholars' Association. KLC and HMB would like to thank the Isaac Newton Institute for Mathematical Sciences, Cambridge, for support and hospitality during the programme Mathematics of movement: an interdisciplinary approach to mutual challenges in animal ecology and cell biology, where work on this paper was undertaken. This work was supported by EPSRC grant EP/R014604/1. All authors acknowledge funding (to MRM, HMB) from the Australia Research Council Discovery Program, grant number DP200102071.

\bmhead{Data statement}

The code used in the current study is available from the corresponding author on request. 


\section*{Declarations}

The authors declare that they have no conflicts of interest.









\begin{appendices}

\section{Parameter estimation}\label{sec: appendix parameters}

The exchange rates, $\pi^{0}_L, \pi^{1}_L$, $\pi^{0}_H$ and $\pi^{1}_H$ are obtained by dividing the respective permeabilities by the human coronary artery tunica intima thickness ($\sim 0.24$ mm \cite{holzapfel2005determination}). We note that the intimal thickness in mice is considerably smaller ($\sim 0.01$mm \cite{thon2018quantitative}), and so murine exchange rates are higher. 

The maximum MDM entry rate, $\sigma_M$, is based on data from monocyte transmigration experiments \cite{williams2009transmigration}. Specifically, Williams \textit{et al.} observed that when human monocytes were placed above stimulated endothelial cells, approximately $20\%$ entered the sub-endothelial layer within $1.5$ hours. Multiplying this rate by $350$ mm$^{-3}$, a typical blood monocyte density \cite{nelson2014immunobiology}, gives $~34000$ mm$^{-3}$ per month. Since approximately half the incoming monocytes do not differentiate into macrophages and leave the lesion within $48$ hours \cite{lee2019sirt1}, we estimate $\sigma_M = 17000$ mm$^{-3}$ per month.

The uptake and efflux rates are obtained from \textit{in vitro} experiments. We note that the free LDL uptake rate, $k_{\text{\scaleto{LDL}{3.5pt}}}$, is calibrated to data for native LDL (data for oxidised LDL yields a slightly higher but comparable estimate), while the larger rLDL uptake rate, $k_\text{r}$, is based on aggregated LDL data. This is reasonable since proteoglycan-bound LDL aggregates into lipid droplets that are more readily taken up by MDMs \textit{in vivo} \cite{oorni2021aggregation}. The apoptotic lipid uptake rate, $k_\text{ap}$, has been scaled down from its \textit{in vitro} estimate by a factor $19$ to account for the $19$-fold deficiency in efferocytosis measured in human atherosclerotic plaques \cite{schrijvers2005phagocytosis}. 

The mediator mass, $\Delta s$, and natural decay rate, $\delta_S$, are based on data for the inflammatory mediator TNF, which has a molecular weight of $17$kDa in its secreted form as a monomer and a $19$ minute half-life \cite{atzeni2013tumor, liu2021cytokines}. Other mediators have a comparable mass and half-life (e.g. $15$ kDa and 20 minutes respectively for the resolving mediator IL-4 \cite{liu2021cytokines}). 

We estimate the rate of rLDL-stimulated mediator production from resident cells, $\alpha$, using reports that MDMs are observed in murine atherosclerotic lesions within 2 weeks of high fat diet treatment \cite{williams2020limited}. The value $\alpha = 3.9 \times 10^{-5}$ per month ensures that the MDM density in our model is approximately $350$ mm$^{-3}$ (approximately blood monocyte concentration) in 2 weeks when $L^{(0)} = 750$ mg/dL.

The MDM phenotypic plasticity, $\chi$, is calibrated to experiments of macrophage phenotype modulation  \textit{in vitro} \cite{tarique2015phenotypic}. Tarique \textit{et al.} observed that macrophages exposed to inflammatory and resolving mediators at $20$ ng/mL underwent phenotypic polarisation in 2 days. We fix $\chi = 6.3\times10^{-4}$ so that phenotype modulation in our model occurs on this timescale (i.e. $\hat{\Phi}_M(t = 2\text{days}) = \pm 1/2$ if $S_{\pm} = 20$ng/mL and $\hat{\Phi}_M(t = 0) = 0$). 

Finally, we note that the lipid uptake/efflux increment, $\Delta a$, and maximal structure indices, $\ell_\text{max}$ and $\phi_{\text{max}}$, are model abstractions. In reality, lipid increments vary substantially depending on the specific interaction. The smallest increment is likely that associated with efflux, given that each HDL particle can store approximately $75$kDa of cholesterol. The largest increment is likely that due to apoptotic or necrotic lipid uptake, which we can assume to be smaller than the endogenous lipid content, $a_0$, since cellular uptake \textit{in vivo} is typically piecemeal rather than involving whole cell engulfment \cite{taefehshokr2021rab}. We fix $\Delta a = 16$pg as a reasonable intermediate value. This choice means that $\ell_\text{max} = \kappa/\Delta a \approx 100$. The value of $\phi_\text{max}$ has an upper bound of $(2\chi)^{-1} \approx 83000$ since the phenotype transition probabilities in Eqs.\eqref{eqn: S+- reactions} must satisfy $p_{\phi}^{\pm} \leq 1$. We fix $\phi_\text{max} = 50$ so that phenotype and lipid content are equally resolved in our model.





\end{appendices}


\bibliography{sn-bibliography}


\begin{thebibliography}{122}
\ifx \bisbn   \undefined \def \bisbn  #1{ISBN #1}\fi
\ifx \binits  \undefined \def \binits#1{#1}\fi
\ifx \bauthor  \undefined \def \bauthor#1{#1}\fi
\ifx \batitle  \undefined \def \batitle#1{#1}\fi
\ifx \bjtitle  \undefined \def \bjtitle#1{#1}\fi
\ifx \bvolume  \undefined \def \bvolume#1{\textbf{#1}}\fi
\ifx \byear  \undefined \def \byear#1{#1}\fi
\ifx \bissue  \undefined \def \bissue#1{#1}\fi
\ifx \bfpage  \undefined \def \bfpage#1{#1}\fi
\ifx \blpage  \undefined \def \blpage #1{#1}\fi
\ifx \burl  \undefined \def \burl#1{\textsf{#1}}\fi
\ifx \doiurl  \undefined \def \doiurl#1{\url{https://doi.org/#1}}\fi
\ifx \betal  \undefined \def \betal{\textit{et al.}}\fi
\ifx \binstitute  \undefined \def \binstitute#1{#1}\fi
\ifx \binstitutionaled  \undefined \def \binstitutionaled#1{#1}\fi
\ifx \bctitle  \undefined \def \bctitle#1{#1}\fi
\ifx \beditor  \undefined \def \beditor#1{#1}\fi
\ifx \bpublisher  \undefined \def \bpublisher#1{#1}\fi
\ifx \bbtitle  \undefined \def \bbtitle#1{#1}\fi
\ifx \bedition  \undefined \def \bedition#1{#1}\fi
\ifx \bseriesno  \undefined \def \bseriesno#1{#1}\fi
\ifx \blocation  \undefined \def \blocation#1{#1}\fi
\ifx \bsertitle  \undefined \def \bsertitle#1{#1}\fi
\ifx \bsnm \undefined \def \bsnm#1{#1}\fi
\ifx \bsuffix \undefined \def \bsuffix#1{#1}\fi
\ifx \bparticle \undefined \def \bparticle#1{#1}\fi
\ifx \barticle \undefined \def \barticle#1{#1}\fi
\bibcommenthead
\ifx \bconfdate \undefined \def \bconfdate #1{#1}\fi
\ifx \botherref \undefined \def \botherref #1{#1}\fi
\ifx \url \undefined \def \url#1{\textsf{#1}}\fi
\ifx \bchapter \undefined \def \bchapter#1{#1}\fi
\ifx \bbook \undefined \def \bbook#1{#1}\fi
\ifx \bcomment \undefined \def \bcomment#1{#1}\fi
\ifx \oauthor \undefined \def \oauthor#1{#1}\fi
\ifx \citeauthoryear \undefined \def \citeauthoryear#1{#1}\fi
\ifx \endbibitem  \undefined \def \endbibitem {}\fi
\ifx \bconflocation  \undefined \def \bconflocation#1{#1}\fi
\ifx \arxivurl  \undefined \def \arxivurl#1{\textsf{#1}}\fi
\csname PreBibitemsHook\endcsname

\bibitem{back2019inflammation}
\begin{barticle}
\bauthor{\bsnm{B{\"a}ck}, \binits{M.}},
\bauthor{\bsnm{Yurdagul~Jr}, \binits{A.}},
\bauthor{\bsnm{Tabas}, \binits{I.}},
\bauthor{\bsnm{{\"O}{\"o}rni}, \binits{K.}},
\bauthor{\bsnm{Kovanen}, \binits{P.T.}}:
\batitle{Inflammation and its resolution in atherosclerosis: mediators and therapeutic opportunities}.
\bjtitle{Nature Reviews Cardiology}
\bvolume{16}(\bissue{7}),
\bfpage{389}--\blpage{406}
(\byear{2019}).
\doiurl{10.1038/s41569-019-0169-2}
\end{barticle}
\endbibitem

\bibitem{willemsen2020macrophage}
\begin{barticle}
\bauthor{\bsnm{Willemsen}, \binits{L.}},
\bauthor{\bparticle{de} \bsnm{Winther}, \binits{M.P.}}:
\batitle{Macrophage subsets in atherosclerosis as defined by single-cell technologies}.
\bjtitle{The Journal of Pathology}
\bvolume{250}(\bissue{5}),
\bfpage{705}--\blpage{714}
(\byear{2020}).
\doiurl{10.1002/path.5392}
\end{barticle}
\endbibitem

\bibitem{kloc2020role}
\begin{barticle}
\bauthor{\bsnm{Kloc}, \binits{M.}},
\bauthor{\bsnm{Uosef}, \binits{A.}},
\bauthor{\bsnm{Kubiak}, \binits{J.Z.}},
\bauthor{\bsnm{Ghobrial}, \binits{R.M.}}:
\batitle{Role of macrophages and rhoa pathway in atherosclerosis}.
\bjtitle{International Journal of Molecular Sciences}
\bvolume{22}(\bissue{1}),
\bfpage{216}
(\byear{2020}).
\doiurl{10.3390/ijms22010216}
\end{barticle}
\endbibitem

\bibitem{tabas2016macrophage}
\begin{barticle}
\bauthor{\bsnm{Tabas}, \binits{I.}},
\bauthor{\bsnm{Bornfeldt}, \binits{K.E.}}:
\batitle{Macrophage phenotype and function in different stages of atherosclerosis}.
\bjtitle{Circulation research}
\bvolume{118}(\bissue{4}),
\bfpage{653}--\blpage{667}
(\byear{2016}).
\doiurl{10.1161/CIRCRESAHA.115.306256}
\end{barticle}
\endbibitem

\bibitem{guyton1996development}
\begin{barticle}
\bauthor{\bsnm{Guyton}, \binits{J.R.}},
\bauthor{\bsnm{Klemp}, \binits{K.F.}}:
\batitle{Development of the lipid-rich core in human atherosclerosis}.
\bjtitle{Arteriosclerosis, thrombosis, and vascular biology}
\bvolume{16}(\bissue{1}),
\bfpage{4}--\blpage{11}
(\byear{1996}).
\doiurl{10.1161/01.atv.16.1.4}
\end{barticle}
\endbibitem

\bibitem{gonzalez2017macrophage}
\begin{barticle}
\bauthor{\bsnm{Gonzalez}, \binits{L.}},
\bauthor{\bsnm{Trigatti}, \binits{B.L.}}:
\batitle{Macrophage apoptosis and necrotic core development in atherosclerosis: a rapidly advancing field with clinical relevance to imaging and therapy}.
\bjtitle{Canadian Journal of Cardiology}
\bvolume{33}(\bissue{3}),
\bfpage{303}--\blpage{312}
(\byear{2017}).
\doiurl{10.1016/j.cjca.2016.12.010}
\end{barticle}
\endbibitem

\bibitem{costopoulos2017plaque}
\begin{barticle}
\bauthor{\bsnm{Costopoulos}, \binits{C.}},
\bauthor{\bsnm{Huang}, \binits{Y.}},
\bauthor{\bsnm{Brown}, \binits{A.J.}},
\bauthor{\bsnm{Calvert}, \binits{P.A.}},
\bauthor{\bsnm{Hoole}, \binits{S.P.}},
\bauthor{\bsnm{West}, \binits{N.E.}},
\bauthor{\bsnm{Gillard}, \binits{J.H.}},
\bauthor{\bsnm{Teng}, \binits{Z.}},
\bauthor{\bsnm{Bennett}, \binits{M.R.}}:
\batitle{Plaque rupture in coronary atherosclerosis is associated with increased plaque structural stress}.
\bjtitle{JACC: Cardiovascular Imaging}
\bvolume{10}(\bissue{12}),
\bfpage{1472}--\blpage{1483}
(\byear{2017}).
\doiurl{10.1016/j.jcmg.2017.04.017}
\end{barticle}
\endbibitem

\bibitem{rothwell2007atherothrombosis}
\begin{botherref}
\oauthor{\bsnm{Rothwell}, \binits{P.}}:
Atherothrombosis and ischaemic stroke.
British Medical Journal Publishing Group
(2007).
\doiurl{10.1136/bmj.38964.489051.80}
\end{botherref}
\endbibitem

\bibitem{leitinger2013phenotypic}
\begin{barticle}
\bauthor{\bsnm{Leitinger}, \binits{N.}},
\bauthor{\bsnm{Schulman}, \binits{I.G.}}:
\batitle{Phenotypic polarization of macrophages in atherosclerosis}.
\bjtitle{Arteriosclerosis, thrombosis, and vascular biology}
\bvolume{33}(\bissue{6}),
\bfpage{1120}--\blpage{1126}
(\byear{2013}).
\doiurl{10.1161/ATVBAHA.112.300173}
\end{barticle}
\endbibitem

\bibitem{barrett2020macrophages}
\begin{barticle}
\bauthor{\bsnm{Barrett}, \binits{T.J.}}:
\batitle{Macrophages in atherosclerosis regression}.
\bjtitle{Arteriosclerosis, thrombosis, and vascular biology}
\bvolume{40}(\bissue{1}),
\bfpage{20}--\blpage{33}
(\byear{2020}).
\doiurl{10.1161/ATVBAHA.119.312802}
\end{barticle}
\endbibitem

\bibitem{lin2021macrophage}
\begin{barticle}
\bauthor{\bsnm{Lin}, \binits{P.}},
\bauthor{\bsnm{Ji}, \binits{H.-H.}},
\bauthor{\bsnm{Li}, \binits{Y.-J.}},
\bauthor{\bsnm{Guo}, \binits{S.-D.}}:
\batitle{Macrophage plasticity and atherosclerosis therapy}.
\bjtitle{Frontiers in molecular biosciences}
\bvolume{8},
\bfpage{679797}
(\byear{2021}).
\doiurl{10.3389/fmolb.2021.679797}
\end{barticle}
\endbibitem

\bibitem{wang2014molecular}
\begin{barticle}
\bauthor{\bsnm{Wang}, \binits{N.}},
\bauthor{\bsnm{Liang}, \binits{H.}},
\bauthor{\bsnm{Zen}, \binits{K.}}:
\batitle{Molecular mechanisms that influence the macrophage m1--m2 polarization balance}.
\bjtitle{Frontiers in immunology}
\bvolume{5},
\bfpage{614}
(\byear{2014}).
\doiurl{10.3389/fimmu.2014.00614}
\end{barticle}
\endbibitem

\bibitem{liu2014oxldl}
\begin{barticle}
\bauthor{\bsnm{Liu}, \binits{W.}},
\bauthor{\bsnm{Yin}, \binits{Y.}},
\bauthor{\bsnm{Zhou}, \binits{Z.}},
\bauthor{\bsnm{He}, \binits{M.}},
\bauthor{\bsnm{Dai}, \binits{Y.}}:
\batitle{Oxldl-induced il-1beta secretion promoting foam cells formation was mainly via cd36 mediated ros production leading to nlrp3 inflammasome activation}.
\bjtitle{Inflammation Research}
\bvolume{63},
\bfpage{33}--\blpage{43}
(\byear{2014}).
\doiurl{10.1007/s00011-013-0667-3}
\end{barticle}
\endbibitem

\bibitem{sachet2017immune}
\begin{barticle}
\bauthor{\bsnm{Sachet}, \binits{M.}},
\bauthor{\bsnm{Liang}, \binits{Y.Y.}},
\bauthor{\bsnm{Oehler}, \binits{R.}}:
\batitle{The immune response to secondary necrotic cells}.
\bjtitle{Apoptosis}
\bvolume{22}(\bissue{10}),
\bfpage{1189}--\blpage{1204}
(\byear{2017}).
\doiurl{10.1007/s10495-017-1413-z}
\end{barticle}
\endbibitem

\bibitem{decker2021pro}
\begin{barticle}
\bauthor{\bsnm{Decker}, \binits{C.}},
\bauthor{\bsnm{Sadhu}, \binits{S.}},
\bauthor{\bsnm{Fredman}, \binits{G.}}:
\batitle{Pro-resolving ligands orchestrate phagocytosis}.
\bjtitle{Frontiers in immunology}
\bvolume{12},
\bfpage{660865}
(\byear{2021}).
\doiurl{10.3389/fimmu.2021.660865}
\end{barticle}
\endbibitem

\bibitem{serhan2018resolvins}
\begin{barticle}
\bauthor{\bsnm{Serhan}, \binits{C.N.}},
\bauthor{\bsnm{Levy}, \binits{B.D.}}, \betal:
\batitle{Resolvins in inflammation: emergence of the pro-resolving superfamily of mediators}.
\bjtitle{The Journal of clinical investigation}
\bvolume{128}(\bissue{7}),
\bfpage{2657}--\blpage{2669}
(\byear{2018}).
\doiurl{10.1172/JCI97943}
\end{barticle}
\endbibitem

\bibitem{parton2016computational}
\begin{barticle}
\bauthor{\bsnm{Parton}, \binits{A.}},
\bauthor{\bsnm{McGilligan}, \binits{V.}},
\bauthor{\bsnm{O’Kane}, \binits{M.}},
\bauthor{\bsnm{Baldrick}, \binits{F.R.}},
\bauthor{\bsnm{Watterson}, \binits{S.}}:
\batitle{Computational modelling of atherosclerosis}.
\bjtitle{Briefings in bioinformatics}
\bvolume{17}(\bissue{4}),
\bfpage{562}--\blpage{575}
(\byear{2016}).
\doiurl{10.1093/bib/bbv081}
\end{barticle}
\endbibitem

\bibitem{avgerinos2019mathematical}
\begin{barticle}
\bauthor{\bsnm{Avgerinos}, \binits{N.A.}},
\bauthor{\bsnm{Neofytou}, \binits{P.}}:
\batitle{Mathematical modelling and simulation of atherosclerosis formation and progress: a review}.
\bjtitle{Annals of Biomedical Engineering}
\bvolume{47},
\bfpage{1764}--\blpage{1785}
(\byear{2019}).
\doiurl{10.1007/s10439-019-02268-3}
\end{barticle}
\endbibitem

\bibitem{cai2021mathematical}
\begin{barticle}
\bauthor{\bsnm{Cai}, \binits{Y.}},
\bauthor{\bsnm{Li}, \binits{Z.}}:
\batitle{Mathematical modeling of plaque progression and associated microenvironment: How far from predicting the fate of atherosclerosis?}
\bjtitle{Computer Methods and Programs in Biomedicine}
\bvolume{211},
\bfpage{106435}
(\byear{2021}).
\doiurl{10.1016/j.cmpb.2021.106435}
\end{barticle}
\endbibitem

\bibitem{mc2022modeling}
\begin{barticle}
\bauthor{\bsnm{Mc~Auley}, \binits{M.T.}}:
\batitle{Modeling cholesterol metabolism and atherosclerosis}.
\bjtitle{WIREs Mechanisms of Disease}
\bvolume{14}(\bissue{3}),
\bfpage{1546}
(\byear{2022}).
\doiurl{10.1002/wsbm.1546}
\end{barticle}
\endbibitem

\bibitem{prosi2005mathematical}
\begin{barticle}
\bauthor{\bsnm{Prosi}, \binits{M.}},
\bauthor{\bsnm{Zunino}, \binits{P.}},
\bauthor{\bsnm{Perktold}, \binits{K.}},
\bauthor{\bsnm{Quarteroni}, \binits{A.}}:
\batitle{Mathematical and numerical models for transfer of low-density lipoproteins through the arterial walls: a new methodology for the model set up with applications to the study of disturbed lumenal flow}.
\bjtitle{Journal of biomechanics}
\bvolume{38}(\bissue{4}),
\bfpage{903}--\blpage{917}
(\byear{2005}).
\doiurl{10.1016/j.jbiomech.2004.04.024}
\end{barticle}
\endbibitem

\bibitem{yang2006modeling}
\begin{barticle}
\bauthor{\bsnm{Yang}, \binits{N.}},
\bauthor{\bsnm{Vafai}, \binits{K.}}:
\batitle{Modeling of low-density lipoprotein (ldl) transport in the artery—effects of hypertension}.
\bjtitle{International Journal of Heat and Mass Transfer}
\bvolume{49}(\bissue{5-6}),
\bfpage{850}--\blpage{867}
(\byear{2006}).
\doiurl{10.1016/j.ijheatmasstransfer.2005.09.019}
\end{barticle}
\endbibitem

\bibitem{yang2008low}
\begin{barticle}
\bauthor{\bsnm{Yang}, \binits{N.}},
\bauthor{\bsnm{Vafai}, \binits{K.}}:
\batitle{Low-density lipoprotein (ldl) transport in an artery--a simplified analytical solution}.
\bjtitle{International Journal of Heat and Mass Transfer}
\bvolume{51}(\bissue{3-4}),
\bfpage{497}--\blpage{505}
(\byear{2008}).
\doiurl{10.1016/j.ijheatmasstransfer.2007.05.023}
\end{barticle}
\endbibitem

\bibitem{fok2012mathematical}
\begin{barticle}
\bauthor{\bsnm{Fok}, \binits{P.-W.}}:
\batitle{Mathematical model of intimal thickening in atherosclerosis: vessel stenosis as a free boundary problem}.
\bjtitle{Journal of theoretical biology}
\bvolume{314},
\bfpage{23}--\blpage{33}
(\byear{2012}).
\doiurl{10.1016/j.jtbi.2012.07.029}
\end{barticle}
\endbibitem

\bibitem{watson2018two}
\begin{barticle}
\bauthor{\bsnm{Watson}, \binits{M.G.}},
\bauthor{\bsnm{Byrne}, \binits{H.M.}},
\bauthor{\bsnm{Macaskill}, \binits{C.}},
\bauthor{\bsnm{Myerscough}, \binits{M.R.}}:
\batitle{A two-phase model of early fibrous cap formation in atherosclerosis}.
\bjtitle{Journal of Theoretical Biology}
\bvolume{456},
\bfpage{123}--\blpage{136}
(\byear{2018}).
\doiurl{10.1016/j.jtbi.2018.08.010}
\end{barticle}
\endbibitem

\bibitem{fok2018media}
\begin{barticle}
\bauthor{\bsnm{Fok}, \binits{P.-W.}},
\bauthor{\bsnm{Lanzer}, \binits{P.}}:
\batitle{Media sclerosis drives and localizes atherosclerosis in peripheral arteries}.
\bjtitle{PLoS One}
\bvolume{13}(\bissue{10}),
\bfpage{0205599}
(\byear{2018}).
\doiurl{10.1371/journal.pone.0205599}
\end{barticle}
\endbibitem

\bibitem{watson2020multiphase}
\begin{barticle}
\bauthor{\bsnm{Watson}, \binits{M.G.}},
\bauthor{\bsnm{Byrne}, \binits{H.M.}},
\bauthor{\bsnm{Macaskill}, \binits{C.}},
\bauthor{\bsnm{Myerscough}, \binits{M.R.}}:
\batitle{A multiphase model of growth factor-regulated atherosclerotic cap formation}.
\bjtitle{Journal of Mathematical Biology}
\bvolume{81}(\bissue{2}),
\bfpage{725}--\blpage{767}
(\byear{2020}).
\doiurl{10.1007/s00285-020-01526-6}
\end{barticle}
\endbibitem

\bibitem{fok2021modeling}
\begin{bchapter}
\bauthor{\bsnm{Fok}, \binits{P.-W.}},
\bauthor{\bsnm{Mirzaei}, \binits{N.M.}}:
\bctitle{Modeling the glagov's compensatory enlargement of human coronary atherosclerotic plaque}.
In: \bbtitle{Biomechanics of Coronary Atherosclerotic Plaque},
pp. \bfpage{107}--\blpage{130}.
\bpublisher{Elsevier}, \blocation{???}
(\byear{2021}).
\doiurl{10.1016/B978-0-12-817195-0.00004-4}
\end{bchapter}
\endbibitem

\bibitem{bulelzai2012long}
\begin{barticle}
\bauthor{\bsnm{Bulelzai}, \binits{M.A.}},
\bauthor{\bsnm{Dubbeldam}, \binits{J.L.}}:
\batitle{Long time evolution of atherosclerotic plaques}.
\bjtitle{Journal of theoretical biology}
\bvolume{297},
\bfpage{1}--\blpage{10}
(\byear{2012}).
\doiurl{10.1016/j.jtbi.2011.11.023}
\end{barticle}
\endbibitem

\bibitem{cohen2014athero}
\begin{barticle}
\bauthor{\bsnm{Cohen}, \binits{A.}},
\bauthor{\bsnm{Myerscough}, \binits{M.R.}},
\bauthor{\bsnm{Thompson}, \binits{R.S.}}:
\batitle{Athero-protective effects of high density lipoproteins (hdl): an ode model of the early stages of atherosclerosis}.
\bjtitle{Bulletin of mathematical biology}
\bvolume{76},
\bfpage{1117}--\blpage{1142}
(\byear{2014}).
\doiurl{10.1007/s11538-014-9948-4}
\end{barticle}
\endbibitem

\bibitem{islam2015mathematical}
\begin{barticle}
\bauthor{\bsnm{Islam}, \binits{M.H.}},
\bauthor{\bsnm{Johnston}, \binits{P.}}:
\batitle{A mathematical model for atherosclerotic plaque formation and arterial wall remodelling}.
\bjtitle{ANZIAM Journal}
\bvolume{57},
\bfpage{320}--\blpage{345}
(\byear{2015}).
\doiurl{10.21914/anziamj.v57i0.10386}
\end{barticle}
\endbibitem

\bibitem{thon2018quantitative}
\begin{barticle}
\bauthor{\bsnm{Thon}, \binits{M.P.}},
\bauthor{\bsnm{Ford}, \binits{H.Z.}},
\bauthor{\bsnm{Gee}, \binits{M.W.}},
\bauthor{\bsnm{Myerscough}, \binits{M.R.}}:
\batitle{A quantitative model of early atherosclerotic plaques parameterized using in vitro experiments}.
\bjtitle{Bulletin of mathematical biology}
\bvolume{80},
\bfpage{175}--\blpage{214}
(\byear{2018}).
\doiurl{10.1007/s11538-017-0367-1}
\end{barticle}
\endbibitem

\bibitem{lui2021modelling}
\begin{barticle}
\bauthor{\bsnm{Lui}, \binits{G.}},
\bauthor{\bsnm{Myerscough}, \binits{M.R.}}:
\batitle{Modelling preferential phagocytosis in atherosclerosis: delineating timescales in plaque development}.
\bjtitle{Bulletin of Mathematical Biology}
\bvolume{83}(\bissue{9}),
\bfpage{96}
(\byear{2021}).
\doiurl{10.1007/s11538-021-00926-z}
\end{barticle}
\endbibitem

\bibitem{xie2022well}
\begin{barticle}
\bauthor{\bsnm{Xie}, \binits{X.}}:
\batitle{Well-posedness of a mathematical model of diabetic atherosclerosis}.
\bjtitle{Journal of mathematical analysis and applications}
\bvolume{505}(\bissue{2}),
\bfpage{125606}
(\byear{2022}).
\doiurl{10.1016/j.jmaa.2021.125606}
\end{barticle}
\endbibitem

\bibitem{calvez2009mathematical}
\begin{bchapter}
\bauthor{\bsnm{Calvez}, \binits{V.}},
\bauthor{\bsnm{Ebde}, \binits{A.}},
\bauthor{\bsnm{Meunier}, \binits{N.}},
\bauthor{\bsnm{Raoult}, \binits{A.}}:
\bctitle{Mathematical modelling of the atherosclerotic plaque formation}.
In: \bbtitle{ESAIM: Proceedings},
vol. \bseriesno{28},
pp. \bfpage{1}--\blpage{12}
(\byear{2009}).
\doiurl{10.1051/proc/2009036}.
\bcomment{EDP Sciences}
\end{bchapter}
\endbibitem

\bibitem{fok2012growth}
\begin{barticle}
\bauthor{\bsnm{Fok}, \binits{P.-W.}}:
\batitle{Growth of necrotic cores in atherosclerotic plaque}.
\bjtitle{Mathematical medicine and biology: a journal of the IMA}
\bvolume{29}(\bissue{4}),
\bfpage{301}--\blpage{327}
(\byear{2012}).
\doiurl{10.1093/imammb/dqr012}
\end{barticle}
\endbibitem

\bibitem{hao2014ldl}
\begin{barticle}
\bauthor{\bsnm{Hao}, \binits{W.}},
\bauthor{\bsnm{Friedman}, \binits{A.}}:
\batitle{The ldl-hdl profile determines the risk of atherosclerosis: a mathematical model}.
\bjtitle{PloS one}
\bvolume{9}(\bissue{3}),
\bfpage{90497}
(\byear{2014}).
\doiurl{10.1371/journal.pone.0090497}
\end{barticle}
\endbibitem

\bibitem{chalmers2015bifurcation}
\begin{barticle}
\bauthor{\bsnm{Chalmers}, \binits{A.D.}},
\bauthor{\bsnm{Cohen}, \binits{A.}},
\bauthor{\bsnm{Bursill}, \binits{C.A.}},
\bauthor{\bsnm{Myerscough}, \binits{M.R.}}:
\batitle{Bifurcation and dynamics in a mathematical model of early atherosclerosis: How acute inflammation drives lesion development}.
\bjtitle{Journal of mathematical biology}
\bvolume{71},
\bfpage{1451}--\blpage{1480}
(\byear{2015}).
\doiurl{10.1007/s00285-015-0864-5}
\end{barticle}
\endbibitem

\bibitem{mukherjee2019reaction}
\begin{barticle}
\bauthor{\bsnm{Mukherjee}, \binits{D.}},
\bauthor{\bsnm{Guin}, \binits{L.N.}},
\bauthor{\bsnm{Chakravarty}, \binits{S.}}:
\batitle{A reaction--diffusion mathematical model on mild atherosclerosis}.
\bjtitle{Modeling Earth Systems and Environment}
\bvolume{5},
\bfpage{1853}--\blpage{1865}
(\byear{2019}).
\doiurl{10.1007/s40808-019-00643-6}
\end{barticle}
\endbibitem

\bibitem{mohammad2020integrated}
\begin{barticle}
\bauthor{\bsnm{Mohammad~Mirzaei}, \binits{N.}},
\bauthor{\bsnm{Weintraub}, \binits{W.S.}},
\bauthor{\bsnm{Fok}, \binits{P.-W.}}:
\batitle{An integrated approach to simulating the vulnerable atherosclerotic plaque}.
\bjtitle{American Journal of Physiology-Heart and Circulatory Physiology}
\bvolume{319}(\bissue{4}),
\bfpage{835}--\blpage{846}
(\byear{2020}).
\doiurl{10.1152/ajpheart.00174.2020}
\end{barticle}
\endbibitem

\bibitem{ahmed2023macrophage}
\begin{barticle}
\bauthor{\bsnm{Ahmed}, \binits{I.U.}},
\bauthor{\bsnm{Byrne}, \binits{H.M.}},
\bauthor{\bsnm{Myerscough}, \binits{M.R.}}:
\batitle{Macrophage anti-inflammatory behaviour in a multiphase model of atherosclerotic plaque development}.
\bjtitle{Bulletin of Mathematical Biology}
\bvolume{85}(\bissue{5}),
\bfpage{37}
(\byear{2023}).
\doiurl{10.1007/s11538-023-01142-7}
\end{barticle}
\endbibitem

\bibitem{corti2020fully}
\begin{barticle}
\bauthor{\bsnm{Corti}, \binits{A.}},
\bauthor{\bsnm{Chiastra}, \binits{C.}},
\bauthor{\bsnm{Colombo}, \binits{M.}},
\bauthor{\bsnm{Garbey}, \binits{M.}},
\bauthor{\bsnm{Migliavacca}, \binits{F.}},
\bauthor{\bsnm{Casarin}, \binits{S.}}:
\batitle{A fully coupled computational fluid dynamics--agent-based model of atherosclerotic plaque development: multiscale modeling framework and parameter sensitivity analysis}.
\bjtitle{Computers in biology and medicine}
\bvolume{118},
\bfpage{103623}
(\byear{2020}).
\doiurl{10.1016/j.compbiomed.2020.103623}
\end{barticle}
\endbibitem

\bibitem{bayani2020spatial}
\begin{barticle}
\bauthor{\bsnm{Bayani}, \binits{A.}},
\bauthor{\bsnm{Dunster}, \binits{J.L.}},
\bauthor{\bsnm{Crofts}, \binits{J.J.}},
\bauthor{\bsnm{Nelson}, \binits{M.R.}}:
\batitle{Spatial considerations in the resolution of inflammation: elucidating leukocyte interactions via an experimentally-calibrated agent-based model}.
\bjtitle{PLoS Computational Biology}
\bvolume{16}(\bissue{11}),
\bfpage{1008413}
(\byear{2020}).
\doiurl{10.1371/journal.pcbi.1008413}
\end{barticle}
\endbibitem

\bibitem{friedman2015mathematical}
\begin{barticle}
\bauthor{\bsnm{Friedman}, \binits{A.}},
\bauthor{\bsnm{Hao}, \binits{W.}}:
\batitle{A mathematical model of atherosclerosis with reverse cholesterol transport and associated risk factors}.
\bjtitle{Bulletin of mathematical biology}
\bvolume{77},
\bfpage{758}--\blpage{781}
(\byear{2015}).
\doiurl{10.1007/s11538-014-0010-3}
\end{barticle}
\endbibitem

\bibitem{bezyaev2020model}
\begin{bchapter}
\bauthor{\bsnm{Bezyaev}, \binits{V.}},
\bauthor{\bsnm{Sadekov}, \binits{N.}},
\bauthor{\bsnm{Volpert}, \binits{V.}}:
\bctitle{A model of chronic inflammation in atherosclerosis}.
In: \bbtitle{ITM Web of Conferences},
vol. \bseriesno{31},
p. \bfpage{04002}
(\byear{2020}).
\doiurl{10.1051/itmconf/20203104002}.
\bcomment{EDP Sciences}
\end{bchapter}
\endbibitem

\bibitem{liu2022macrophage}
\begin{barticle}
\bauthor{\bsnm{Liu}, \binits{M.}},
\bauthor{\bsnm{Cai}, \binits{Y.}},
\bauthor{\bsnm{Pan}, \binits{J.}},
\bauthor{\bsnm{Peter}, \binits{K.}},
\bauthor{\bsnm{Li}, \binits{Z.}}:
\batitle{Macrophage polarization as a potential therapeutic target for atherosclerosis: a dynamic stochastic modelling study}.
\bjtitle{Royal Society Open Science}
\bvolume{9}(\bissue{8}),
\bfpage{220239}
(\byear{2022}).
\doiurl{10.1098/rsos.220239}
\end{barticle}
\endbibitem

\bibitem{chalmers2017nonlinear}
\begin{barticle}
\bauthor{\bsnm{Chalmers}, \binits{A.D.}},
\bauthor{\bsnm{Bursill}, \binits{C.A.}},
\bauthor{\bsnm{Myerscough}, \binits{M.R.}}:
\batitle{Nonlinear dynamics of early atherosclerotic plaque formation may determine the efficacy of high density lipoproteins (hdl) in plaque regression}.
\bjtitle{PloS one}
\bvolume{12}(\bissue{11}),
\bfpage{0187674}
(\byear{2017}).
\doiurl{10.1371/journal.pone.0187674}
\end{barticle}
\endbibitem

\bibitem{silva2020modeling}
\begin{barticle}
\bauthor{\bsnm{Silva}, \binits{T.}},
\bauthor{\bsnm{J{\"a}ger}, \binits{W.}},
\bauthor{\bsnm{Neuss-Radu}, \binits{M.}},
\bauthor{\bsnm{Sequeira}, \binits{A.}}:
\batitle{Modeling of the early stage of atherosclerosis with emphasis on the regulation of the endothelial permeability}.
\bjtitle{Journal of Theoretical Biology}
\bvolume{496},
\bfpage{110229}
(\byear{2020}).
\doiurl{10.1016/j.jtbi.2020.110229}
\end{barticle}
\endbibitem

\bibitem{ford2019lipid}
\begin{barticle}
\bauthor{\bsnm{Ford}, \binits{H.Z.}},
\bauthor{\bsnm{Byrne}, \binits{H.M.}},
\bauthor{\bsnm{Myerscough}, \binits{M.R.}}:
\batitle{A lipid-structured model for macrophage populations in atherosclerotic plaques}.
\bjtitle{Journal of Theoretical Biology}
\bvolume{479},
\bfpage{48}--\blpage{63}
(\byear{2019}).
\doiurl{10.1016/j.jtbi.2019.07.003}
\end{barticle}
\endbibitem

\bibitem{meunier2019mathematical}
\begin{barticle}
\bauthor{\bsnm{Meunier}, \binits{N.}},
\bauthor{\bsnm{Muller}, \binits{N.}}:
\batitle{Mathematical study of an inflammatory model for atherosclerosis: a nonlinear renewal equation}.
\bjtitle{Acta Applicandae Mathematicae}
\bvolume{161},
\bfpage{107}--\blpage{126}
(\byear{2019}).
\doiurl{10.1007/s10440-018-0206-x}
\end{barticle}
\endbibitem

\bibitem{chambers2022lipid}
\begin{barticle}
\bauthor{\bsnm{Chambers}, \binits{K.L.}},
\bauthor{\bsnm{Watson}, \binits{M.G.}},
\bauthor{\bsnm{Myerscough}, \binits{M.R.}}:
\batitle{A lipid-structured mathematical model of atherosclerosis with macrophage proliferation}.
\bjtitle{arXiv preprint arXiv:2205.04715}
(\byear{2022}).
\doiurl{10.48550/arXiv.2205.04715}
\end{barticle}
\endbibitem

\bibitem{chambers2023new}
\begin{botherref}
\oauthor{\bsnm{Chambers}, \binits{K.L.}},
\oauthor{\bsnm{Myerscough}, \binits{M.R.}},
\oauthor{\bsnm{Byrne}, \binits{H.M.}}:
A new lipid-structured model to investigate the opposing effects of ldl and hdl on atherosclerotic plaque macrophages.
Mathematical Biosciences,
108971
(2023).
\doiurl{10.1016/j.mbs.2023.108971}
\end{botherref}
\endbibitem

\bibitem{watson2023lipid}
\begin{barticle}
\bauthor{\bsnm{Watson}, \binits{M.G.}},
\bauthor{\bsnm{Chambers}, \binits{K.L.}},
\bauthor{\bsnm{Myerscough}, \binits{M.R.}}:
\batitle{A lipid-structured model of atherosclerotic plaque macrophages with lipid-dependent kinetics}.
\bjtitle{Bulletin of Mathematical Biology}
\bvolume{85}(\bissue{9}),
\bfpage{85}
(\byear{2023}).
\doiurl{10.1007/s11538-023-01193-w}
\end{barticle}
\endbibitem

\bibitem{dib2023lipid}
\begin{barticle}
\bauthor{\bsnm{Dib}, \binits{L.}},
\bauthor{\bsnm{Koneva}, \binits{L.A.}},
\bauthor{\bsnm{Edsfeldt}, \binits{A.}},
\bauthor{\bsnm{Zurke}, \binits{Y.-X.}},
\bauthor{\bsnm{Sun}, \binits{J.}},
\bauthor{\bsnm{Nitulescu}, \binits{M.}},
\bauthor{\bsnm{Attar}, \binits{M.}},
\bauthor{\bsnm{Lutgens}, \binits{E.}},
\bauthor{\bsnm{Schmidt}, \binits{S.}},
\bauthor{\bsnm{Lindholm}, \binits{M.W.}}, \betal:
\batitle{Lipid-associated macrophages transition to an inflammatory state in human atherosclerosis, increasing the risk of cerebrovascular complications}.
\bjtitle{Nature cardiovascular research}
\bvolume{2}(\bissue{7}),
\bfpage{656}--\blpage{672}
(\byear{2023}).
\doiurl{10.1038/s44161-023-00295-x}
\end{barticle}
\endbibitem

\bibitem{bernard2003analysis}
\begin{barticle}
\bauthor{\bsnm{Bernard}, \binits{S.}},
\bauthor{\bsnm{Pujo-Menjouet}, \binits{L.}},
\bauthor{\bsnm{Mackey}, \binits{M.C.}}:
\batitle{Analysis of cell kinetics using a cell division marker: mathematical modeling of experimental data}.
\bjtitle{Biophysical journal}
\bvolume{84}(\bissue{5}),
\bfpage{3414}--\blpage{3424}
(\byear{2003}).
\doiurl{10.1016/S0006-3495(03)70063-0}
\end{barticle}
\endbibitem

\bibitem{doumic2007analysis}
\begin{barticle}
\bauthor{\bsnm{Doumic}, \binits{M.}}:
\batitle{Analysis of a population model structured by the cells molecular content}.
\bjtitle{Mathematical Modelling of Natural Phenomena}
\bvolume{2}(\bissue{3}),
\bfpage{121}--\blpage{152}
(\byear{2007}).
\doiurl{10.1051/mmnp:2007006}
\end{barticle}
\endbibitem

\bibitem{laroche2016threshold}
\begin{barticle}
\bauthor{\bsnm{Laroche}, \binits{B.}},
\bauthor{\bsnm{Perasso}, \binits{A.}}:
\batitle{Threshold behaviour of a si epidemiological model with two structuring variables}.
\bjtitle{Journal of Evolution Equations}
\bvolume{16}(\bissue{2}),
\bfpage{293}--\blpage{315}
(\byear{2016}).
\doiurl{10.1007/s00028-015-0303-5}
\end{barticle}
\endbibitem

\bibitem{hodgkinson2019spatio}
\begin{barticle}
\bauthor{\bsnm{Hodgkinson}, \binits{A.}},
\bauthor{\bsnm{Le~Cam}, \binits{L.}},
\bauthor{\bsnm{Trucu}, \binits{D.}},
\bauthor{\bsnm{Radulescu}, \binits{O.}}:
\batitle{Spatio-genetic and phenotypic modelling elucidates resistance and re-sensitisation to treatment in heterogeneous melanoma}.
\bjtitle{Journal of theoretical biology}
\bvolume{466},
\bfpage{84}--\blpage{105}
(\byear{2019}).
\doiurl{10.1016/j.jtbi.2018.11.037}
\end{barticle}
\endbibitem

\bibitem{kang2020nonlinear}
\begin{barticle}
\bauthor{\bsnm{Kang}, \binits{H.}},
\bauthor{\bsnm{Huo}, \binits{X.}},
\bauthor{\bsnm{Ruan}, \binits{S.}}:
\batitle{Nonlinear physiologically structured population models with two internal variables}.
\bjtitle{Journal of Nonlinear Science}
\bvolume{30},
\bfpage{2847}--\blpage{2884}
(\byear{2020}).
\doiurl{10.1007/s00332-020-09638-5}
\end{barticle}
\endbibitem

\bibitem{williams2019cytokine}
\begin{barticle}
\bauthor{\bsnm{Williams}, \binits{J.W.}},
\bauthor{\bsnm{Huang}, \binits{L.-h.}},
\bauthor{\bsnm{Randolph}, \binits{G.J.}}:
\batitle{Cytokine circuits in cardiovascular disease}.
\bjtitle{Immunity}
\bvolume{50}(\bissue{4}),
\bfpage{941}--\blpage{954}
(\byear{2019}).
\doiurl{10.1016/j.immuni.2019.03.007}
\end{barticle}
\endbibitem

\bibitem{williams2020limited}
\begin{barticle}
\bauthor{\bsnm{Williams}, \binits{J.W.}},
\bauthor{\bsnm{Zaitsev}, \binits{K.}},
\bauthor{\bsnm{Kim}, \binits{K.-W.}},
\bauthor{\bsnm{Ivanov}, \binits{S.}},
\bauthor{\bsnm{Saunders}, \binits{B.T.}},
\bauthor{\bsnm{Schrank}, \binits{P.R.}},
\bauthor{\bsnm{Kim}, \binits{K.}},
\bauthor{\bsnm{Elvington}, \binits{A.}},
\bauthor{\bsnm{Kim}, \binits{S.H.}},
\bauthor{\bsnm{Tucker}, \binits{C.G.}}, \betal:
\batitle{Limited proliferation capacity of aortic intima resident macrophages requires monocyte recruitment for atherosclerotic plaque progression}.
\bjtitle{Nature immunology}
\bvolume{21}(\bissue{10}),
\bfpage{1194}--\blpage{1204}
(\byear{2020}).
\doiurl{10.1038/s41590-020-0768-4}
\end{barticle}
\endbibitem

\bibitem{williams2005lipoprotein}
\begin{botherref}
\oauthor{\bsnm{Williams}, \binits{K.J.}},
\oauthor{\bsnm{Tabas}, \binits{I.}}:
Lipoprotein retention—and clues for atheroma regression.
Am Heart Assoc
(2005).
\doiurl{10.1161/01.ATV.0000174795.62387.d3}
\end{botherref}
\endbibitem

\bibitem{allen2022ldl}
\begin{barticle}
\bauthor{\bsnm{Allen}, \binits{R.M.}},
\bauthor{\bsnm{Michell}, \binits{D.L.}},
\bauthor{\bsnm{Cavnar}, \binits{A.B.}},
\bauthor{\bsnm{Zhu}, \binits{W.}},
\bauthor{\bsnm{Makhijani}, \binits{N.}},
\bauthor{\bsnm{Contreras}, \binits{D.M.}},
\bauthor{\bsnm{Raby}, \binits{C.A.}},
\bauthor{\bsnm{Semler}, \binits{E.M.}},
\bauthor{\bsnm{DeJulius}, \binits{C.}},
\bauthor{\bsnm{Castleberry}, \binits{M.}}, \betal:
\batitle{Ldl delivery of microbial small rnas drives atherosclerosis through macrophage tlr8}.
\bjtitle{Nature Cell Biology}
\bvolume{24}(\bissue{12}),
\bfpage{1701}--\blpage{1713}
(\byear{2022}).
\doiurl{10.1038/s41556-022-01030-7}
\end{barticle}
\endbibitem

\bibitem{chen2015oxidized}
\begin{barticle}
\bauthor{\bsnm{Chen}, \binits{C.}},
\bauthor{\bsnm{Khismatullin}, \binits{D.B.}}:
\batitle{Oxidized low-density lipoprotein contributes to atherogenesis via co-activation of macrophages and mast cells}.
\bjtitle{PloS one}
\bvolume{10}(\bissue{3}),
\bfpage{0123088}
(\byear{2015}).
\doiurl{10.1371/journal.pone.0123088}
\end{barticle}
\endbibitem

\bibitem{dalli2012specific}
\begin{barticle}
\bauthor{\bsnm{Dalli}, \binits{J.}},
\bauthor{\bsnm{Serhan}, \binits{C.N.}}:
\batitle{Specific lipid mediator signatures of human phagocytes: microparticles stimulate macrophage efferocytosis and pro-resolving mediators}.
\bjtitle{Blood, The Journal of the American Society of Hematology}
\bvolume{120}(\bissue{15}),
\bfpage{60}--\blpage{72}
(\byear{2012}).
\doiurl{10.1182/blood-2012-04-423525}
\end{barticle}
\endbibitem

\bibitem{kadomoto2021macrophage}
\begin{barticle}
\bauthor{\bsnm{Kadomoto}, \binits{S.}},
\bauthor{\bsnm{Izumi}, \binits{K.}},
\bauthor{\bsnm{Mizokami}, \binits{A.}}:
\batitle{Macrophage polarity and disease control}.
\bjtitle{International Journal of Molecular Sciences}
\bvolume{23}(\bissue{1}),
\bfpage{144}
(\byear{2021}).
\doiurl{10.3390/ijms23010144}
\end{barticle}
\endbibitem

\bibitem{liu2021cytokines}
\begin{barticle}
\bauthor{\bsnm{Liu}, \binits{C.}},
\bauthor{\bsnm{Chu}, \binits{D.}},
\bauthor{\bsnm{Kalantar-Zadeh}, \binits{K.}},
\bauthor{\bsnm{George}, \binits{J.}},
\bauthor{\bsnm{Young}, \binits{H.A.}},
\bauthor{\bsnm{Liu}, \binits{G.}}:
\batitle{Cytokines: From clinical significance to quantification}.
\bjtitle{Advanced Science}
\bvolume{8}(\bissue{15}),
\bfpage{2004433}
(\byear{2021}).
\doiurl{10.1002/advs.202004433}
\end{barticle}
\endbibitem

\bibitem{lee2012characteristics}
\begin{barticle}
\bauthor{\bsnm{Lee}, \binits{J.-G.}},
\bauthor{\bsnm{Koh}, \binits{S.J.}},
\bauthor{\bsnm{Yoo}, \binits{S.Y.}},
\bauthor{\bsnm{Yu}, \binits{J.R.}},
\bauthor{\bsnm{Lee}, \binits{S.A.}},
\bauthor{\bsnm{Koh}, \binits{G.}},
\bauthor{\bsnm{Lee}, \binits{D.}}:
\batitle{Characteristics of subjects with very low serum low-density lipoprotein cholesterol and the risk for intracerebral hemorrhage}.
\bjtitle{The Korean journal of internal medicine}
\bvolume{27}(\bissue{3}),
\bfpage{317}
(\byear{2012}).
\doiurl{10.3904/kjim.2012.27.3.317}
\end{barticle}
\endbibitem

\bibitem{orlova1999three}
\begin{barticle}
\bauthor{\bsnm{Orlova}, \binits{E.V.}},
\bauthor{\bsnm{Sherman}, \binits{M.B.}},
\bauthor{\bsnm{Chiu}, \binits{W.}},
\bauthor{\bsnm{Mowri}, \binits{H.}},
\bauthor{\bsnm{Smith}, \binits{L.C.}},
\bauthor{\bsnm{Gotto~Jr}, \binits{A.M.}}:
\batitle{Three-dimensional structure of low density lipoproteins by electron cryomicroscopy}.
\bjtitle{Proceedings of the National Academy of Sciences}
\bvolume{96}(\bissue{15}),
\bfpage{8420}--\blpage{8425}
(\byear{1999}).
\doiurl{10.1073/pnas.96.15.8420}
\end{barticle}
\endbibitem

\bibitem{madsen2017extreme}
\begin{barticle}
\bauthor{\bsnm{Madsen}, \binits{C.M.}},
\bauthor{\bsnm{Varbo}, \binits{A.}},
\bauthor{\bsnm{Nordestgaard}, \binits{B.G.}}:
\batitle{Extreme high high-density lipoprotein cholesterol is paradoxically associated with high mortality in men and women: two prospective cohort studies}.
\bjtitle{European heart journal}
\bvolume{38}(\bissue{32}),
\bfpage{2478}--\blpage{2486}
(\byear{2017}).
\doiurl{10.1093/eurheartj/ehx163}
\end{barticle}
\endbibitem

\bibitem{nielsen1996transfer}
\begin{barticle}
\bauthor{\bsnm{Nielsen}, \binits{L.B.}}:
\batitle{Transfer of low density lipoprotein into the arterial wall and risk of atherosclerosis}.
\bjtitle{Atherosclerosis}
\bvolume{123}(\bissue{1-2}),
\bfpage{1}--\blpage{15}
(\byear{1996}).
\doiurl{10.1016/0021-9150(96)05802-9}
\end{barticle}
\endbibitem

\bibitem{holzapfel2005determination}
\begin{barticle}
\bauthor{\bsnm{Holzapfel}, \binits{G.A.}},
\bauthor{\bsnm{Sommer}, \binits{G.}},
\bauthor{\bsnm{Gasser}, \binits{C.T.}},
\bauthor{\bsnm{Regitnig}, \binits{P.}}:
\batitle{Determination of layer-specific mechanical properties of human coronary arteries with nonatherosclerotic intimal thickening and related constitutive modeling}.
\bjtitle{American Journal of Physiology-Heart and Circulatory Physiology}
\bvolume{289}(\bissue{5}),
\bfpage{2048}--\blpage{2058}
(\byear{2005}).
\doiurl{10.1152/ajpheart.00934.2004}
\end{barticle}
\endbibitem

\bibitem{stender1981transfer}
\begin{barticle}
\bauthor{\bsnm{Stender}, \binits{S.}},
\bauthor{\bsnm{Zilversmit}, \binits{D.}}:
\batitle{Transfer of plasma lipoprotein components and of plasma proteins into aortas of cholesterol-fed rabbits. molecular size as a determinant of plasma lipoprotein influx.}
\bjtitle{Arteriosclerosis: An Official Journal of the American Heart Association, Inc.}
\bvolume{1}(\bissue{1}),
\bfpage{38}--\blpage{49}
(\byear{1981}).
\doiurl{10.1161/01.atv.1.1.38}
\end{barticle}
\endbibitem

\bibitem{smith1982plasma}
\begin{barticle}
\bauthor{\bsnm{Smith}, \binits{E.B.}},
\bauthor{\bsnm{Staples}, \binits{E.M.}}:
\batitle{Plasma protein concentrations in interstitial fluid from human aortas}.
\bjtitle{Proceedings of the Royal Society of London. Series B. Biological Sciences}
\bvolume{217}(\bissue{1206}),
\bfpage{59}--\blpage{75}
(\byear{1982}).
\doiurl{10.1098/rspb.1982.0094}
\end{barticle}
\endbibitem

\bibitem{penn1994relative}
\begin{barticle}
\bauthor{\bsnm{Penn}, \binits{M.S.}},
\bauthor{\bsnm{Saidel}, \binits{G.M.}},
\bauthor{\bsnm{Chisolm}, \binits{G.M.}}:
\batitle{Relative significance of endothelium and internal elastic lamina in regulating the entry of macromolecules into arteries in vivo.}
\bjtitle{Circulation research}
\bvolume{74}(\bissue{1}),
\bfpage{74}--\blpage{82}
(\byear{1994}).
\doiurl{10.1161/01.res.74.1.74}
\end{barticle}
\endbibitem

\bibitem{wight2018role}
\begin{barticle}
\bauthor{\bsnm{Wight}, \binits{T.N.}}:
\batitle{A role for proteoglycans in vascular disease}.
\bjtitle{Matrix Biology}
\bvolume{71},
\bfpage{396}--\blpage{420}
(\byear{2018}).
\doiurl{10.1016/j.matbio.2018.02.019}
\end{barticle}
\endbibitem

\bibitem{guyton1989lipid}
\begin{barticle}
\bauthor{\bsnm{Guyton}, \binits{J.R.}},
\bauthor{\bsnm{Klemp}, \binits{K.F.}}:
\batitle{The lipid-rich core region of human atherosclerotic fibrous plaques. prevalence of small lipid droplets and vesicles by electron microscopy.}
\bjtitle{The American journal of pathology}
\bvolume{134}(\bissue{3}),
\bfpage{705}
(\byear{1989})
\end{barticle}
\endbibitem

\bibitem{liu2023co}
\begin{barticle}
\bauthor{\bsnm{Liu}, \binits{M.}},
\bauthor{\bsnm{Samant}, \binits{S.}},
\bauthor{\bsnm{Vasa}, \binits{C.H.}},
\bauthor{\bsnm{Pedrigi}, \binits{R.M.}},
\bauthor{\bsnm{Oguz}, \binits{U.M.}},
\bauthor{\bsnm{Ryu}, \binits{S.}},
\bauthor{\bsnm{Wei}, \binits{T.}},
\bauthor{\bsnm{Anderson}, \binits{D.R.}},
\bauthor{\bsnm{Agrawal}, \binits{D.K.}},
\bauthor{\bsnm{Chatzizisis}, \binits{Y.S.}}:
\batitle{Co-culture models of endothelial cells, macrophages, and vascular smooth muscle cells for the study of the natural history of atherosclerosis}.
\bjtitle{PloS one}
\bvolume{18}(\bissue{1}),
\bfpage{0280385}
(\byear{2023}).
\doiurl{10.1371/journal.pone.0280385}
\end{barticle}
\endbibitem

\bibitem{bancells2009high}
\begin{barticle}
\bauthor{\bsnm{Bancells}, \binits{C.}},
\bauthor{\bsnm{Ben{\'\i}tez}, \binits{S.}},
\bauthor{\bsnm{Jauhiainen}, \binits{M.}},
\bauthor{\bsnm{Ord{\'o}{\~n}ez-Llanos}, \binits{J.}},
\bauthor{\bsnm{Kovanen}, \binits{P.T.}},
\bauthor{\bsnm{Villegas}, \binits{S.}},
\bauthor{\bsnm{S{\'a}nchez-Quesada}, \binits{J.L.}},
\bauthor{\bsnm{Katariina}, \binits{O.}}, \betal:
\batitle{High binding affinity of electronegative ldl to human aortic proteoglycans depends on its aggregation level}.
\bjtitle{Journal of lipid research}
\bvolume{50}(\bissue{3}),
\bfpage{446}--\blpage{455}
(\byear{2009}).
\doiurl{10.1194/jlr.M800318-JLR200}
\end{barticle}
\endbibitem

\bibitem{williams2009transmigration}
\begin{barticle}
\bauthor{\bsnm{Williams}, \binits{M.R.}},
\bauthor{\bsnm{Sakurai}, \binits{Y.}},
\bauthor{\bsnm{Zughaier}, \binits{S.M.}},
\bauthor{\bsnm{Eskin}, \binits{S.G.}},
\bauthor{\bsnm{McIntire}, \binits{L.V.}}:
\batitle{Transmigration across activated endothelium induces transcriptional changes, inhibits apoptosis, and decreases antimicrobial protein expression in human monocytes}.
\bjtitle{Journal of leukocyte biology}
\bvolume{86}(\bissue{6}),
\bfpage{1331}--\blpage{1343}
(\byear{2009}).
\doiurl{10.1189/jlb.0209062}
\end{barticle}
\endbibitem

\bibitem{nelson2014immunobiology}
\begin{bbook}
\bauthor{\bsnm{Nelson}, \binits{D.S.}}:
\bbtitle{Immunobiology of the Macrophage}.
\bpublisher{Academic Press}, \blocation{???}
(\byear{2014})
\end{bbook}
\endbibitem

\bibitem{lee2019sirt1}
\begin{barticle}
\bauthor{\bsnm{Lee}, \binits{S.J.}},
\bauthor{\bsnm{Baek}, \binits{S.E.}},
\bauthor{\bsnm{Jang}, \binits{M.A.}},
\bauthor{\bsnm{Kim}, \binits{C.D.}}:
\batitle{Sirt1 inhibits monocyte adhesion to the vascular endothelium by suppressing mac-1 expression on monocytes}.
\bjtitle{Experimental \& Molecular Medicine}
\bvolume{51}(\bissue{4}),
\bfpage{1}--\blpage{12}
(\byear{2019}).
\doiurl{10.1038/s12276-019-0239-x}
\end{barticle}
\endbibitem

\bibitem{yona2013fate}
\begin{barticle}
\bauthor{\bsnm{Yona}, \binits{S.}},
\bauthor{\bsnm{Kim}, \binits{K.-W.}},
\bauthor{\bsnm{Wolf}, \binits{Y.}},
\bauthor{\bsnm{Mildner}, \binits{A.}},
\bauthor{\bsnm{Varol}, \binits{D.}},
\bauthor{\bsnm{Breker}, \binits{M.}},
\bauthor{\bsnm{Strauss-Ayali}, \binits{D.}},
\bauthor{\bsnm{Viukov}, \binits{S.}},
\bauthor{\bsnm{Guilliams}, \binits{M.}},
\bauthor{\bsnm{Misharin}, \binits{A.}}, \betal:
\batitle{Fate mapping reveals origins and dynamics of monocytes and tissue macrophages under homeostasis}.
\bjtitle{Immunity}
\bvolume{38}(\bissue{1}),
\bfpage{79}--\blpage{91}
(\byear{2013}).
\doiurl{10.1016/j.immuni.2012.12.001}
\end{barticle}
\endbibitem

\bibitem{williams2018limited}
\begin{barticle}
\bauthor{\bsnm{Williams}, \binits{J.W.}},
\bauthor{\bsnm{Martel}, \binits{C.}},
\bauthor{\bsnm{Potteaux}, \binits{S.}},
\bauthor{\bsnm{Esaulova}, \binits{E.}},
\bauthor{\bsnm{Ingersoll}, \binits{M.A.}},
\bauthor{\bsnm{Elvington}, \binits{A.}},
\bauthor{\bsnm{Saunders}, \binits{B.T.}},
\bauthor{\bsnm{Huang}, \binits{L.-H.}},
\bauthor{\bsnm{Habenicht}, \binits{A.J.}},
\bauthor{\bsnm{Zinselmeyer}, \binits{B.H.}}, \betal:
\batitle{Limited macrophage positional dynamics in progressing or regressing murine atherosclerotic plaques—brief report}.
\bjtitle{Arteriosclerosis, thrombosis, and vascular biology}
\bvolume{38}(\bissue{8}),
\bfpage{1702}--\blpage{1710}
(\byear{2018}).
\doiurl{10.1161/ATVBAHA.118.311319}
\end{barticle}
\endbibitem

\bibitem{saraste2000morphologic}
\begin{barticle}
\bauthor{\bsnm{Saraste}, \binits{A.}},
\bauthor{\bsnm{Pulkki}, \binits{K.}}:
\batitle{Morphologic and biochemical hallmarks of apoptosis}.
\bjtitle{Cardiovascular research}
\bvolume{45}(\bissue{3}),
\bfpage{528}--\blpage{537}
(\byear{2000}).
\doiurl{10.1016/S0008-6363(99)00384-3}
\end{barticle}
\endbibitem

\bibitem{sokol1991changes}
\begin{barticle}
\bauthor{\bsnm{Sokol}, \binits{R.}},
\bauthor{\bsnm{Wales}, \binits{J.}},
\bauthor{\bsnm{Hudson}, \binits{G.}},
\bauthor{\bsnm{Goldstein}, \binits{D.}},
\bauthor{\bsnm{James}, \binits{N.}}:
\batitle{Changes in cellular dry mass during macrophage development}.
\bjtitle{Cells Tissues Organs}
\bvolume{142}(\bissue{3}),
\bfpage{246}--\blpage{248}
(\byear{1991}).
\doiurl{10.1159/000147197}
\end{barticle}
\endbibitem

\bibitem{cooper2022cell}
\begin{bbook}
\bauthor{\bsnm{Cooper}, \binits{G.}},
\bauthor{\bsnm{Adams}, \binits{K.}}:
\bbtitle{The Cell: a Molecular Approach}.
\bpublisher{Oxford University Press}, \blocation{???}
(\byear{2022})
\end{bbook}
\endbibitem

\bibitem{ford2019efferocytosis}
\begin{barticle}
\bauthor{\bsnm{Ford}, \binits{H.Z.}},
\bauthor{\bsnm{Zeboudj}, \binits{L.}},
\bauthor{\bsnm{Purvis}, \binits{G.S.}},
\bauthor{\bsnm{Ten~Bokum}, \binits{A.}},
\bauthor{\bsnm{Zarebski}, \binits{A.E.}},
\bauthor{\bsnm{Bull}, \binits{J.A.}},
\bauthor{\bsnm{Byrne}, \binits{H.M.}},
\bauthor{\bsnm{Myerscough}, \binits{M.R.}},
\bauthor{\bsnm{Greaves}, \binits{D.R.}}:
\batitle{Efferocytosis perpetuates substance accumulation inside macrophage populations}.
\bjtitle{Proceedings of the Royal Society B}
\bvolume{286}(\bissue{1904}),
\bfpage{20190730}
(\byear{2019}).
\doiurl{10.1098/rspb.2019.0730}
\end{barticle}
\endbibitem

\bibitem{sanda2021aggregated}
\begin{barticle}
\bauthor{\bsnm{Sanda}, \binits{G.M.}},
\bauthor{\bsnm{Stancu}, \binits{C.S.}},
\bauthor{\bsnm{Deleanu}, \binits{M.}},
\bauthor{\bsnm{Toma}, \binits{L.}},
\bauthor{\bsnm{Niculescu}, \binits{L.S.}},
\bauthor{\bsnm{Sima}, \binits{A.V.}}:
\batitle{Aggregated ldl turn human macrophages into foam cells and induce mitochondrial dysfunction without triggering oxidative or endoplasmic reticulum stress}.
\bjtitle{Plos one}
\bvolume{16}(\bissue{1}),
\bfpage{0245797}
(\byear{2021}).
\doiurl{10.1371/journal.pone.0245797}
\end{barticle}
\endbibitem

\bibitem{taruc2018quantification}
\begin{botherref}
\oauthor{\bsnm{Taruc}, \binits{K.}},
\oauthor{\bsnm{Yin}, \binits{C.}},
\oauthor{\bsnm{Wootton}, \binits{D.G.}},
\oauthor{\bsnm{Heit}, \binits{B.}}:
Quantification of efferocytosis by single-cell fluorescence microscopy.
JoVE (Journal of Visualized Experiments)
(138),
58149
(2018).
\doiurl{10.3791/58149}
\end{botherref}
\endbibitem

\bibitem{schrijvers2005phagocytosis}
\begin{barticle}
\bauthor{\bsnm{Schrijvers}, \binits{D.M.}},
\bauthor{\bsnm{De~Meyer}, \binits{G.R.}},
\bauthor{\bsnm{Kockx}, \binits{M.M.}},
\bauthor{\bsnm{Herman}, \binits{A.G.}},
\bauthor{\bsnm{Martinet}, \binits{W.}}:
\batitle{Phagocytosis of apoptotic cells by macrophages is impaired in atherosclerosis}.
\bjtitle{Arteriosclerosis, thrombosis, and vascular biology}
\bvolume{25}(\bissue{6}),
\bfpage{1256}--\blpage{1261}
(\byear{2005}).
\doiurl{10.1161/01.ATV.0000166517.18801.a7}
\end{barticle}
\endbibitem

\bibitem{brouckaert2004phagocytosis}
\begin{barticle}
\bauthor{\bsnm{Brouckaert}, \binits{G.}},
\bauthor{\bsnm{Kalai}, \binits{M.}},
\bauthor{\bsnm{Krysko}, \binits{D.V.}},
\bauthor{\bsnm{Saelens}, \binits{X.}},
\bauthor{\bsnm{Vercammen}, \binits{D.}},
\bauthor{\bsnm{Ndlovu}, \binits{M.}},
\bauthor{\bsnm{Haegeman}, \binits{G.}},
\bauthor{\bsnm{D'Herde}, \binits{K.}},
\bauthor{\bsnm{Vandenabeele}, \binits{P.}}:
\batitle{Phagocytosis of necrotic cells by macrophages is phosphatidylserine dependent and does not induce inflammatory cytokine production}.
\bjtitle{Molecular biology of the cell}
\bvolume{15}(\bissue{3}),
\bfpage{1089}--\blpage{1100}
(\byear{2004}).
\doiurl{.1091/mbc.E03– 09 – 0668}
\end{barticle}
\endbibitem

\bibitem{kritharides1998cholesterol}
\begin{barticle}
\bauthor{\bsnm{Kritharides}, \binits{L.}},
\bauthor{\bsnm{Christian}, \binits{A.}},
\bauthor{\bsnm{Stoudt}, \binits{G.}},
\bauthor{\bsnm{Morel}, \binits{D.}},
\bauthor{\bsnm{Rothblat}, \binits{G.H.}}:
\batitle{Cholesterol metabolism and efflux in human thp-1 macrophages}.
\bjtitle{Arteriosclerosis, thrombosis, and vascular biology}
\bvolume{18}(\bissue{10}),
\bfpage{1589}--\blpage{1599}
(\byear{1998}).
\doiurl{10.1161/01.ATV.18.10.1589}
\end{barticle}
\endbibitem

\bibitem{woudberg2018pharmacological}
\begin{barticle}
\bauthor{\bsnm{Woudberg}, \binits{N.J.}},
\bauthor{\bsnm{Pedretti}, \binits{S.}},
\bauthor{\bsnm{Lecour}, \binits{S.}},
\bauthor{\bsnm{Schulz}, \binits{R.}},
\bauthor{\bsnm{Vuilleumier}, \binits{N.}},
\bauthor{\bsnm{James}, \binits{R.W.}},
\bauthor{\bsnm{Frias}, \binits{M.A.}}:
\batitle{Pharmacological intervention to modulate hdl: what do we target?}
\bjtitle{Frontiers in pharmacology}
\bvolume{8},
\bfpage{989}
(\byear{2018}).
\doiurl{10.3389/fphar.2017.00989}
\end{barticle}
\endbibitem

\bibitem{o2015pro}
\begin{barticle}
\bauthor{\bsnm{O’Carroll}, \binits{S.J.}},
\bauthor{\bsnm{Kho}, \binits{D.T.}},
\bauthor{\bsnm{Wiltshire}, \binits{R.}},
\bauthor{\bsnm{Nelson}, \binits{V.}},
\bauthor{\bsnm{Rotimi}, \binits{O.}},
\bauthor{\bsnm{Johnson}, \binits{R.}},
\bauthor{\bsnm{Angel}, \binits{C.E.}},
\bauthor{\bsnm{Graham}, \binits{E.S.}}:
\batitle{Pro-inflammatory tnf$\alpha$ and il-1$\beta$ differentially regulate the inflammatory phenotype of brain microvascular endothelial cells}.
\bjtitle{Journal of neuroinflammation}
\bvolume{12},
\bfpage{1}--\blpage{18}
(\byear{2015}).
\doiurl{10.1186/s12974-015-0346-0}
\end{barticle}
\endbibitem

\bibitem{pugin1993lipopolysaccharide}
\begin{barticle}
\bauthor{\bsnm{Pugin}, \binits{J.}},
\bauthor{\bsnm{Sch{\"u}rer-Maly}, \binits{C.}},
\bauthor{\bsnm{Leturcq}, \binits{D.}},
\bauthor{\bsnm{Moriarty}, \binits{A.}},
\bauthor{\bsnm{Ulevitch}, \binits{R.J.}},
\bauthor{\bsnm{Tobias}, \binits{P.S.}}:
\batitle{Lipopolysaccharide activation of human endothelial and epithelial cells is mediated by lipopolysaccharide-binding protein and soluble cd14.}
\bjtitle{Proceedings of the National Academy of Sciences}
\bvolume{90}(\bissue{7}),
\bfpage{2744}--\blpage{2748}
(\byear{1993}).
\doiurl{10.1073/pnas.90.7.2744}
\end{barticle}
\endbibitem

\bibitem{sha2015interleukin}
\begin{barticle}
\bauthor{\bsnm{Sha}, \binits{X.}},
\bauthor{\bsnm{Meng}, \binits{S.}},
\bauthor{\bsnm{Li}, \binits{X.}},
\bauthor{\bsnm{Xi}, \binits{H.}},
\bauthor{\bsnm{Maddaloni}, \binits{M.}},
\bauthor{\bsnm{Pascual}, \binits{D.W.}},
\bauthor{\bsnm{Shan}, \binits{H.}},
\bauthor{\bsnm{Jiang}, \binits{X.}},
\bauthor{\bsnm{Wang}, \binits{H.}},
\bauthor{\bsnm{Yang}, \binits{X.-f.}}:
\batitle{Interleukin-35 inhibits endothelial cell activation by suppressing mapk-ap-1 pathway}.
\bjtitle{Journal of Biological Chemistry}
\bvolume{290}(\bissue{31}),
\bfpage{19307}--\blpage{19318}
(\byear{2015}).
\doiurl{10.1074/jbc.M115.663286}
\end{barticle}
\endbibitem

\bibitem{atzeni2013tumor}
\begin{botherref}
\oauthor{\bsnm{Atzeni}, \binits{F.}},
\oauthor{\bsnm{Sarzi-Puttini}, \binits{P.}}:
Tumor necrosis factor.
Brenner's Encyclopedia of Genetics,
229--231
(2013)
\end{botherref}
\endbibitem

\bibitem{watanabe1988continuous}
\begin{barticle}
\bauthor{\bsnm{Watanabe}, \binits{N.}},
\bauthor{\bsnm{Kuriyama}, \binits{H.}},
\bauthor{\bsnm{Sone}, \binits{H.}},
\bauthor{\bsnm{Neda}, \binits{H.}},
\bauthor{\bsnm{Yamauchi}, \binits{N.}},
\bauthor{\bsnm{Maeda}, \binits{M.}},
\bauthor{\bsnm{Niitsu}, \binits{Y.}}:
\batitle{Continuous internalization of tumor necrosis factor receptors in a human myosarcoma cell line.}
\bjtitle{Journal of Biological Chemistry}
\bvolume{263}(\bissue{21}),
\bfpage{10262}--\blpage{10266}
(\byear{1988})
\end{barticle}
\endbibitem

\bibitem{niitsu1988analysis}
\begin{barticle}
\bauthor{\bsnm{Niitsu}, \binits{Y.}},
\bauthor{\bsnm{Watanabe}, \binits{N.}},
\bauthor{\bsnm{Sone}, \binits{H.}},
\bauthor{\bsnm{Neda}, \binits{H.}},
\bauthor{\bsnm{Yamauchi}, \binits{N.}},
\bauthor{\bsnm{Maeda}, \binits{M.}},
\bauthor{\bsnm{Urushizaki}, \binits{I.}}:
\batitle{Analysis of the tnf receptor on kym cells by binding assay and affinity cross-linking}.
\bjtitle{Journal of Immunotherapy}
\bvolume{7}(\bissue{3}),
\bfpage{276}--\blpage{282}
(\byear{1988})
\end{barticle}
\endbibitem

\bibitem{schutte2009cytokine}
\begin{barticle}
\bauthor{\bsnm{Schutte}, \binits{R.J.}},
\bauthor{\bsnm{Parisi-Amon}, \binits{A.}},
\bauthor{\bsnm{Reichert}, \binits{W.M.}}:
\batitle{Cytokine profiling using monocytes/macrophages cultured on common biomaterials with a range of surface chemistries}.
\bjtitle{Journal of Biomedical Materials Research Part A: An Official Journal of The Society for Biomaterials, The Japanese Society for Biomaterials, and The Australian Society for Biomaterials and the Korean Society for Biomaterials}
\bvolume{88}(\bissue{1}),
\bfpage{128}--\blpage{139}
(\byear{2009}).
\doiurl{10.1002/jbm.a.31863}
\end{barticle}
\endbibitem

\bibitem{tarique2015phenotypic}
\begin{barticle}
\bauthor{\bsnm{Tarique}, \binits{A.A.}},
\bauthor{\bsnm{Logan}, \binits{J.}},
\bauthor{\bsnm{Thomas}, \binits{E.}},
\bauthor{\bsnm{Holt}, \binits{P.G.}},
\bauthor{\bsnm{Sly}, \binits{P.D.}},
\bauthor{\bsnm{Fantino}, \binits{E.}}:
\batitle{Phenotypic, functional, and plasticity features of classical and alternatively activated human macrophages}.
\bjtitle{American journal of respiratory cell and molecular biology}
\bvolume{53}(\bissue{5}),
\bfpage{676}--\blpage{688}
(\byear{2015}).
\doiurl{10.1165/rcmb.2015-0012OC}
\end{barticle}
\endbibitem

\bibitem{kontush2007preferential}
\begin{barticle}
\bauthor{\bsnm{Kontush}, \binits{A.}},
\bauthor{\bsnm{Therond}, \binits{P.}},
\bauthor{\bsnm{Zerrad}, \binits{A.}},
\bauthor{\bsnm{Couturier}, \binits{M.}},
\bauthor{\bsnm{N{\'e}gre-Salvayre}, \binits{A.}},
\bauthor{\bparticle{de} \bsnm{Souza}, \binits{J.A.}},
\bauthor{\bsnm{Chantepie}, \binits{S.}},
\bauthor{\bsnm{Chapman}, \binits{M.J.}}:
\batitle{Preferential sphingosine-1-phosphate enrichment and sphingomyelin depletion are key features of small dense hdl3 particles: relevance to antiapoptotic and antioxidative activities}.
\bjtitle{Arteriosclerosis, thrombosis, and vascular biology}
\bvolume{27}(\bissue{8}),
\bfpage{1843}--\blpage{1849}
(\byear{2007}).
\doiurl{10.1161/ATVBAHA.107.145672}
\end{barticle}
\endbibitem

\bibitem{taefehshokr2021rab}
\begin{botherref}
\oauthor{\bsnm{Taefehshokr}, \binits{N.}},
\oauthor{\bsnm{Yin}, \binits{C.}},
\oauthor{\bsnm{Heit}, \binits{B.}}:
Rab gtpases in the differential processing of phagocytosed pathogens versus efferocytosed apoptotic cells
(2021).
\doiurl{10.14670/HH-18-252}
\end{botherref}
\endbibitem

\bibitem{schoeneck2021effects}
\begin{barticle}
\bauthor{\bsnm{Schoeneck}, \binits{M.}},
\bauthor{\bsnm{Iggman}, \binits{D.}}:
\batitle{The effects of foods on ldl cholesterol levels: A systematic review of the accumulated evidence from systematic reviews and meta-analyses of randomized controlled trials}.
\bjtitle{Nutrition, Metabolism and Cardiovascular Diseases}
\bvolume{31}(\bissue{5}),
\bfpage{1325}--\blpage{1338}
(\byear{2021}).
\doiurl{10.1016/j.numecd.2020.12.032}
\end{barticle}
\endbibitem

\bibitem{lewis2023capacity}
\begin{barticle}
\bauthor{\bsnm{Lewis}, \binits{E.A.}},
\bauthor{\bsnm{Mu{\~n}iz-Anquela}, \binits{R.}},
\bauthor{\bsnm{Redondo-Angulo}, \binits{I.}},
\bauthor{\bsnm{Gonz{\'a}lez-Cintado}, \binits{L.}},
\bauthor{\bsnm{Labrador-Cantarero}, \binits{V.}},
\bauthor{\bsnm{Bentzon}, \binits{J.F.}}:
\batitle{Capacity for ldl (low-density lipoprotein) retention predicts the course of atherogenesis in the murine aortic arch}.
\bjtitle{Arteriosclerosis, Thrombosis, and Vascular Biology}
\bvolume{43}(\bissue{5}),
\bfpage{637}--\blpage{649}
(\byear{2023}).
\doiurl{10.1161/ATVBAHA.122.318573}
\end{barticle}
\endbibitem

\bibitem{yoneda2002biosynthesis}
\begin{barticle}
\bauthor{\bsnm{Yoneda}, \binits{S.}},
\bauthor{\bsnm{Shibata}, \binits{S.}},
\bauthor{\bsnm{Yamashita}, \binits{Y.}},
\bauthor{\bsnm{Yanagishita}, \binits{M.}}:
\batitle{Biosynthesis of versican by rat dental pulp cells in culture}.
\bjtitle{Archives of oral biology}
\bvolume{47}(\bissue{6}),
\bfpage{435}--\blpage{442}
(\byear{2002}).
\doiurl{10.1016/s0003-9969(02)00029-8}
\end{barticle}
\endbibitem

\bibitem{perez2022macrophage}
\begin{barticle}
\bauthor{\bsnm{P{\'e}rez}, \binits{S.}},
\bauthor{\bsnm{Rius-P{\'e}rez}, \binits{S.}}:
\batitle{Macrophage polarization and reprogramming in acute inflammation: a redox perspective}.
\bjtitle{Antioxidants}
\bvolume{11}(\bissue{7}),
\bfpage{1394}
(\byear{2022}).
\doiurl{10.3390/antiox11071394}
\end{barticle}
\endbibitem

\bibitem{insull2009pathology}
\begin{barticle}
\bauthor{\bsnm{Insull~Jr}, \binits{W.}}:
\batitle{The pathology of atherosclerosis: plaque development and plaque responses to medical treatment}.
\bjtitle{The American journal of medicine}
\bvolume{122}(\bissue{1}),
\bfpage{3}--\blpage{14}
(\byear{2009}).
\doiurl{10.1016/j.amjmed.2008.10.013}
\end{barticle}
\endbibitem

\bibitem{sansbury2016resolution}
\begin{barticle}
\bauthor{\bsnm{Sansbury}, \binits{B.E.}},
\bauthor{\bsnm{Spite}, \binits{M.}}:
\batitle{Resolution of acute inflammation and the role of resolvins in immunity, thrombosis, and vascular biology}.
\bjtitle{Circulation research}
\bvolume{119}(\bissue{1}),
\bfpage{113}--\blpage{130}
(\byear{2016}).
\doiurl{10.1161/CIRCRESAHA.116.307308}
\end{barticle}
\endbibitem

\bibitem{daskalopoulos2015role}
\begin{bchapter}
\bauthor{\bsnm{Daskalopoulos}, \binits{E.P.}},
\bauthor{\bsnm{Hermans}, \binits{K.C.}},
\bauthor{\bparticle{van} \bsnm{Delft}, \binits{L.}},
\bauthor{\bsnm{Altara}, \binits{R.}},
\bauthor{\bsnm{Blankesteijn}, \binits{W.M.}}:
\bctitle{The role of inflammation in myocardial infarction}.
In: \bbtitle{Inflammation in Heart Failure},
pp. \bfpage{39}--\blpage{65}.
\bpublisher{Elsevier}, \blocation{???}
(\byear{2015}).
\doiurl{10.1016/B978-0-12-800039-7.00003-7}
\end{bchapter}
\endbibitem

\bibitem{brown2009macrophage}
\begin{barticle}
\bauthor{\bsnm{Brown}, \binits{B.N.}},
\bauthor{\bsnm{Valentin}, \binits{J.E.}},
\bauthor{\bsnm{Stewart-Akers}, \binits{A.M.}},
\bauthor{\bsnm{McCabe}, \binits{G.P.}},
\bauthor{\bsnm{Badylak}, \binits{S.F.}}:
\batitle{Macrophage phenotype and remodeling outcomes in response to biologic scaffolds with and without a cellular component}.
\bjtitle{Biomaterials}
\bvolume{30}(\bissue{8}),
\bfpage{1482}--\blpage{1491}
(\byear{2009}).
\doiurl{10.1016/j.biomaterials.2008.11.040}
\end{barticle}
\endbibitem

\bibitem{schulz2019depth}
\begin{barticle}
\bauthor{\bsnm{Schulz}, \binits{D.}},
\bauthor{\bsnm{Severin}, \binits{Y.}},
\bauthor{\bsnm{Zanotelli}, \binits{V.R.T.}},
\bauthor{\bsnm{Bodenmiller}, \binits{B.}}:
\batitle{In-depth characterization of monocyte-derived macrophages using a mass cytometry-based phagocytosis assay}.
\bjtitle{Scientific reports}
\bvolume{9}(\bissue{1}),
\bfpage{1925}
(\byear{2019}).
\doiurl{10.1038/s41598-018-38127-9}
\end{barticle}
\endbibitem

\bibitem{cui2018distinct}
\begin{barticle}
\bauthor{\bsnm{Cui}, \binits{K.}},
\bauthor{\bsnm{Ardell}, \binits{C.L.}},
\bauthor{\bsnm{Podolnikova}, \binits{N.P.}},
\bauthor{\bsnm{Yakubenko}, \binits{V.P.}}:
\batitle{Distinct migratory properties of m1, m2, and resident macrophages are regulated by $\alpha$d$\beta$2 and $\alpha$m$\beta$2 integrin-mediated adhesion}.
\bjtitle{Frontiers in immunology}
\bvolume{9},
\bfpage{2650}
(\byear{2018}).
\doiurl{10.3389/fimmu.2018.02650}
\end{barticle}
\endbibitem

\bibitem{kim2018transcriptome}
\begin{barticle}
\bauthor{\bsnm{Kim}, \binits{K.}},
\bauthor{\bsnm{Shim}, \binits{D.}},
\bauthor{\bsnm{Lee}, \binits{J.S.}},
\bauthor{\bsnm{Zaitsev}, \binits{K.}},
\bauthor{\bsnm{Williams}, \binits{J.W.}},
\bauthor{\bsnm{Kim}, \binits{K.-W.}},
\bauthor{\bsnm{Jang}, \binits{M.-Y.}},
\bauthor{\bsnm{Seok~Jang}, \binits{H.}},
\bauthor{\bsnm{Yun}, \binits{T.J.}},
\bauthor{\bsnm{Lee}, \binits{S.H.}}, \betal:
\batitle{Transcriptome analysis reveals nonfoamy rather than foamy plaque macrophages are proinflammatory in atherosclerotic murine models}.
\bjtitle{Circulation research}
\bvolume{123}(\bissue{10}),
\bfpage{1127}--\blpage{1142}
(\byear{2018}).
\doiurl{10.1161/CIRCRESAHA.118.312804}
\end{barticle}
\endbibitem

\bibitem{celora2023spatio}
\begin{barticle}
\bauthor{\bsnm{Celora}, \binits{G.L.}},
\bauthor{\bsnm{Byrne}, \binits{H.M.}},
\bauthor{\bsnm{Kevrekidis}, \binits{P.}}:
\batitle{Spatio-temporal modelling of phenotypic heterogeneity in tumour tissues and its impact on radiotherapy treatment}.
\bjtitle{Journal of Theoretical Biology}
\bvolume{556},
\bfpage{111248}
(\byear{2023}).
\doiurl{10.1016/j.jtbi.2022.111248}
\end{barticle}
\endbibitem

\bibitem{fiandaca2022phenotype}
\begin{barticle}
\bauthor{\bsnm{Fiandaca}, \binits{G.}},
\bauthor{\bsnm{Bernardi}, \binits{S.}},
\bauthor{\bsnm{Scianna}, \binits{M.}},
\bauthor{\bsnm{Delitala}, \binits{M.E.}}:
\batitle{A phenotype-structured model to reproduce the avascular growth of a tumor and its interaction with the surrounding environment}.
\bjtitle{Journal of Theoretical Biology}
\bvolume{535},
\bfpage{110980}
(\byear{2022}).
\doiurl{10.1016/j.jtbi.2021.110980}
\end{barticle}
\endbibitem

\bibitem{pan2022propagation}
\begin{barticle}
\bauthor{\bsnm{Pan}, \binits{Y.}}:
\batitle{Propagation dynamics for an age-structured population model in time-space periodic habitat}.
\bjtitle{Journal of Mathematical Biology}
\bvolume{84}(\bissue{3}),
\bfpage{19}
(\byear{2022}).
\doiurl{10.1007/s00285-022-01721-7}
\end{barticle}
\endbibitem

\bibitem{boulouz2022spatially}
\begin{botherref}
\oauthor{\bsnm{Boulouz}, \binits{A.}}:
A spatially and size-structured population model with unbounded birth process.
Discrete \& Continuous Dynamical Systems-Series B
\textbf{27}(12)
(2022).
\doiurl{10.3934/dcdsb.2022038}
\end{botherref}
\endbibitem

\bibitem{hu2019spatial}
\begin{barticle}
\bauthor{\bsnm{Hu}, \binits{W.}}:
\batitle{Spatial--temporal patterns of a two age structured population model with spatial non-locality}.
\bjtitle{Computers \& Mathematics with Applications}
\bvolume{78}(\bissue{1}),
\bfpage{123}--\blpage{135}
(\byear{2019}).
\doiurl{10.1016/j.camwa.2019.02.030}
\end{barticle}
\endbibitem

\bibitem{liu2015hopf}
\begin{barticle}
\bauthor{\bsnm{Liu}, \binits{Z.}},
\bauthor{\bsnm{Tang}, \binits{H.}},
\bauthor{\bsnm{Magal}, \binits{P.}}:
\batitle{Hopf bifurcation for a spatially and age structured population dynamics model}.
\bjtitle{Discr. Cont. Dyn. Syst. B}
\bvolume{20},
\bfpage{1735}--\blpage{1757}
(\byear{2015}).
\doiurl{10.3934/dcdsb.2015.20.1735}
\end{barticle}
\endbibitem

\bibitem{oorni2021aggregation}
\begin{barticle}
\bauthor{\bsnm{{\"O}{\"o}rni}, \binits{K.}},
\bauthor{\bsnm{Kovanen}, \binits{P.T.}}:
\batitle{Aggregation susceptibility of low-density lipoproteins—a novel modifiable biomarker of cardiovascular risk}.
\bjtitle{Journal of clinical medicine}
\bvolume{10}(\bissue{8}),
\bfpage{1769}
(\byear{2021}).
\doiurl{10.3390/jcm10081769}
\end{barticle}
\endbibitem

\end{thebibliography}


\end{document}